\documentclass[preprint,12pt,authoryear]{elsarticle}
\usepackage[a4paper, top=2cm, bottom=2cm, left=2cm, right=2cm]{geometry}
\usepackage{framed,multirow}\setcitestyle{square,numbers,sort&compress}
\biboptions{numbers,sort&compress}
\usepackage{amssymb}
\usepackage{amsmath}
\usepackage{latexsym}
\usepackage{bm}
\usepackage{amssymb}
\usepackage{subfigure}
\usepackage{booktabs}
\usepackage{url}
\usepackage[colorlinks=true,linkcolor=black,citecolor=blue,urlcolor=blue,]{hyperref}
\usepackage[doipre={doi:~}]{uri}
\usepackage{xcolor}
\usepackage{listings}
\usepackage{threeparttable}

\begin{document}

\begin{frontmatter}
\title{Generalized Gauss-Jacobi rules for discrete velocity method in Multiscale Flow Simulations} 

\author[inst1]{Lu Wang}
\ead{wanglu@hdu.edu.cn}

\author[inst1]{Lingyun Deng}
\author[inst1]{Guanqing Wang}

\author[inst1]{Hong Liang\corref{cor1}}
\ead{lianghongstefanie@hdu.edu.cn}

\author[inst2]{Jiangrong Xu\corref{cor1}}
\ead{jrxu@hdu.edu.cn}

\cortext[cor1]{Corresponding author.}
\affiliation[inst1]{organization={Department of Physics, Hangzhou Dianzi University},
	city={Hangzhou},
	postcode={310018},
	country={China}}
\affiliation[inst2]{organization={College of Energy Environment and Safety Engineering, China Jiliang University},
	city={Hangzhou},
	postcode={310018},
	country={China}}
\begin{abstract}
	
The discrete velocity method (DVM) is a powerful framework for simulating gas flows across continuum to rarefied regimes, yet its efficiency remains limited by existing quadrature rules. Conventional infinite-domain quadratures, such as Gauss–Hermite, distribute velocity nodes globally and perform well near equilibrium but fail under strong nonequilibrium conditions. In contrast, finite-interval quadratures, such as Newton–Cotes, enable local refinement but lose efficiency near equilibrium. To overcome these limitations, we propose a generalized Gauss–Jacobi quadrature (GGJQ) for DVM, built upon a new class of adjustable weight functions. This framework systematically constructs one- to three-dimensional quadratures and maps the velocity space into polar or spherical coordinates, enabling flexible and adaptive discretization. The GGJQ accurately captures both near-equilibrium and highly rarefied regimes, as well as low- and high-Mach flows, achieving superior computational efficiency without compromising accuracy. Numerical experiments over a broad range of Knudsen numbers confirm that GGJQ consistently outperforms traditional Newton–Cotes and Gauss–Hermite schemes, offering a robust and efficient quadrature strategy for multiscale kinetic simulations.

\end{abstract}
\end{frontmatter}

\section{Introduction}

The Boltzmann equation provides a fundamental theoretical framework for describing multiscale gas dynamics, with broad applications in micro-electromechanical systems (MEMS) \citep{MEMS}, aerospace engineering \citep{aerospace}, microelectronics cooling \citep{cooling}, and rarefied gas transport under vacuum conditions \citep{vacuum}. Among the numerical strategies developed for its solution, the stochastic approach represented by the Direct Simulation Monte Carlo (DSMC) method \citep{DSMC1, DSMC2} has long been regarded as the benchmark for rarefied gas simulations. By accurately capturing non-equilibrium effects, DSMC performs well in highly rarefied regimes. However, its reliance on statistical sampling introduces fluctuations in macroscopic quantities. Furthermore, as the time step and cell size must remain smaller than the mean collision time and mean free path, respectively, computational efficiency deteriorates rapidly in near-continuum regimes.

Deterministic methods provide an alternative framework for directly solving the Boltzmann equation. A representative example is the discrete velocity method (DVM) or discrete ordinate method (DOM) \citep{DVM1, DVM2, DOM}, which discretizes the molecular velocity space and offers a noise-free, systematic approach for modeling gas transport across a wide range of Knudsen numbers. Within the DVM framework, the kinetic equation is discretized in both physical and velocity spaces, allowing a deterministic description of gas transport processes. Over the years, numerous numerical schemes have been developed to enhance accuracy, stability, and multiscale adaptability. Notable examples include the Improved Discrete Velocity Method (IDVM) \citep{IDVM}, the General Synthetic Iterative Scheme (GSIS) \citep{GSIS}, the Gas Kinetic Unified Algorithm (GKUA) \citep{GKUA}, the Unified Gas-Kinetic Scheme (UGKS) \citep{UGKS1, UGKS2}, and the Discrete Unified Gas-Kinetic Scheme (DUGKS) \citep{DUGKS}.

Among various numerical frameworks, the UGKS has emerged as a well-established deterministic multiscale method, whose core idea lies in coupling molecular transport and collision processes within a single time step, thereby enabling a seamless numerical transition between the continuum and rarefied regimes. The UGKS possesses a rigorous asymptotic-preserving (AP) property \citep{AP}, ensuring smooth transitions across flow regimes without altering the underlying algorithmic structure. Owing to its strong physical fidelity and numerical stability, the UGKS has demonstrated great potential in engineering applications such as MEMS \citep{applicationUGKS1} and spacecraft design \citep{applicationUGKS2}. Building upon this foundation, Guo \emph{et al.} \citep{dugks2013, dugks2015} incorporated the multiscale modeling capability of the UGKS into the Lattice Boltzmann Method (LBM) framework \citep{LBM}, leading to the development of the DUGKS. The DUGKS preserves the AP nature of the UGKS while substantially enhancing computational efficiency through a simplified yet coupled flux evaluation. With its advantageous balance of accuracy, efficiency, and robustness, The DUGKS has become a powerful tool for studying a wide range of problems, including multiphase flows \citep{multiphaseDUGKS}, multi-species rarefied flows \citep{multispeciesDUGKS},  and various multiscale transport phenomena such as phonon heat transfer \citep{phononDUGKS} and radiative transport \citep{radiativeDUGKS}.

Overall, advances within the DVM framework have substantially extended the applicability of kinetic theory and established a solid foundation for unified numerical modeling capable of bridging continuum and rarefied regimes seamlessly. Despite differences in formulation and implementation, all deterministic solvers fundamentally rely on the construction of discrete velocity sets and corresponding quadrature rules. The design of these velocity discretizations directly governs the trade-off among accuracy, stability, and computational efficiency, and thus remains a central concern in the development of DVM-based methods. Depending on the integration domain, existing quadrature rules can generally be categorized into two classes: those defined over infinite or semi-infinite intervals, and those over finite intervals.
The first class includes the Gauss–Hermite \citep{GH1, GH2, GH3}, half-range Gauss–Hermite \citep{HRGH1, HRGH2, HRGH3}, and Gauss–Laguerre quadratures \citep{GL1, GL2, GL3}. Both the Gauss–Hermite and half-range Gauss–Hermite quadratures employ the weight function $\exp(-\xi^2)$, which is highly consistent with the Maxwellian distribution. This intrinsic compatibility makes them particularly advantageous for near-equilibrium flow simulations and explains their widespread use in methods such as the LBM, UGKS, and DUGKS. The Gauss–Laguerre and half-range Gauss–Hermite quadratures are both defined on the interval $[0,+\infty)$, but the Gauss–Laguerre quadrature employs an exponential weight function $\exp(-\xi)$, leading to more dispersed discrete velocity nodes compared to the half-range Gauss–Hermite rule. Ambruş and Sofonea \citep{HRGH2} systematically compared these three types of quadratures and further explored mixed quadrature strategies by combining different rules along distinct velocity directions.
The second class comprises quadrature rules defined on finite intervals, such as the Gauss–Legendre \citep{Legendre}, Gauss–Chebyshev \citep{differentGaussian}, and Gauss–Jacobi \citep{GJ1, GJ2} rules over $[-1,1]$, as well as non-Gaussian schemes like the Newton–Cotes rule \citep{NC1, NC2} defined over arbitrary finite domains $[a,b]$. In theory, particle velocities are distributed over an infinite domain, and thus finite-interval quadratures inevitably introduce truncation errors. Moreover, the determination of an appropriate truncation range is itself a nontrivial task. Hu and Li \citep{Hu}, using the GKUA, conducted a comparative study of the Gauss–Legendre, Gauss–Chebyshev, Newton–Cotes, and Gauss–Hermite rules across flow regimes from free-molecular to continuum limits.

The aforementioned quadrature formulations are primarily constructed in one-dimensional space and are typically extended to higher dimensions via tensor-product combinations. Alternatively, transformations to polar (2D) or spherical (3D) coordinates can be employed \citep{spherical1, spherical2}, in which distinct quadrature rules are applied to the radial, polar, and azimuthal directions. Such coordinate-based approaches have demonstrated superior performance in multiscale simulations from rarefied to continuum regimes. Furthermore, Zhao \emph{et al.} \citep{zhao2020}, Chen \emph{et al.} \citep{Chen2019}, and Yang \emph{et al.} \citep{yang2022} introduced unstructured discretization techniques in velocity space, enabling local refinement in regions where the distribution function exhibits strong gradients.
In general, quadrature rules defined over infinite or semi-infinite intervals produce fixed, globally distributed velocity nodes, which are well suited for near-equilibrium flow simulations but struggle to capture strong nonequilibrium effects. Conversely, finite-interval and unstructured quadratures allow local refinement of discrete velocities, achieving higher efficiency in rarefied flow regimes but becoming inefficient near equilibrium. To reconcile this contradiction, Wang \emph{et al.} \citep{wang2025} proposed a parametric Gauss-type quadrature rule that allows the velocity distribution of an infinite-domain quadrature to be adjusted via tunable parameters. They demonstrated the efficiency and reliability of this approach through simulations ranging from continuum to rarefied regimes. In their formulation, the weight function remains Maxwellian, and the discrete velocities are mapped into polar or spherical coordinates through coordinate transformations. However, the application of this approach to three-dimensional problems faces severe challenges: the Maxwellian weight function introduces logarithmic terms under spherical transformations, making the determination of discrete velocities and corresponding weights exceedingly difficult. Moreover, the flexibility of the discrete velocity distribution remains limited under a fixed weight function.

To further enhance the adaptability of velocity discretization and to overcome the difficulties associated with three-dimensional implementations in \citep{wang2025}, the present study constructs a new class of adjustable weight functions and, based on this foundation, develops a more flexible and easily implementable discrete velocity approach for kinetic modeling. The arrangement of this paper is as follows: Section~\ref{sec2} introduces the BGK–Shakhov model, providing the kinetic framework for our study. Section~\ref{sec3} develops the generalized Gauss–Jacobi quadrature (GGJQ) from one to three dimensions (Sections~\ref{sec3.1}–\ref{sec3.3}) and presents a comprehensive analysis of its properties (Section~\ref{sec3.4}). Section~\ref{sec4} validates the proposed GGJQ using six benchmark problems, including one-dimensional shock-tube and shock-structure tests, two-dimensional thermally driven cavity and supersonic cylinder flows, and three-dimensional lid-driven cavity and spherical Fourier flows, thus demonstrating the accuracy and versatility of the GGJQ in simulating multiscale gas-dynamic phenomena.

\section{BGK-Shakhov model}
\label{sec2}

The starting point of the discrete velocity method considered in this paper is the Boltzmann equation with the BGK-Shakhov collision model \citep{Shakhov1968}, which describes the time evolution of the particle distribution function. The dimensionless form is expressed as,
\begin{equation}
	\frac{\partial f}{\partial t}+\boldsymbol{\xi }\cdot \nabla f=\frac{f^s-f}{\tau},
	\label{eq:boltzmann}
\end{equation}
where \( f = f\left( \mathbf{x}, \boldsymbol{\xi}, \boldsymbol{\eta}, \boldsymbol{\zeta}, t \right) \) is the distribution function defined over the D-dimensional physical space \( \boldsymbol{x} = (x_1, \dots, x_D) \), microscopic velocity components \( (\boldsymbol{\xi}, \boldsymbol{\eta}, \boldsymbol{\zeta}) \), and time \( t \). Here, \( \boldsymbol{\xi} = (\xi_1, \dots, \xi_D) \) denotes the particle velocity in the phase velocity space, \( \boldsymbol{\eta} = (\xi_{D+1}, \dots, \xi_3) \) accounts for the remaining velocity components (if any), and \( \boldsymbol{\zeta} = (\zeta_1, \dots, \zeta_D) \) represents the internal degrees of freedom. The local mean collision time \( \tau \) is defined as the ratio of the dynamic viscosity \( \mu \) to the local pressure \( p \), i.e., \( \tau = \mu / p \), where the dynamic viscosity follows a temperature-dependent power law: 
\begin{equation}
	\mu = \mu_{\infty} \left( \frac{T}{T_{\infty}} \right)^{\omega_{\infty}},
\end{equation}
with \( T \) and \( T_{\infty} \) denoting the local and reference temperatures, respectively, \( \omega_{\infty} \) the temperature exponent, and \( \mu_0 \) the reference viscosity. The reference viscosity is given by
\begin{equation}
	\mu_0 = \frac{5 (\alpha_0 + 1)(\alpha_0 + 2) \sqrt{\pi}}{4 \alpha_0 (5 - 2 \omega_0)(7 - 2 \omega_0)} Kn,
\end{equation}
where \( \alpha_0 \) and \( \omega_0 \) are coefficients related to the inter-molecular interaction models, and \( Kn \) the Knudsen number evaluated at the reference state.

The post-collision distribution \( f^s \) is defined by the Shakhov model to adjust the Prandtl number \(Pr\):
\begin{equation}
	f^s=f^{eq}\left[ 1+\small{\left( 1-Pr \right) \small{\frac{2\boldsymbol{c}\cdot \boldsymbol{q}}{5pT}}\left( \small{\frac{2c^2+2\eta ^2}{T}}-5 \right)} \right],
	\label{eq:fs}
\end{equation}
\begin{equation}
	f^{eq}=\small{\frac{\small{\rho}}{\left( \pi T \right) ^{{{\left( 3+N \right)}/{2}}}}}\exp \left( -\small{\frac{c^2+\eta ^2+\zeta ^2}{T}} \right),
	\label{eq:feq}
\end{equation}
where \( f^{eq} \) is the local Maxwellian equilibrium distribution, \( \boldsymbol{c} = \boldsymbol{\xi} - \boldsymbol{u} \) is the peculiar velocity, and \( c = |\boldsymbol{c}| \), \( \eta = |\boldsymbol{\eta}| \), \( \zeta = |\boldsymbol{\zeta}| \) are the magnitudes of the corresponding vectors, \( \boldsymbol{u} \) and \( \boldsymbol{q} \) are the macroscopic velocity and heat flux, respectively.

To simplify the treatment of internal degrees of freedom, the following reduced distribution functions are defined:
\begin{equation}
	g_n\left( \mathbf{x}, \boldsymbol{\xi}, t \right) = \int \left( \eta^2 + \zeta^2 \right)^n f\left( \mathbf{x}, \boldsymbol{\xi}, \boldsymbol{\eta}, \boldsymbol{\zeta}, t \right) \, d\boldsymbol{\eta} d\boldsymbol{\zeta},
	\label{eq:gnd}
\end{equation}
where \(n = 0, 1\). By integrating Eq.~(\ref{eq:boltzmann}) with respect to \( \left( \eta^2 + \zeta^2 \right)^n \), the evolution equations for \( g_n \) can be obtained as \citep{ReducedLi, NC2, ReducedYang, GJ2},
\begin{equation}
	\small{\frac{\partial g_n}{\partial t}}+\boldsymbol{\xi }\cdot \nabla g_n=\small{\frac{g_{n}^{s}-g_n}{\tau}},
	\label{eq:gne}
\end{equation}
where the post-collision form \( g_n^s \) is given by
\begin{equation}
	g_{n}^{s}=g^{eq}\left\{ 1+\small{\left( 1-Pr \right) \small{\frac{2\boldsymbol{c}\cdot \boldsymbol{q}}{5pT}}\left[ \small{\frac{2c^2}{T}}-D-2\left( \small{\frac{N}{3-D+N}} \right) ^n \right]} \right\} \left[ \small{\frac{(3-D+N)T}{2}} \right] ^n,
	\label{eq:gns}
\end{equation}
and the equilibrium function \( g^{eq} \) reads
\begin{equation}
	g^{eq}\small{=\frac{\small{\rho}}{\left( \pi T \right) ^{{{D}/{2}}}}}\exp \left( -\small{\frac{c^2}{T}} \right).
	\label{eq:geq}
\end{equation}

Using the simplified notation \(g_c=c^2g_0+g_1\), once the reduced distributions \( g_0 \) and \( g_1 \) are obtained, the macroscopic variables can be integrated as follows:
\begin{equation}
	\rho =\int{g_0d\boldsymbol{\xi }}, ~~
	\rho \boldsymbol{u}=\int{\boldsymbol{\xi }g_0d\boldsymbol{\xi }},~~
	\rho T=\small{\frac{2}{3+N}}\int{g_c d\boldsymbol{\xi }},~~
	\boldsymbol{q}=\frac{1}{2}\int{\boldsymbol{c}g_c d\boldsymbol{\xi }}.
	\label{eq:irutq}
\end{equation}

In the framework of the DVM, the velocity space integrals appearing in Eq.~(\ref{eq:irutq}) are approximated using a suitable numerical quadrature rule. By introducing a discrete set of velocity points $\boldsymbol{\xi}_i$ along with corresponding weights $w_i$, the continuous integrals over the distribution functions can be replaced by weighted summations. This transformation yields the the discrete expressions for macroscopic quantities such as density, momentum, temperature, and heat flux,
\begin{equation}
	\rho =\sum_i{w_ig_0\left( \boldsymbol{\xi }_i \right)},~~
	\rho \boldsymbol{u}=\sum_i{w_i\boldsymbol{\xi }_ig_0\left( \boldsymbol{\xi }_i \right)},~~
	\rho T=\small{\frac{2}{3+N}\sum_i{w_ig_c\left( \boldsymbol{\xi }_i \right)}},~~
	\boldsymbol{q}=\frac{1}{2}\sum_i{w_i\boldsymbol{c}_ig_c\left( \boldsymbol{\xi }_i \right)}.
	\label{eq:srutq}
\end{equation}

Usually, the discrete velocity set is chosen as the set of abscissae corresponding to specific quadrature rules, such as the Gauss-Hermite or Newton-Cotes formulas. In multi-scale simulations, the choice of quadrature rule plays a critical role in determining both computational complexity and numerical accuracy. This issue forms the core of the present investigation.

\section{Generalized Gauss-Jacobi quadrature}
\label{sec3}

We consider the problem of numerical integration over the $D$-dimensional velocity space $\boldsymbol{R}^D$, and aim to construct a generalized Gauss–Jacobi quadrature rule of the following form:
\begin{equation}
	I(F) = \int_{\boldsymbol{R}^D} w(\boldsymbol{\xi}) F(\boldsymbol{\xi}) \, d\boldsymbol{\xi} = \sum_i \mathcal{W}_i F(\boldsymbol{\xi}_i),
	\label{eq:ggjq}
\end{equation}
where \( F(\boldsymbol{\xi}) \) is the integrand, \( w(\boldsymbol{\xi}) \) is a weight function, \( \boldsymbol{\xi}_i \), \( \mathcal{W}_i \) are the quadrature points and corresponding weights to be determined. 

In existing studies, the Gaussian function \( e^{-\frac{\boldsymbol{\xi }^{2}}{T_0}} \) is commonly adopted as the weight function, giving rise to the classical Gauss--Hermite or half-range Gauss--Hermite quadrature rules. 
In contrast to these conventional approaches, the present work introduces a novel class of weight functions that preserve a Gaussian-like profile while offering enhanced flexibility. 
Based on this new weight function, a corresponding quadrature rule is developed to meet the specific requirements of multi-scale flow simulations within the DVM. 
For integration in \( D \)-dimensional space, the proposed weight function is expressed as:
\begin{equation}
	w\left( \boldsymbol{\xi } \right) =\left[ 1-\tanh \left( \chi _{\xi} \right) \right] ^{\beta}\left[ 1+\tanh \left( \chi _{\xi} \right) \right] \left[ \frac{\tanh \left( \chi _{\xi} \right)}{\chi _{\xi}} \right] ^{\small{\frac{D}{2}-1}},
	\label{eq:w_alphaBeta}
\end{equation}
where \( \chi_{\xi} = \frac{\boldsymbol{\xi}^2}{\alpha T_0} \), under the assumption that particle velocities are concentrated in the vicinity of the zero-velocity region, and where \( \alpha \) and \( \beta \) are adjustable parameters satisfying \( \alpha, \beta > 0 \). 

It is clear that Eq.~(\ref{eq:w_alphaBeta}) differs substantially from the Gaussian function; however, we will later show that it remains closely related to the Gaussian form through its limiting behavior. At first glance, the weight function in Eq.~(\ref{eq:w_alphaBeta}) may appear complicated due to the presence of fractional terms. This complexity, however, is deliberately introduced—following the treatment in \citep{wang2025}—to eliminate the logarithmic terms that would otherwise arise after the integral transformation. As a result, the derived quadrature rule attains a remarkably simple and explicit structure, where both the nodes and weights can be directly obtained from standard scientific computing libraries. This design ensures that the proposed method is straightforward to implement and highly efficient in practice.

In the following, we construct quadrature rules in one to three dimensions based on the weight function in Eq.~(\ref{eq:w_alphaBeta}) and examine their fundamental properties.
To enable the application of these quadrature rules over infinite domains, we employ a transformation that maps the integral from an unbounded interval to a finite one. 
To this end, we first introduce a radial function \( R(r) \) and its derivative \( R^{\prime}(r) \), where \( r \in (0, 1) \):
\begin{equation}	
	R(r) =\sqrt{\alpha T_0\mathrm{arc}\tanh \left( r \right)},~~and~~
	R^{\prime}(r) =\small{\frac{\sqrt{\alpha T_0}}{2\left( 1-r^2 \right) \sqrt{\mathrm{arc}\tanh \left( r \right)}}}.
	\label{eq:RII}
\end{equation}

\subsection{One-dimensional GGJQ}
\label{sec3.1}

According to Eqs.~(\ref{eq:w_alphaBeta}), the weight function in the one-dimensional case is given by  
\begin{equation}
	w\left( \xi \right) =\left[ 1-\tanh \left( \chi _{\xi} \right) \right] ^{\beta}\left[ 1+\tanh \left( \chi _{\xi} \right) \right] \left[ \frac{\tanh \left( \chi _{\xi} \right)}{\chi _{\xi}} \right] ^{-\small{\frac{1}{2}}},
	\label{eq:w_alphaBeta1d}
\end{equation}
where \( \chi_{\xi} = \frac{\xi^{2}}{\alpha T_0} \).  
It is evident that \( w(\xi) \) in Eq.~(\ref{eq:w_alphaBeta1d}) is an even function, i.e., \( w(-\xi) = w(\xi) \).  
Therefore, the integral over the entire real line can be equivalently expressed as:
\begin{equation}
	I(F) = \int_{-\infty}^{+\infty} w(\xi) F(\xi) \, d\xi 
	= \int_0^{+\infty} w(\xi) \left[ F(\xi) + F(-\xi) \right] \, d\xi.
	\label{eq:int_F}
\end{equation}

Equation~(\ref{eq:int_F}) transforms the integral into the semi-infinite interval \((0, +\infty)\); however, standard quadrature rules are not directly applicable over this domain. 
Therefore, an additional transformation is required.
For the weight function in Eq.~(\ref{eq:w_alphaBeta1d}), let \(\xi = R(r)\), so that \(d\xi = R^{\prime}(r)\,dr\). 
Substituting \(\xi\) into the expression for \(\chi_{\xi}\) yields
$\chi_ {\xi}=\mathrm{arc}\tanh \left( r \right) $, implying that $\tanh \left (\chi_ {\xi} \right)=r $. Thus, $w\left( \xi \right) d\xi =\frac{\sqrt{\alpha T_0}}{2}r^{\small{-\small{\frac{1}{2}}}}\left( 1-r \right)^{\small{\beta -1}}dr$. As a result, the integral in Eq.~(\ref{eq:int_F}) takes the form:
\begin{equation}
	I\left( F \right) =\small{\frac{\sqrt{\alpha T_0}}{2}}\small{\int_0^1{r^{\small{-\small{\frac{1}{2}}}}\left( 1-r \right) ^{\beta -1}\left[ F\left( r \right) +F\left( -r \right) \right] dr}}=\sum_i{W_i\left[ F\left( r_i \right) +F\left( -r_i \right) \right]}.
	\label{eq:HTF1d}
\end{equation}

In Eq.~(\ref{eq:HTF1d}), \( r_i \) and \( W_i \) denote the abscissae and weights, respectively, of the Gauss--Jacobi quadrature rule defined over the interval \((0, 1)\), corresponding to the weight function  $w(r) = r^{-{\frac{1}{2}}} \left( 1 - r \right)^{\beta - 1}.$  
For the numerical integration of the distribution function \( g_n \) introduced in Section~\ref{sec2}, the abscissae and associated weights obtained from these two quadrature methods are listed in Table~\ref{tab:AandW}.

\subsection{Two-dimensional GGJQ}
\label{sec3.2}

According to Eq.~(\ref{eq:w_alphaBeta}), the corresponding two-dimensional weight function is given by  
\begin{equation}
	w\left( \boldsymbol{\xi } \right) =\left[ 1-\tanh \left( \chi _{\xi} \right) \right] ^{\beta}\left[ 1+\tanh \left( \chi _{\xi} \right) \right],
	\label{eq:w_alphaBeta2d}
\end{equation}
where \(\chi_{\xi} = \frac{\xi_x^2 + \xi_y^2}{\alpha T_0}\).  
An effective approach for constructing Eq.~(\ref{eq:ggjq}) in two-dimensional space is to employ a polar coordinate transformation. Let

\begin{equation}
	\xi _x=R(r)\cos \theta,~~and ~\xi _y=R(r)\sin \theta,
	\label{eq:polat}
\end{equation}
where $r \in (0, 1)$ and $\theta \in (0, 2\pi)$. The Jacobian determinant of this transformation is
\begin{equation}
	J=\left| \frac{\partial \left( \xi _x,\xi _y \right)}{\partial \left( r,\theta \right)} \right|=R(r)R^{\prime}(r).
	\label{eq:jacobi2D}
\end{equation}

When the radial mapping function is defined as in Eq.~(\ref{eq:RII}), we obtain from Section~\ref{sec3.1} that $\chi _{\xi}=\mathrm{arc}\tanh \left( r \right)$ and $\tanh \left (\chi _ {\xi} \right)=r $.  
Accordingly, the associated Jacobian determinant and weight function are given by:

\begin{equation}
	J=\small{\frac{\alpha T_0}{2\left( 1-r^2 \right)}},~~
	w\left( \boldsymbol{\xi } \right) =\left( 1-r \right) ^{\beta}\left( 1+r \right).
	\label{eq:JWII2d}
\end{equation}
In this case, the two-dimensional integral becomes  
\begin{equation}
	I\left( F \right) =\int_{\boldsymbol{R}^2}{w\left( \xi _x,\xi _y \right) F\left( \xi _x,\xi _y \right) d\xi _xd\xi _y}=\small{\frac{\alpha T_0}{2}}\int_0^1{\left( 1-r \right) ^{\beta -1}\left[ \int_0^{2\pi}{F\left( r,\theta \right) d\theta} \right] dr}.
	\label{eq:HTF2D}
\end{equation}

The radial integral in Eq.~(\ref{eq:HTF2D}) is evaluated using the Gauss--Jacobi quadrature over the interval \((0, 1)\), with the corresponding weight function  
$w(r) = \left( 1 - r \right)^{\beta - 1}.$ For the angular component \(\theta\), a periodic trapezoidal rule with \(N_{\theta}\) points is employed. The quadrature points and weights are given by:
\begin{equation}
	\theta _j=\theta _{j,0}+\small{\frac{2j\pi}{N_{\theta}}},~
	w_j=\small{\frac{2\pi}{N_{\theta}}},~
	for~\small{j}=1,2,\cdots ,N_{\theta}.
	\label{eq:ptr}
\end{equation}

\subsection{Three-dimensional GGJQ}
\label{sec3.3}

This section extends the quadrature rule to triple integration.  
The corresponding three-dimensional weight function associated with Eq.~(\ref{eq:w_alphaBeta}) is given by  
\begin{equation}
	w\left( \boldsymbol{\xi } \right) =\left[ 1-\tanh \left( \chi _{\xi} \right) \right] ^{\beta}\left[ 1+\tanh \left( \chi _{\xi} \right) \right] \left[ \frac{\tanh \left( \chi _{\xi} \right)}{\chi _{\xi}} \right] ^{\small{\frac{1}{2}}},
	\label{eq:w_alphaBeta3d}
\end{equation}
where $\chi_{\xi} = \frac{\xi_x^2 + \xi_y^2 + \xi_z^2}{\alpha T_0}.$  
To facilitate integration over an unbounded domain, we apply the spherical coordinate transformation to Eq.~(\ref{eq:ggjq}), defined as follows:  
\begin{equation}
	\xi _x=R(r)\cos \theta \sqrt{1-\varPhi _{\varphi}^{2}},~
	\xi _y=R(r)\sin \theta \sqrt{1-\varPhi _{\varphi}^{2}},~
	\xi _z=R(r)\varPhi _{\varphi},
	\label{eq:sphere}
\end{equation}
where \( r \in (0, 1) \), \( \theta \in (0, 2\pi) \), and \( \varphi \in (-1, 1) \).  
The polar function is defined by \(\varPhi_{\varphi} = \varphi^{\phi}\). Without loss of generality, we assume the parameter \(\phi\) to be an odd integer.  
The Jacobian determinant corresponding to this transformation is given by  
\begin{equation}
	J=\left| \frac{\partial \left( \xi _x,\xi _y,\xi _z \right)}{\partial \left( r,\theta ,\varphi \right)} \right|=R(r)^{2}R^{\prime}(r)\varPhi _{\varphi}^{\prime}.
	\label{eq:jacobi3d}
\end{equation}

When the radial mapping function is given by Eq.~(\ref{eq:RII}), we have \(\chi_{\xi} = \mathrm{arctanh}(r)\), which implies \(\tanh(\chi_{\xi}) = r\).  
Substituting this relation into the Jacobian determinant in Eq.~(\ref{eq:jacobi3d}), the Jacobian and the corresponding weight function are expressed as  
\begin{equation}
	J=\small{\frac{\left(\alpha T_0\right)^{\small{\frac{3}{2}}}}{2\left( 1-r^2 \right)}\sqrt{\mathrm{arc}\tanh \left( r \right)}}\left( \phi \varphi ^{\phi -1} \right),~~
	w\left( \xi \right) =\left( 1-r \right) ^{\beta}\left( 1+r \right) \sqrt{\frac{r}{\mathrm{arc}\tanh \left( r \right)}}.
	\label{eq:JWII3d}
\end{equation}
Accordingly, the original triple integral can be expressed as:
\begin{align}
	I\left( F \right) &=\int_{\boldsymbol{R}^3}{w\left( \xi _x,\xi _y,\xi _z \right) F\left( \xi _x,\xi _y,\xi _z \right) d\xi _xd\xi _yd\xi _z}
	\nonumber \\
	&=\frac{\left(\alpha T_0\right)^{\small{\frac{3}{2}}}}{2}\int_0^1{r^{\frac{1}{2}}\left( 1-r \right) ^{\beta -1}\left\{ \int_0^{2\pi}{\left[ \int_{-1}^1{\phi\varphi ^{\phi -1}F\left( r,\theta ,\varphi \right) d\varphi} \right] d\theta} \right\} dr}.
	\label{eq:HTF3D}
\end{align}

The integration over the angular variable \(\theta\) in Eq.~(\ref{eq:HTF3D}) is performed using the \(N_\theta\)-point periodic trapezoidal rule, as described in Eq.~(\ref{eq:ptr}).  
For the radial direction \(r \in (0, 1)\), the Gauss--Jacobi quadrature rule is applied with the weight function $w(r) = r^{\frac{1}{2}} (1 - r)^{\beta - 1}.$

As discussed earlier, the polar function adopts the form \(\varPhi_\varphi = \varphi^{\phi}\), where \(\phi\) is assumed to be an odd positive integer.  
When \(\phi = 1\), the associated weight function reduces to \(w(\varphi) = 1\), and the integral over \(\varphi \in (-1, 1)\) can be evaluated using the standard Gauss--Legendre quadrature rule.  
For the more general case \(\phi \neq 1\), the corresponding weight function is \(w(\varphi) = \varphi^{\phi - 1}\).  
Since \(\phi\) is odd, this function is even over the symmetric interval \((-1, 1)\), and the integral can be symmetrized as  
\begin{equation}
	\int_{-1}^1 \varphi^{\phi - 1} F(\varphi) \, d\varphi = \int_0^1 \varphi^{\phi - 1} \big[ F(\varphi) + F(-\varphi) \big] \, d\varphi.
	\label{eq:phiint}
\end{equation}

Therefore, for the integration over \(\varphi\), the Gauss--Jacobi quadrature rule on the interval \((0, 1)\) is adopted, with the weight function \(w(\varphi) = \varphi^{\phi - 1}\).

\subsection{Analysis of the GGJQ}
\label{sec3.4}

In this section, we further analyze the properties of the GGJQ, with particular focus on the asymptotic behavior of the proposed weight function as \(\alpha \rightarrow +\infty\).  
Clearly, in this limit, \(\chi_{\xi}\rightarrow 0\).  
Under this condition, the last two components of the weight function exhibit the following limiting behavior:
\[
\text{(I)}~\frac{\tanh \left( \chi _{\xi} \right)}{\chi _{\xi}}\rightarrow 1;~
\text{(II)}~1+\tanh \left( \chi _{\xi} \right) \rightarrow 1.
\]
The first component of the weight function is associated with another parameter \(\beta\).  
If \(\beta\) is asymptotically equivalent to \(\alpha\) in the limit \(\alpha \to +\infty\),  
that is, $\beta = \alpha + c,$ where \(c\) is a finite constant, then the following limit holds:
\[
\text{(III)}~\left[ 1-\tanh \left( \chi _{\xi} \right) \right] ^{\beta}\rightarrow e^{-{\frac{\boldsymbol{\xi }^2}{T_0}}};~
\text{(IV)}~w(\boldsymbol{\xi})\rightarrow e^{-{\frac{\boldsymbol{\xi }^2}{T_0}}}.
\]
While (I) and (II) follow from straightforward analysis, conclusion (III) and (IV) requires more careful examination. As \(\beta = \alpha + c\), we focus on the asymptotic form of the term :
\begin{equation}
	\left[ 1-\tanh \left( \chi _{\xi} \right) \right] ^{\beta}=\small{\left( \frac{1+e^{\small{2\chi _{\xi}}}}{2} \right)}^{-\left( \alpha +c \right)}.
\end{equation}
Expanding the exponential \(e^{2\chi_{\xi}}\) via a Taylor series yields:  
\begin{equation}
	\frac{1+e^{\small{2\chi _{\xi}}}}{2}=1+\chi _{\xi}+o\left( \chi _{\xi}^{2} \right).
	\label{eq:Taylorex}
\end{equation}
By using the identity $\small {\left (1+{\small {\frac {1} {x}}} \right)} ^ x \rightarrow e $, as $x \rightarrow+\infty $, we have 
\begin{equation}
	\small{\left( \frac{1+e^{\small{2\chi _{\xi}}}}{2} \right)}^{-\left( \alpha +c \right)}\approx \small{\left[ \left( 1+\frac{\boldsymbol{\xi }^2}{\alpha T_0} \right) ^{\frac{\alpha T_0}{\boldsymbol{\xi }^2}} \right]}^{-\frac{\boldsymbol{\xi }^2}{T_0}}\left( 1+\frac{\boldsymbol{\xi }^2}{\alpha T_0} \right) ^{-c}\rightarrow e^{-\frac{\boldsymbol{\xi }^2}{T_0}}.
\end{equation}
As a result, the weight function in Eq.~(\ref{eq:w_alphaBeta}) converges to the Gaussian form:  
\begin{equation}
	w(\boldsymbol{\xi}) \rightarrow e^{-{\frac{\boldsymbol{\xi }^2}{T_0}}}, \quad \text{as} \quad \alpha \rightarrow +\infty.
\end{equation}

\begin{figure*}[!th]
	\centering
	\includegraphics[width=8cm,height=6cm]{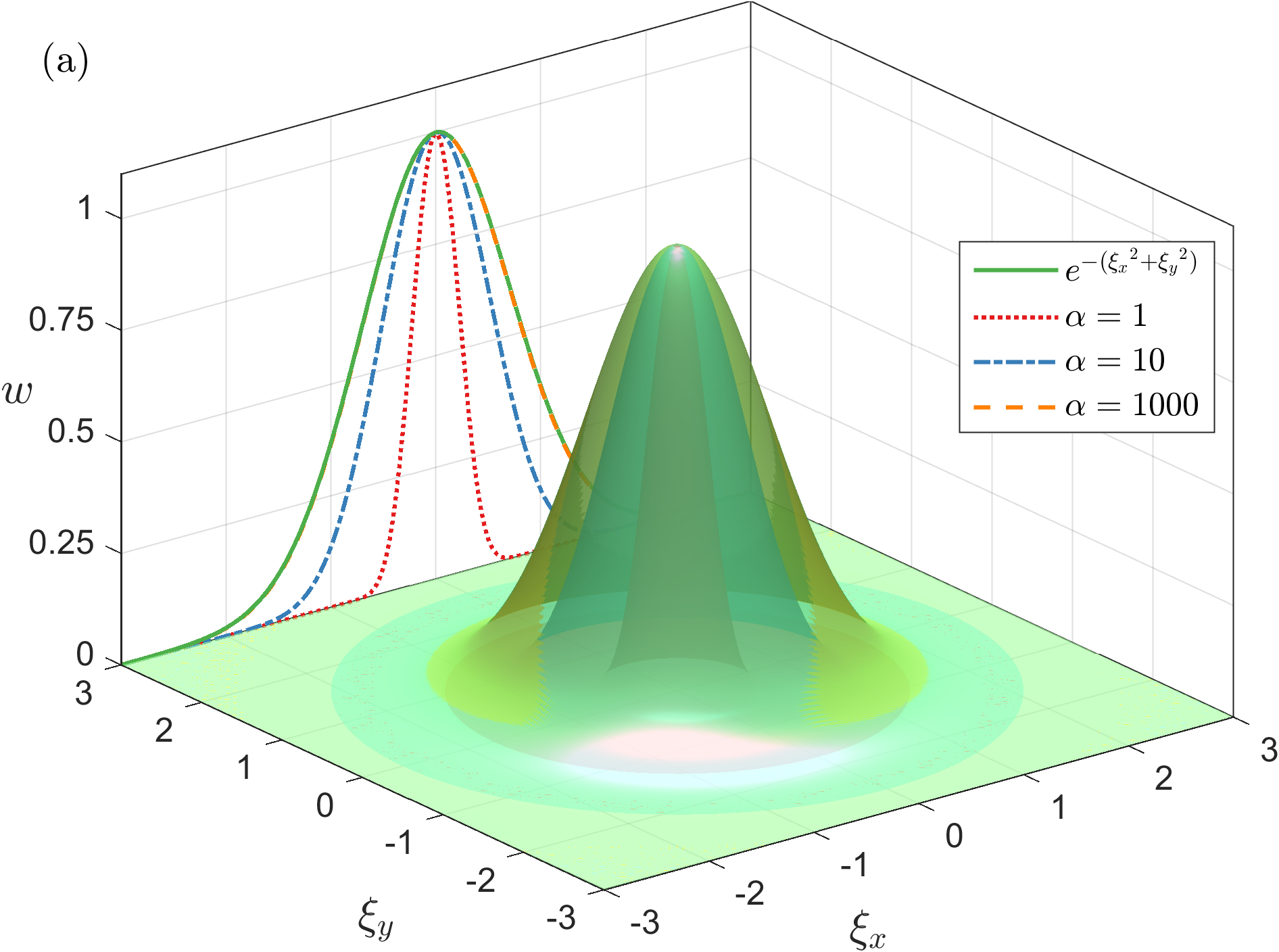}
	\includegraphics[width=8cm,height=6cm]{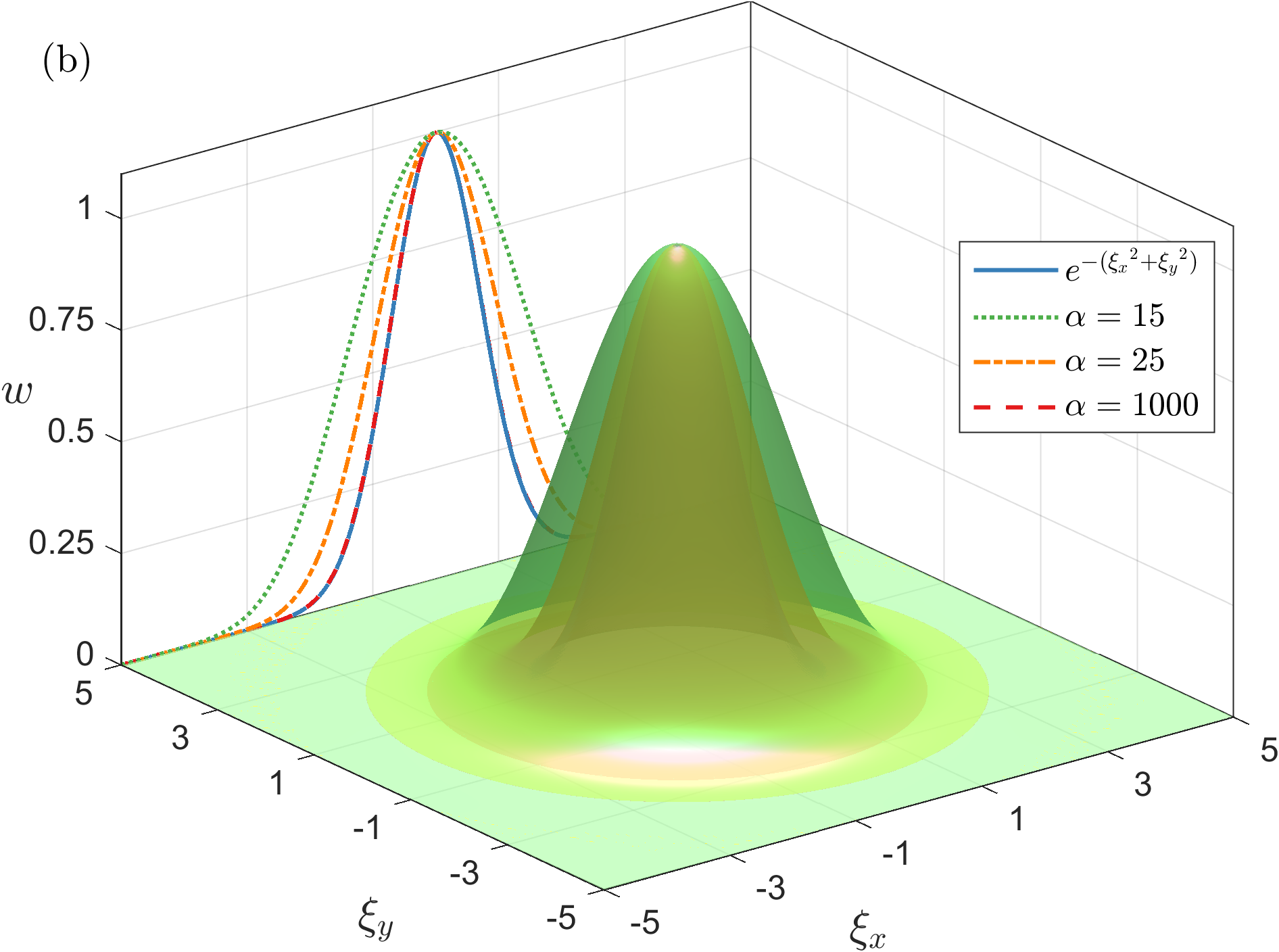}
	\caption{\label{fig:wsurface} \centering  Bell-shaped surfaces and their projections of 2D weight functions. (a)$\beta =\alpha +10$; (b)$\beta =\alpha -10$.}
\end{figure*}

Figure~\ref{fig:wsurface} visualizes the surface distributions of the weight functions under various parameter settings.  
As observed, these functions exhibit bell-shaped profiles.  
With increasing \(\alpha\), the shape of the weight function in Eq.~(\ref{eq:w_alphaBeta2d}) approaches that of the Gaussian distribution.  
When \(\beta > \alpha\), for example \(\beta = \alpha + 10\), the function becomes more peaked and concentrated near the origin.  
Conversely, when \(\beta < \alpha\), such as \(\beta = \alpha - 10\), the function becomes flatter with heavier tails.

This adjustable structure allows the velocity distribution to be tailored to the specific requirements of different flow problems.  
In high Knudsen number flows, where particle velocities are concentrated within a narrow region, a smaller \(\alpha\) with \(\beta > \alpha\) is advantageous.  
In contrast, for high Mach number supersonic flows, where velocities span a broader range, a larger \(\alpha\) with \(\beta < \alpha\) is more appropriate. This tunability represents a key strength of the GGJQ formulation.

\section{Numerical tests}
\label{sec4}

In this section, six benchmark test cases are presented to validate the GGJQ proposed in Section~\ref{sec3}, with the DUGKS employed for numerical implementation. The shock tube and shock structure problems are employed to assess the accuracy of the one-dimensional (1D) GGJQ in capturing flows with varying Knudsen and Mach numbers.  
The performance of the two-dimensional (2D) GGJQ is examined through simulations of thermally driven cavity flow and supersonic flow around a cylinder.  
Finally, the three-dimensional (3D) GGJQ is evaluated using lid-driven cavity flow and spherical Fourier flow.

\subsection{1D shock tube}
\label{sec4.1}

\begin{figure*}[!th]
	\centering
	\includegraphics[width=8cm,height=3.5cm]{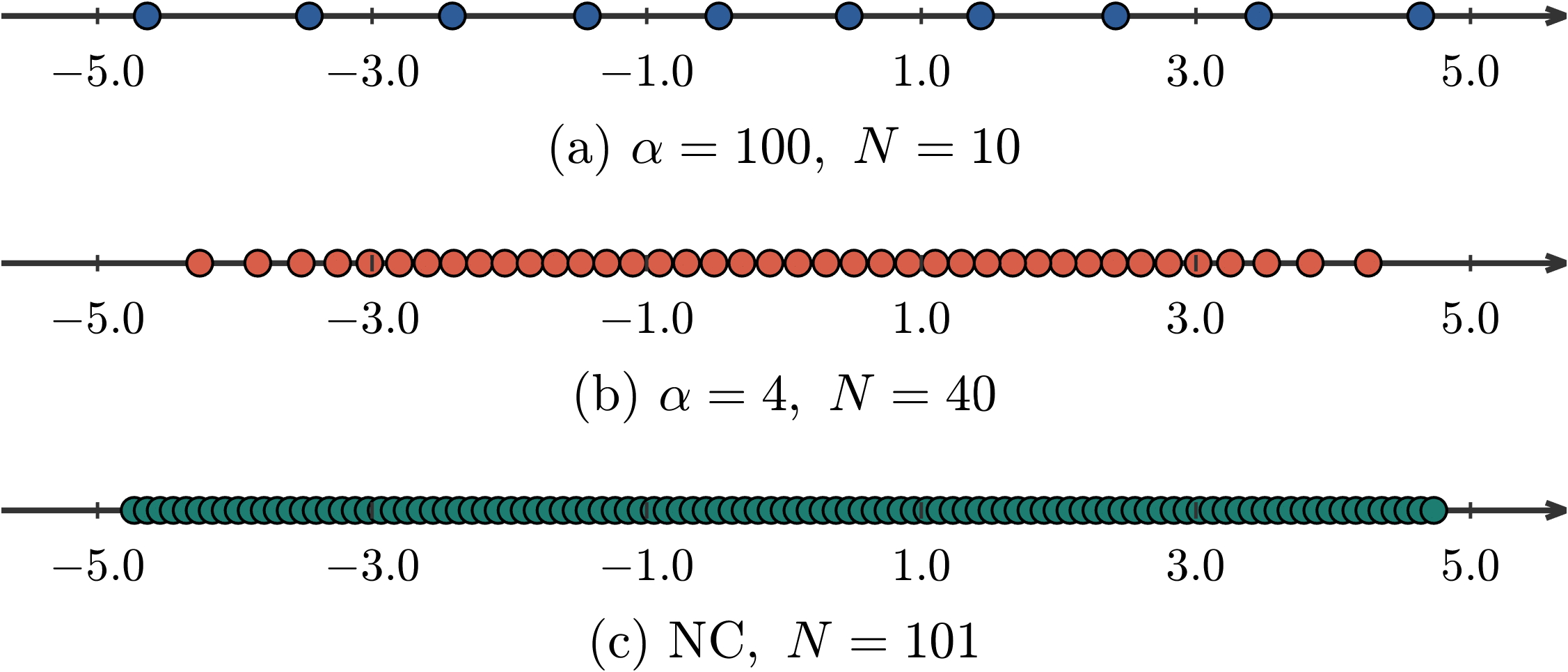}
	\caption{\label{multi_modal_fig} \centering  Illustration of discrete velocity distribution for 1D shock tube.}
\end{figure*}

We first verify the effectiveness of the 1D GGJQ in simulating flows under different Knudsen numbers using the shock tube problem. The initial conditions are given by
\begin{align}
	(\rho_l, u_l, p_l) &= (1.0, 0.0, 1.0), \quad \text{for } x \leq 0, 
	\nonumber \\ 
	(\rho_r, u_r, p_r) &= (0.125, 0.0, 0.1), \quad \text{for } x > 0.
	\label{eq:icShock}
\end{align}

The test configuration follows that in Xu \textit{et al.} \citep{UGKS1}, where the Prandtl number and specific heat ratio are set to \( \text{Pr} = 1 \) and \( \gamma = 1.4 \), respectively. The reference values are chosen as \( (\rho_0, u_0, T_0) = (\rho_l, u_l, T_l) \), and the dynamic viscosity is modeled as \( \mu = \mu_0 (T/T_0)^{0.5} \), where \( \mu_0 \) varies from \(10^{-5}\) to \(10\), covering the full spectrum from continuum to free-molecular regimes.

The spatial domain is discretized using 100 uniform cells, while the velocity space is discretized using both the classical Newton-Cotes (NC) method and the proposed GGJQ. In low Mach number regimes, where particle velocities deviate slightly from the thermal speed, the velocity truncation is set as \( \xi_{\max} = 4\sqrt{\gamma RT_0} \). For the NC rule, 101 discrete velocities points distributed uniformly in \([-\xi_{\max}, \xi_{\max}]\). For the GGJQ, different quadrature parameters are adopted depending on the degree of rarefaction. For rarefied flows (\( \mu_0 = 1 \) and \(10\)), a smaller parameter \( \alpha = \beta = 4 \) with 40 quadrature points are used. For continuum or near-continuum flows (\( \mu_0 = 10^{-3} \) and \(10^{-5}\)), a larger parameter \( \alpha = \beta = 100 \) with 10 points are employed. This ensures that the maximum quadrature node approaches \( \xi_{\max} \), as illustrated in Figure~\ref{multi_modal_fig}.

\begin{figure*}[!th]
	\centering
	\includegraphics[width=18cm,height=5cm]{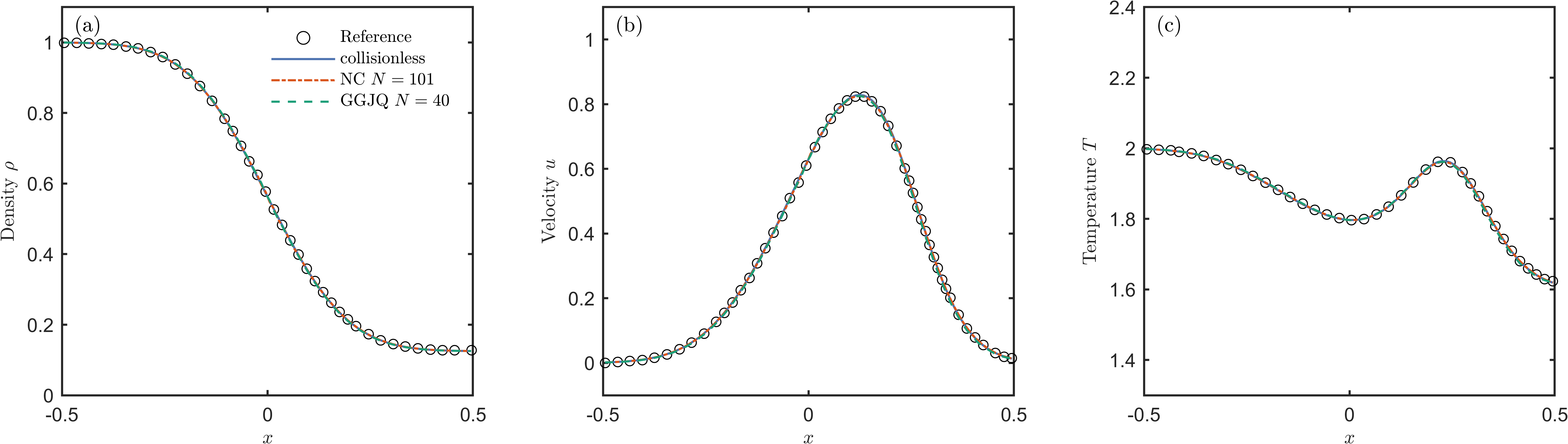}
	\caption{\label{figrho_u_T_10} \centering  Density, velocity, and temperature profiles of the 1D shock tube test ($\mu_0=10$).}
\end{figure*}

\begin{figure*}[!th]
	\centering
	\includegraphics[width=18cm,height=5cm]{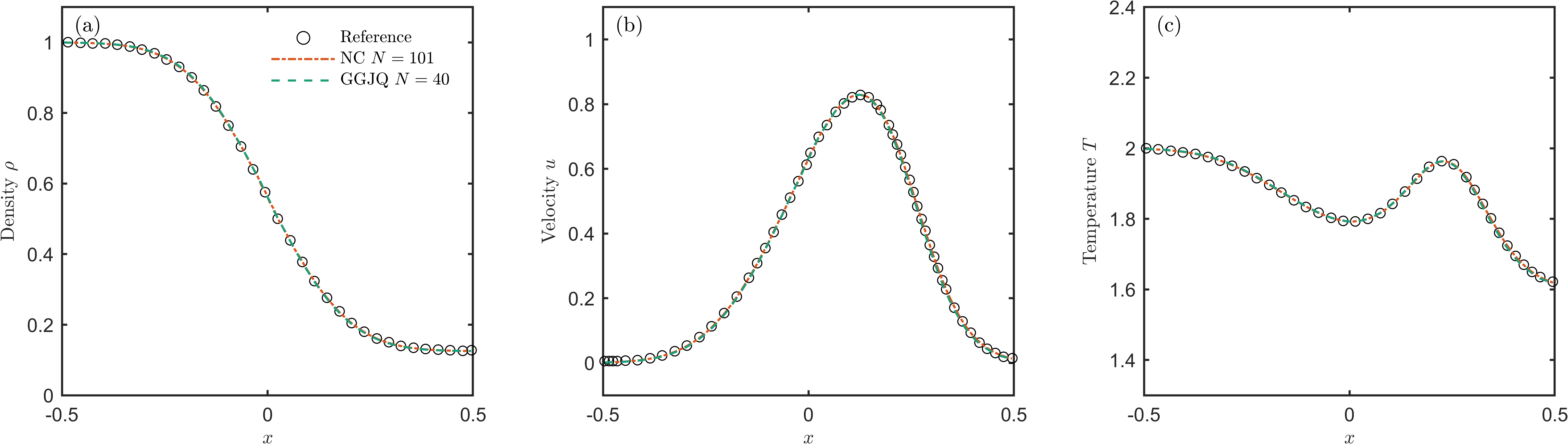}
	\caption{\label{figrho_u_T_1} \centering  Density, velocity, and temperature profiles of the 1D shock tube test ($\mu_0=1$).}
\end{figure*}

\begin{figure*}[!th]
	\centering
	\includegraphics[width=18cm,height=5cm]{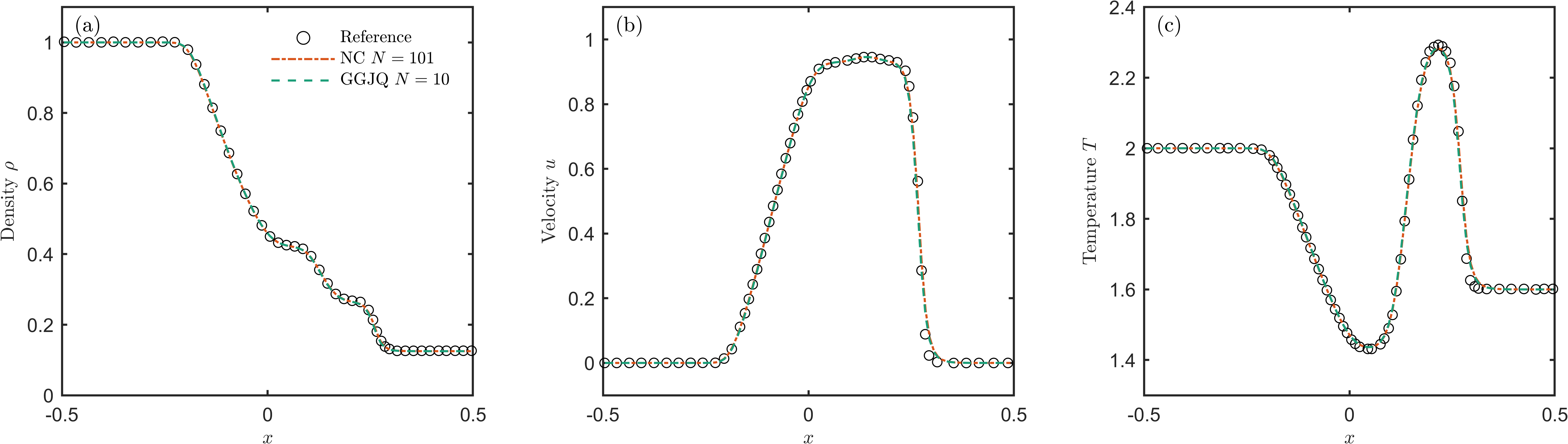}
	\caption{\label{figrho_u_T_E-3} \centering  Density, velocity, and temperature profiles of the 1D shock tube test ($\mu_0=10^{-3}$).}
\end{figure*}

\begin{figure*}[!th]
	\centering
	\includegraphics[width=18cm,height=5cm]{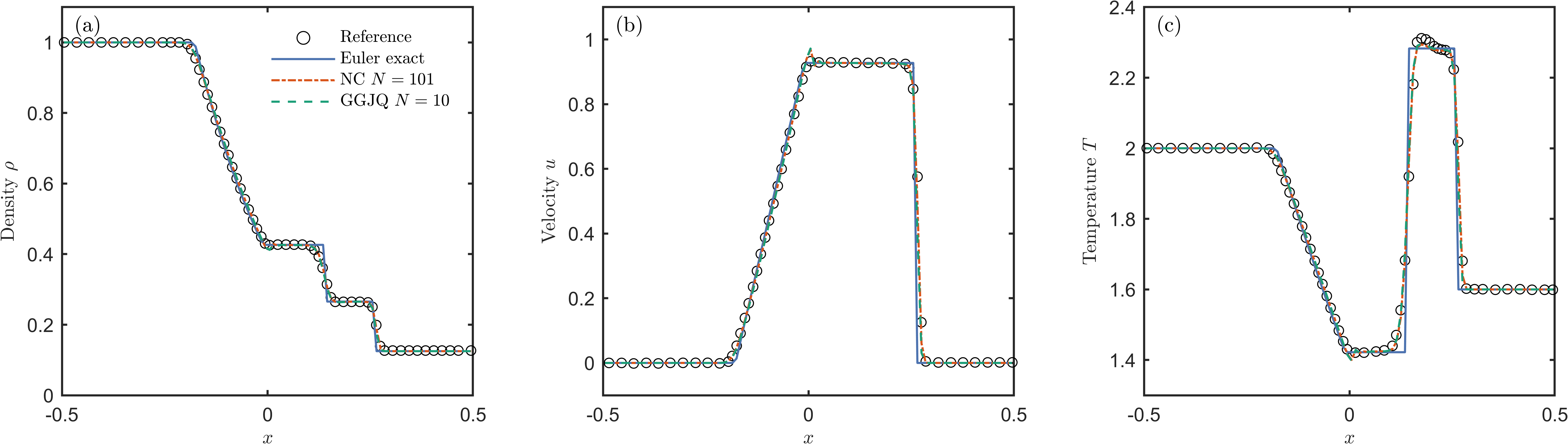}
	\caption{\label{figrho_u_T_E-5} \centering  Density, velocity, and temperature profiles of the 1D shock tube test ($\mu_0=10^{-5}$).}
\end{figure*}

Figure~\ref{figrho_u_T_10} to Figure~\ref{figrho_u_T_E-5} show the numerical results at \( t = 0.15 \). The GGJQ results agree excellently with both the NC method and the reference solution reported by Xu \textit{et al.}, demonstrating the applicability of the proposed method across all flow regimes in shock tube simulations. Moreover, compared with the NC rule, the GGJQ achieves the same accuracy with significantly reduced computational cost and memory usage.

In addition, Table~\ref{tab:comparison} presents a quantitative comparison of the $L_2$ errors for the two methods, using the analytical solution under collisionless conditions and the exact Euler solution as benchmarks. In the rarefied and continuum flow regimes, the GGJQ method requires only \textbf{40\%} and \textbf{10\%} as many nodes as the NC method, respectively.

\begin{table}[htbp]
	\centering
	\caption{ Comparison of numerical $L_2$ errors for 1D shock tube.}
	\label{tab:comparison}
	\begin{tabular}{cccccc}
		\hline
		$\mu_{\mathrm{ref}}$ & Method & Nodes & $E(\rho)$ & $E(u)$ & $E(T)$ \\
		\hline
		\multirow{2}{*}{$10^{-5}$} 
		& NC      & 101 & $1.75\times 10^{-2}$ & $6.94\times 10^{-2}$ & $2.91\times 10^{-2}$ \\
		& GGJQ &  10 & $1.80\times 10^{-2}$ & $9.85\times 10^{-2}$ & $3.11\times 10^{-2}$ \\
		\hline
		\multirow{2}{*}{$10$} 
		& NC      & 101 & $1.99\times 10^{-3}$ & $8.43\times 10^{-3}$ & $1.12\times 10^{-3}$ \\
		& GGJQ &  40 & $2.12\times 10^{-3}$ & $4.86\times 10^{-3}$ & $4.57\times 10^{-4}$ \\
		\hline
	\end{tabular}
\end{table}

\subsection{1D shock structure}
\label{sec4.2}

In this subsection, we assess the performance of the GGJQ scheme in simulating shock structures under different Mach numbers. The computational domain is defined as \( x \in [-25, 25] \), uniformly discretized into 100 cells. The density and temperature ratios across the shock are determined by the following relations:
\[
\frac{\rho_2}{\rho_1} = \frac{(\gamma + 1) Ma^2}{(\gamma - 1) Ma^2 + 2}, \quad
\frac{T_2}{T_1} = \frac{\left(1 + \frac{\gamma - 1}{2} Ma^2\right) \left(\frac{2\gamma}{\gamma - 1} Ma^2 - 1\right)}{Ma^2 \left( \frac{2\gamma}{\gamma - 1} + \frac{\gamma - 1}{2} \right)},
\]
where \( Ma \) is the upstream Mach number. The downstream Mach number is given by
\[
Ma' = \sqrt{\frac{(\gamma - 1) Ma^2 + 2}{2\gamma Ma^2 - (\gamma - 1)}}.
\]

In all cases, the upstream density and temperature are set to \( \rho_1 = 1 \) and \( T_1 = 1 \), while the flow velocity is determined by the specified Mach number \( Ma \). The Prandtl number is fixed at \( \text{Pr} = 1/3 \), and the specific heat ratio is \( \gamma = 5/3 \).
Due to the distinct characteristic velocities and local sound speeds in the upstream and downstream regions, a sufficiently large velocity interval is required to capture the entire distribution function. For reference, Ref.~\cite{dugks2015} employed the Newton–Cotes rule with 101 uniformly spaced nodes over the range \([-15, 15]\). In the present work, the GGJQ method is applied using only 28 nodes, and the results are compared with those obtained from a half-range Gauss–Hermite (GH) quadrature of the same node count. To ensure adequate coverage of the velocity domain, the quadrature parameters of the GGJQ are tuned such that the maximal discrete velocity satisfies \( \xi_{\max} \approx u_{\max} + 4\sqrt{RT} \). Specifically, for \( Ma = 1.2 \), we set \( \alpha = \beta = 20 \), yielding \( \xi_{\max} \approx 4 \); for \( Ma = 3.0 \), \( \alpha = 200 \) and \( \beta = \alpha - 10 \), resulting in \( \xi_{\max} \approx 6 \); and for \( Ma = 8.0 \), \( \alpha = 1000 \) and \( \beta = \alpha - 850 \), giving \( \xi_{\max} \approx 15 \).

\begin{figure*}[!th]
	\centering
	\includegraphics[width=5.5cm,height=4.5cm]{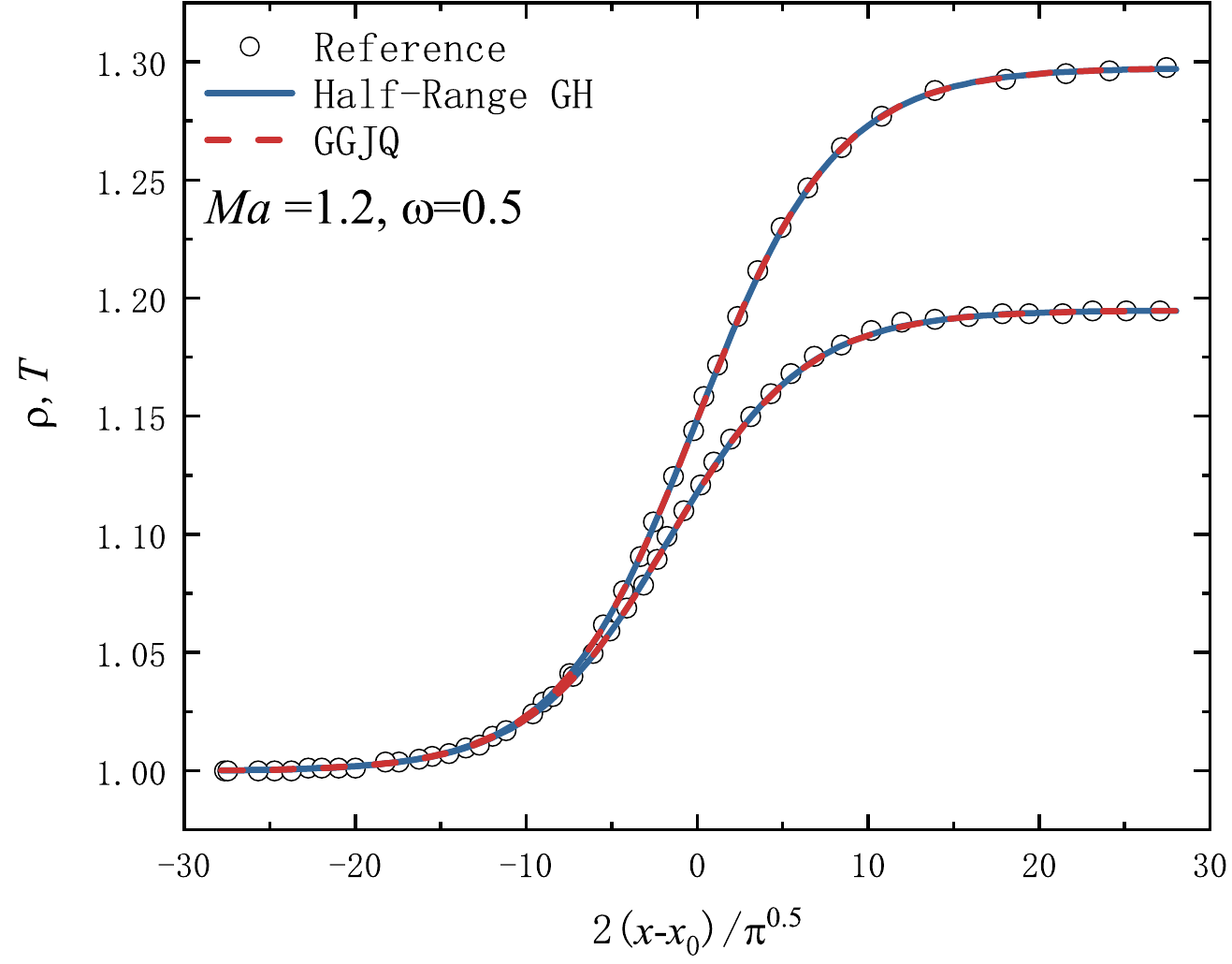}
	\includegraphics[width=5.5cm,height=4.5cm]{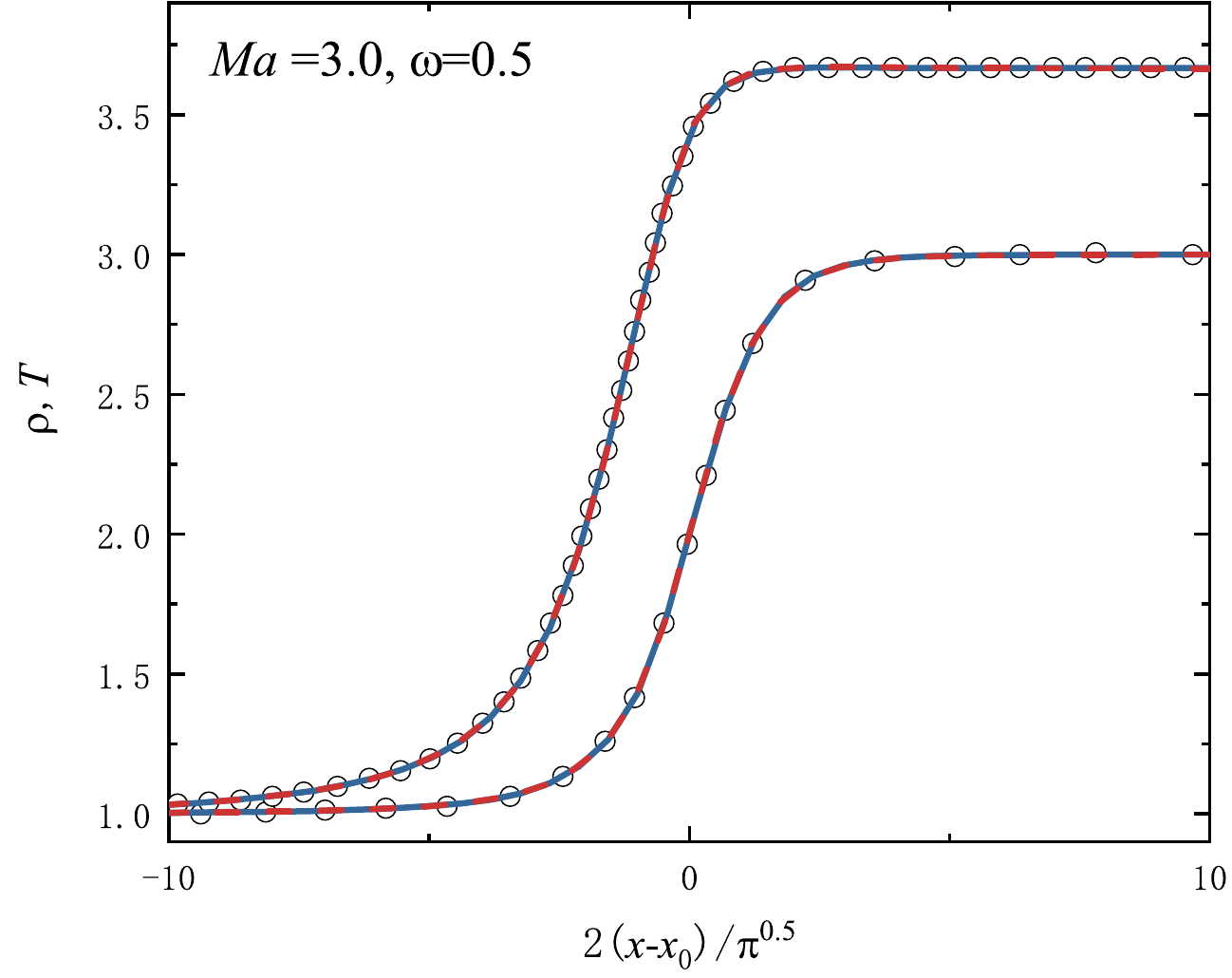}
	\includegraphics[width=5.5cm,height=4.5cm]{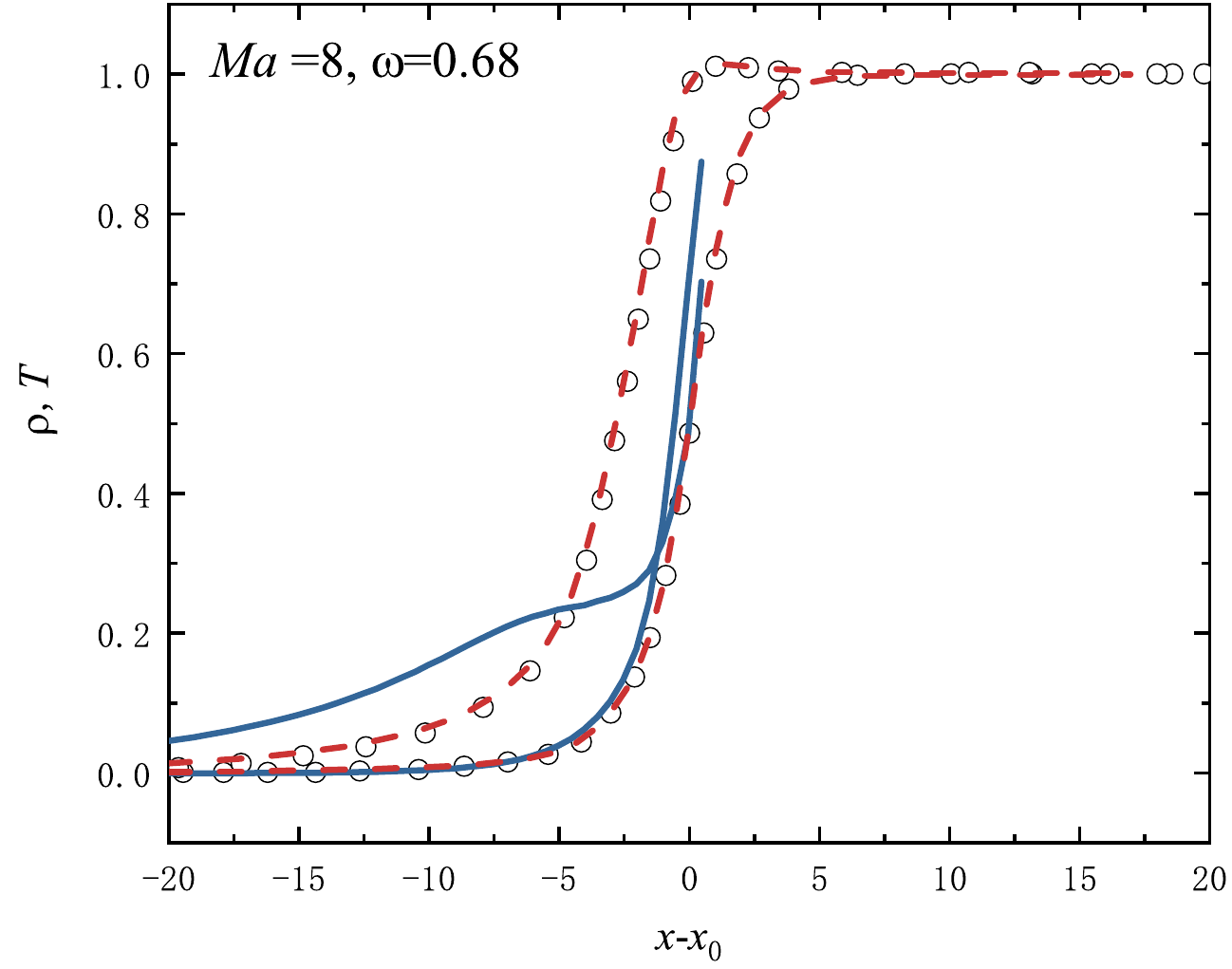}
	\caption{\label{strrho_T} \centering  Density and temperature profiles of the 1D shock structure.}
\end{figure*}

\begin{figure*}[!th]
	\centering
	\includegraphics[width=5.5cm,height=4.5cm]{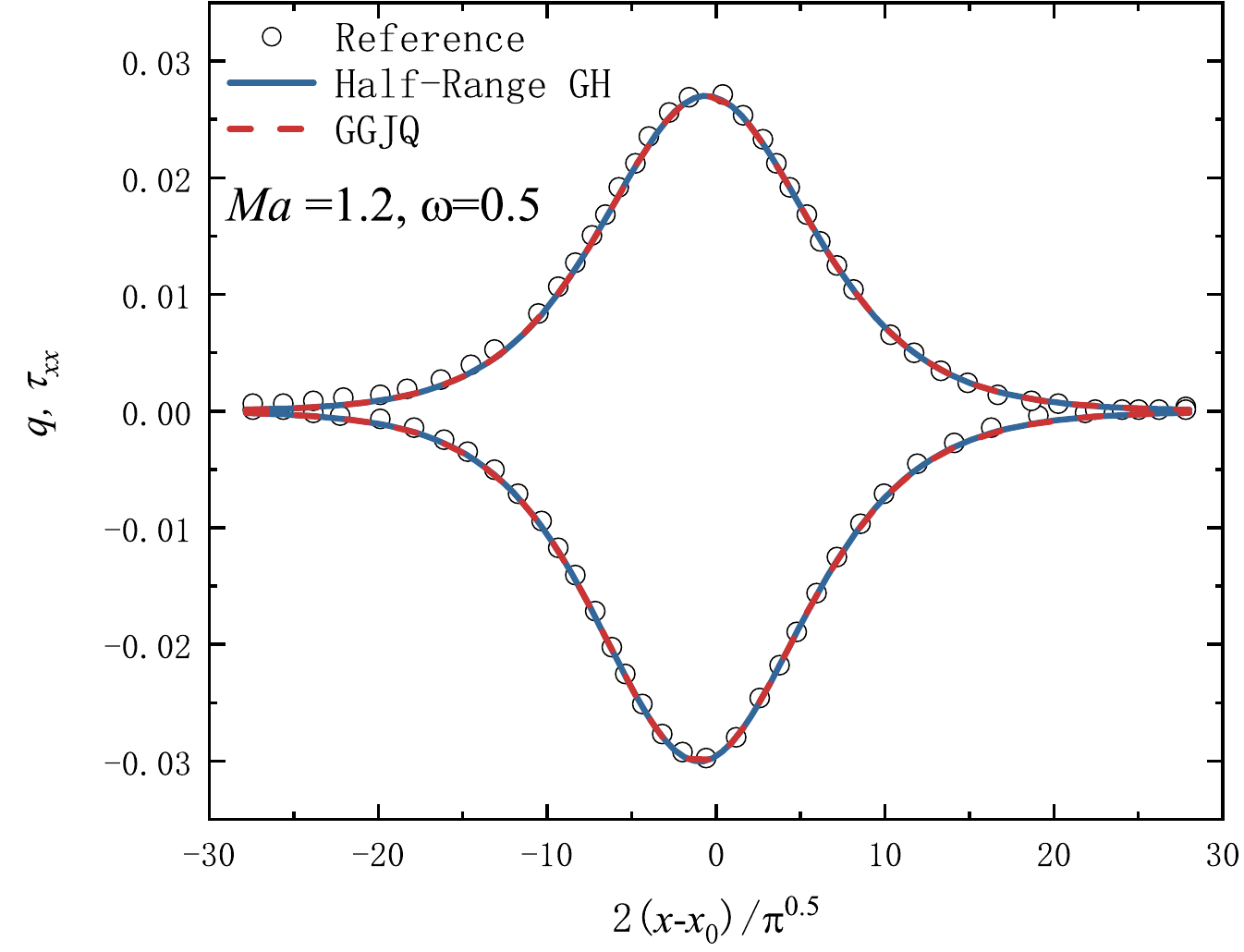}
	\includegraphics[width=5.5cm,height=4.5cm]{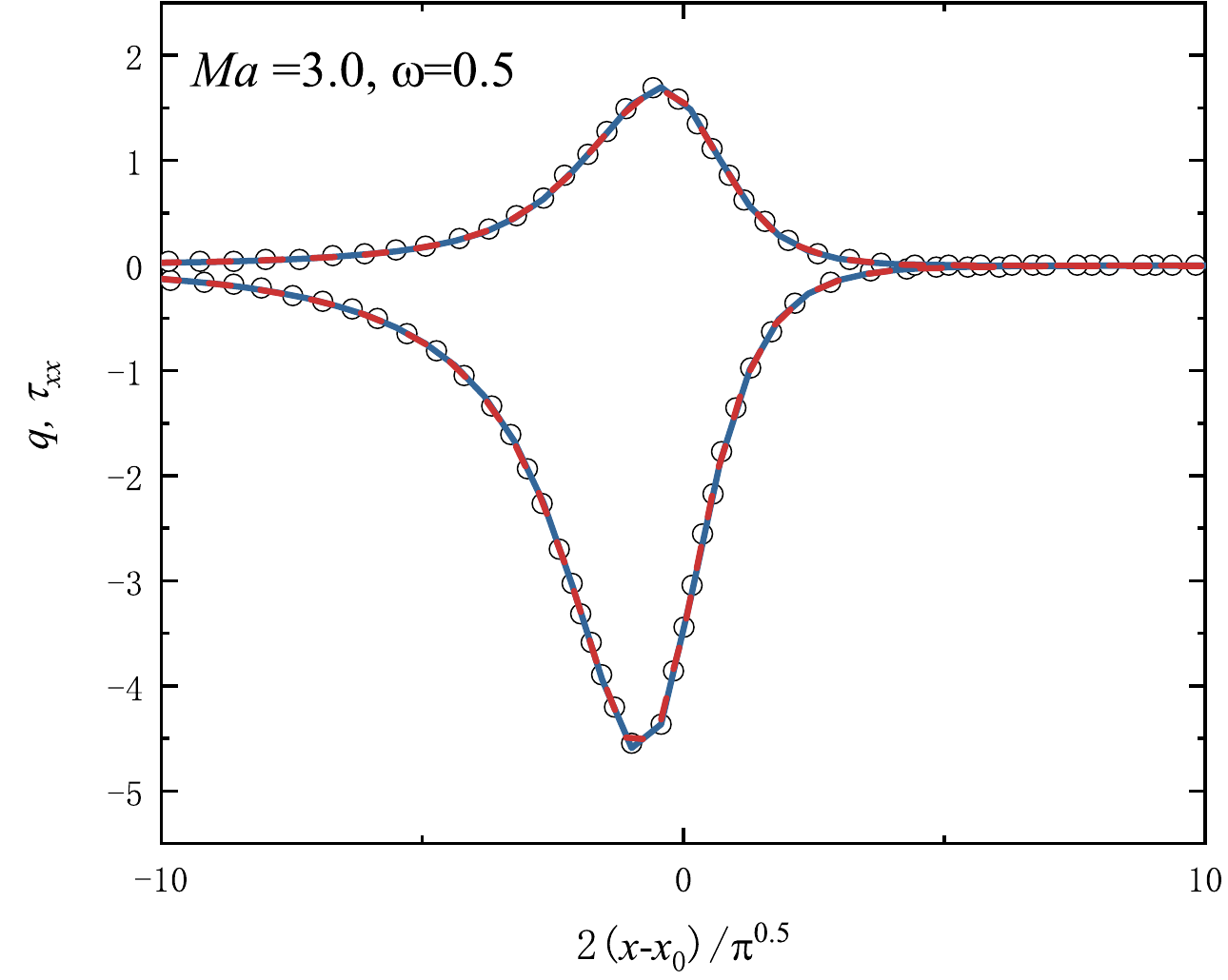}
	\includegraphics[width=5.5cm,height=4.5cm]{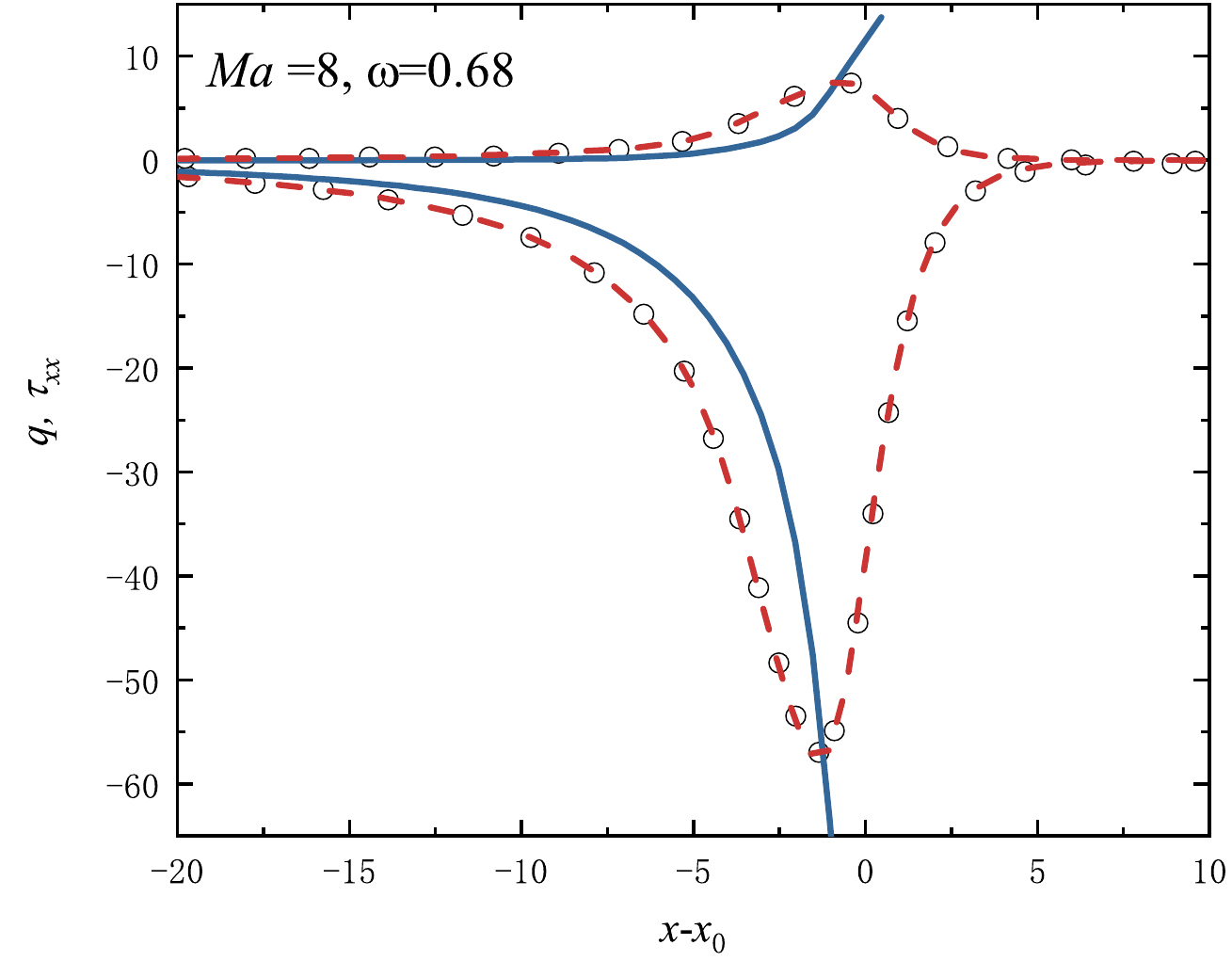}
	\caption{\label{strq_tau} \centering  Heat flux and stress profiles of the 1D shock structure.}
\end{figure*}

Figure~\ref{strrho_T} and Figure~\ref{strq_tau} present the comparison of numerical results for the three Mach numbers. The GGJQ results agree nearly perfectly with the reference data, exhibiting negligible discrepancies. In contrast, the Half-range GH quadrature performs well at low Mach numbers but fails to capture the shock structure accurately at higher Mach numbers. These results demonstrate that, compared with the Half-range GH rule, the GGJQ provides a more robust and efficient approach for simulating shock structures across a wide range of Mach numbers, with significantly lower computational cost.

\subsection{2D thermally driven cavity flow.}
\label{sec4.3}

\begin{figure*}[!th]
	\centering
	\subfigure[Kn=0.001]{
		\label{tdi0.001}
		\includegraphics[width=7cm,height=7cm]{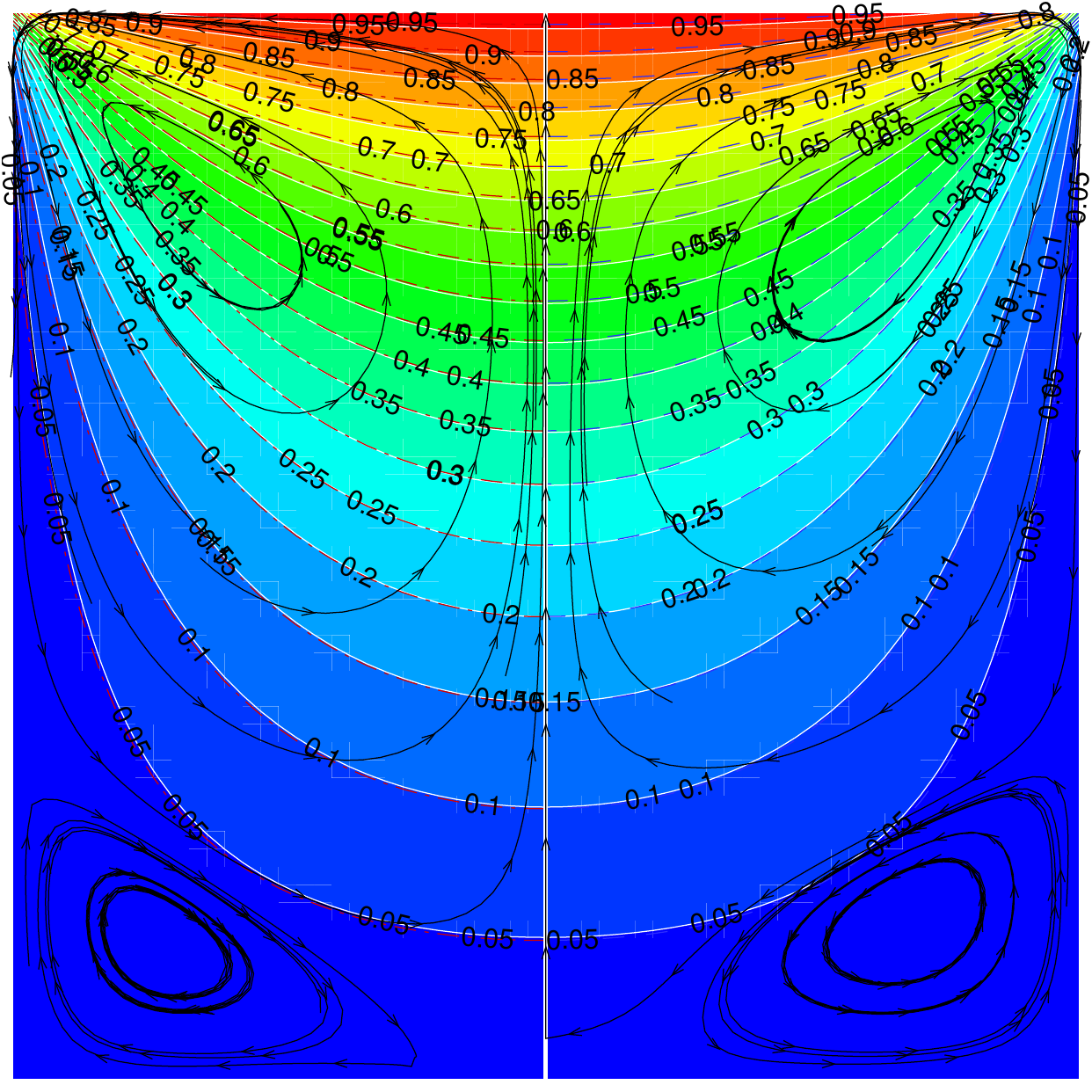}}
	\subfigure[Kn=0.1]{
		\label{tdi0.1}
		\includegraphics[width=7cm,height=7cm]{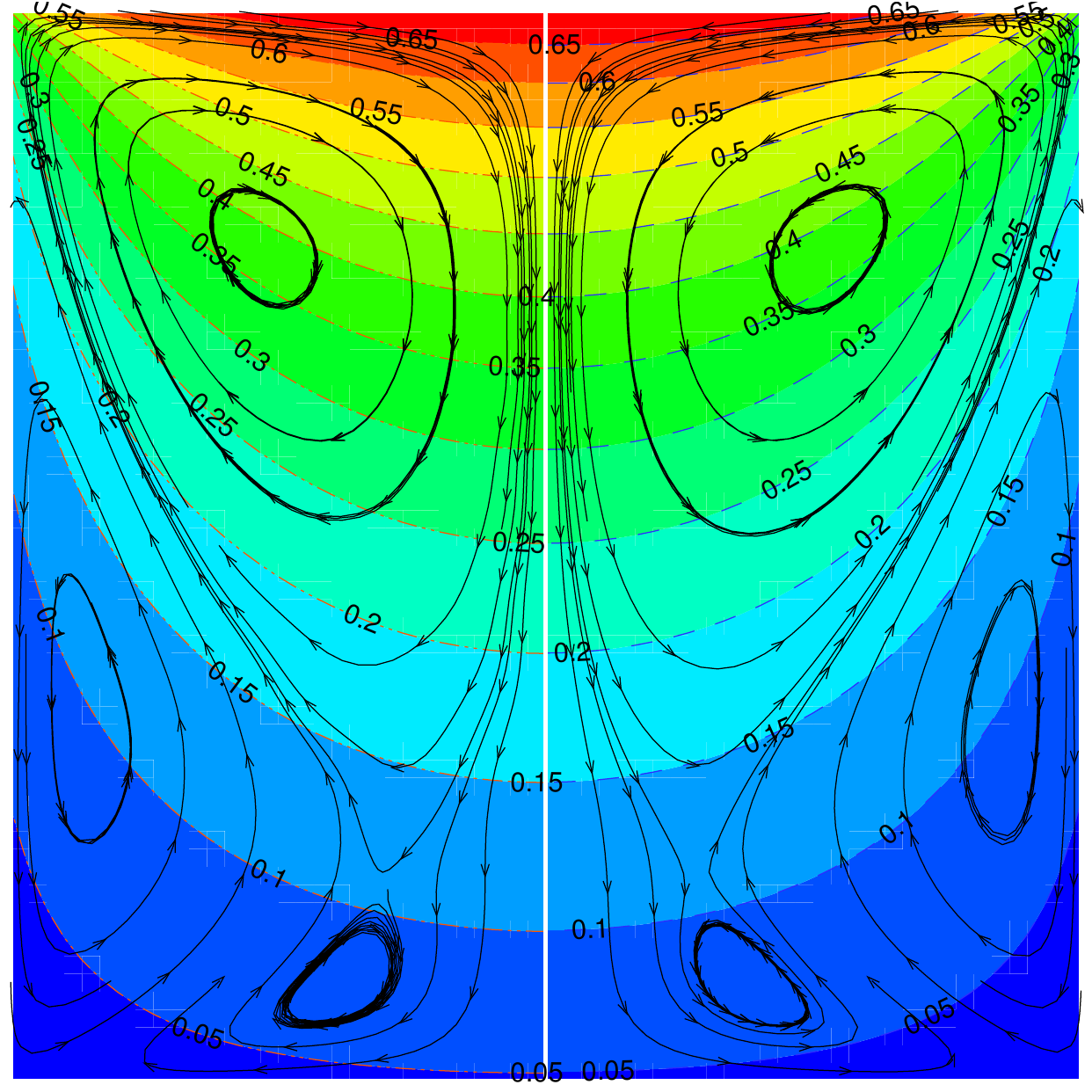}}
	\subfigure[Kn=1.0]{
		\label{tdi1.0}
		\includegraphics[width=7cm,height=7cm]{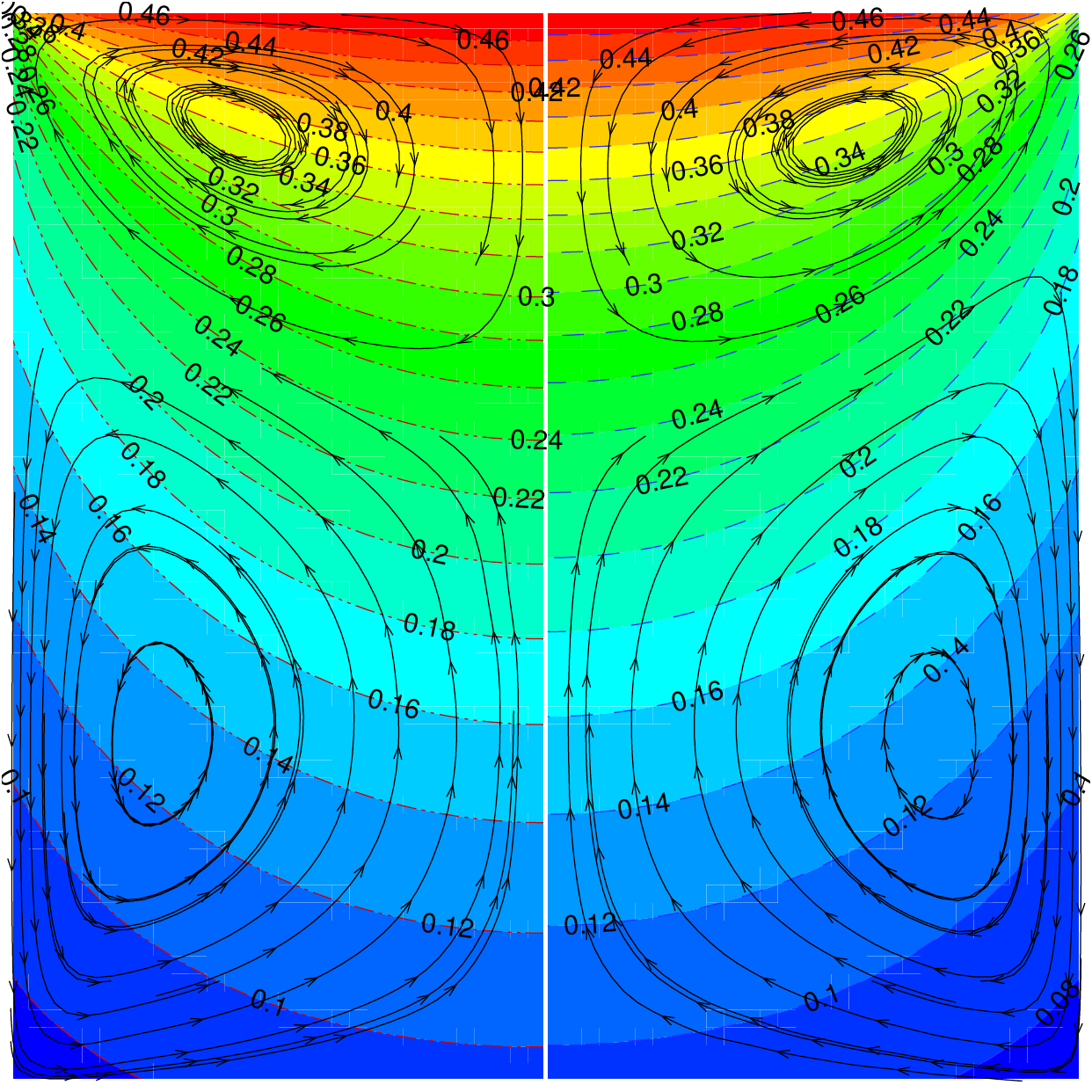}}
	\subfigure[Kn=10.0]{
		\label{tdi10.0}
		\includegraphics[width=7cm,height=7cm]{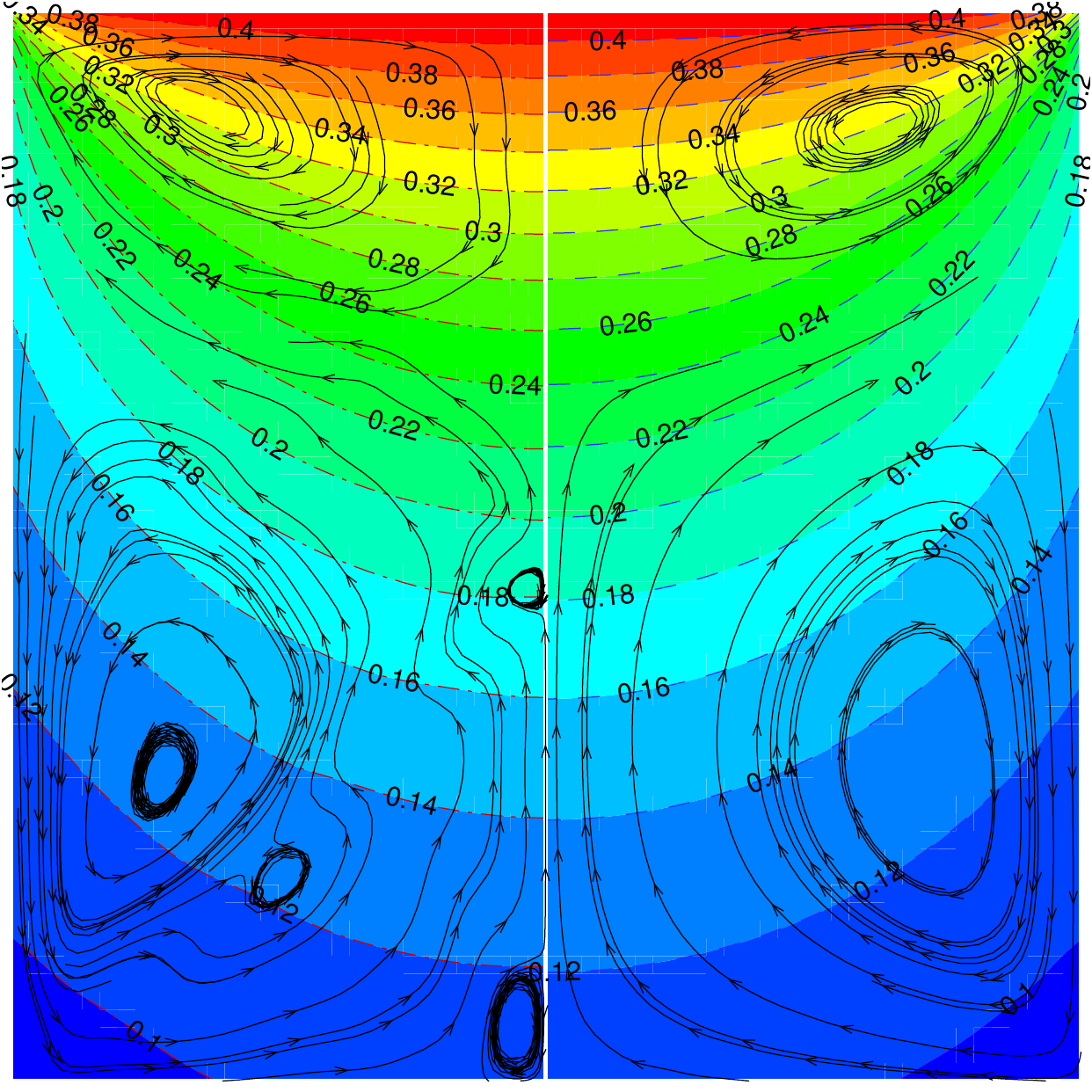}}
	\caption{\label{TCavityTU} \centering  Temperature field and streamlines for the thermally driven cavity flow. In panels (a) and (b), the left half shows the results obtained with the Half-range GH rule, whereas in panels (c) and (d), the left half displays the results obtained with the NC rule. The right half of all panels presents the results computed using the proposed GGJQ rule.}	
\end{figure*}
\begin{figure*}[!th]
	\centering
	\includegraphics[width=7cm,height=5cm]{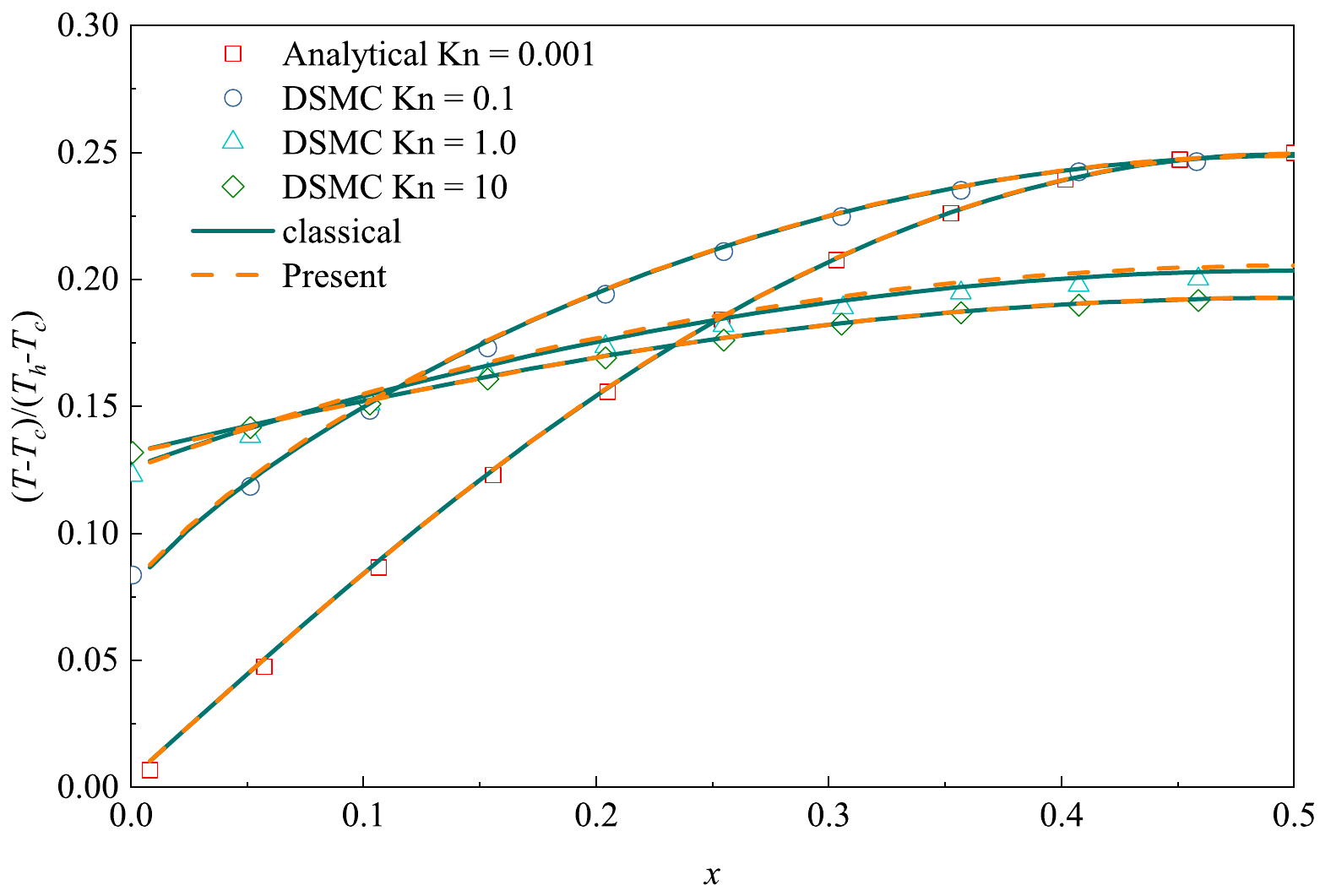}
	\includegraphics[width=7cm,height=5cm]{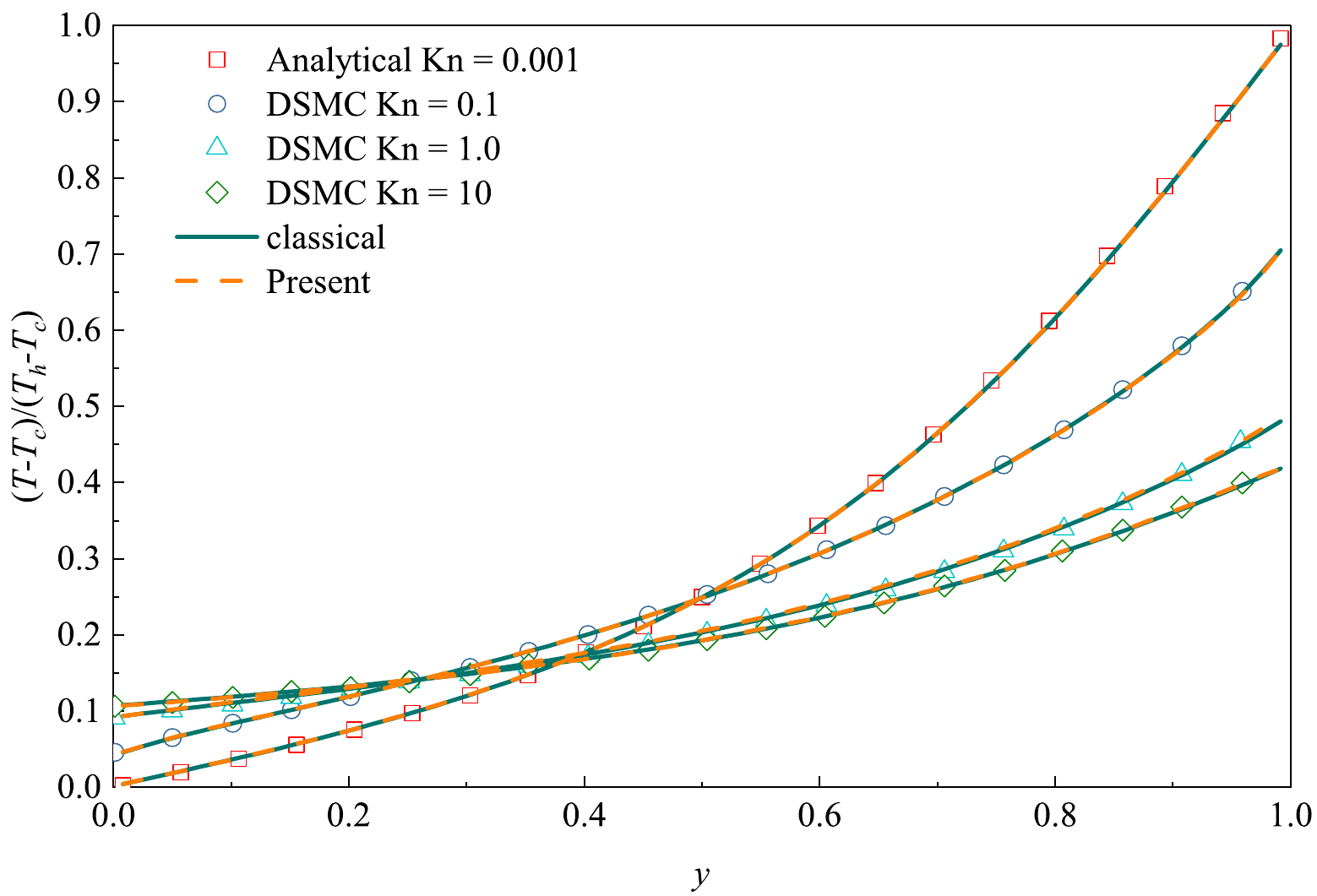}
	\includegraphics[width=7cm,height=5cm]{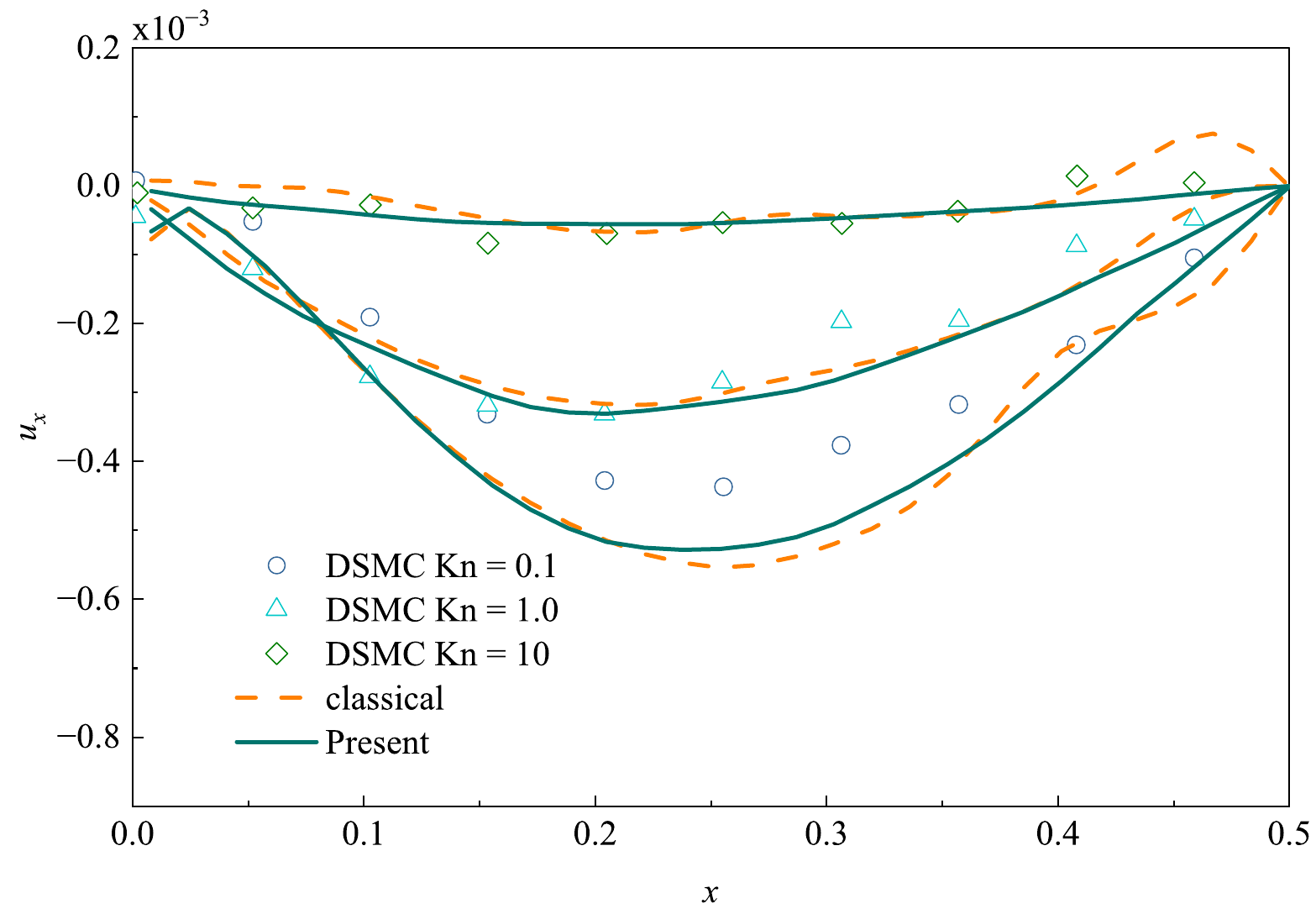}
	\includegraphics[width=7cm,height=5cm]{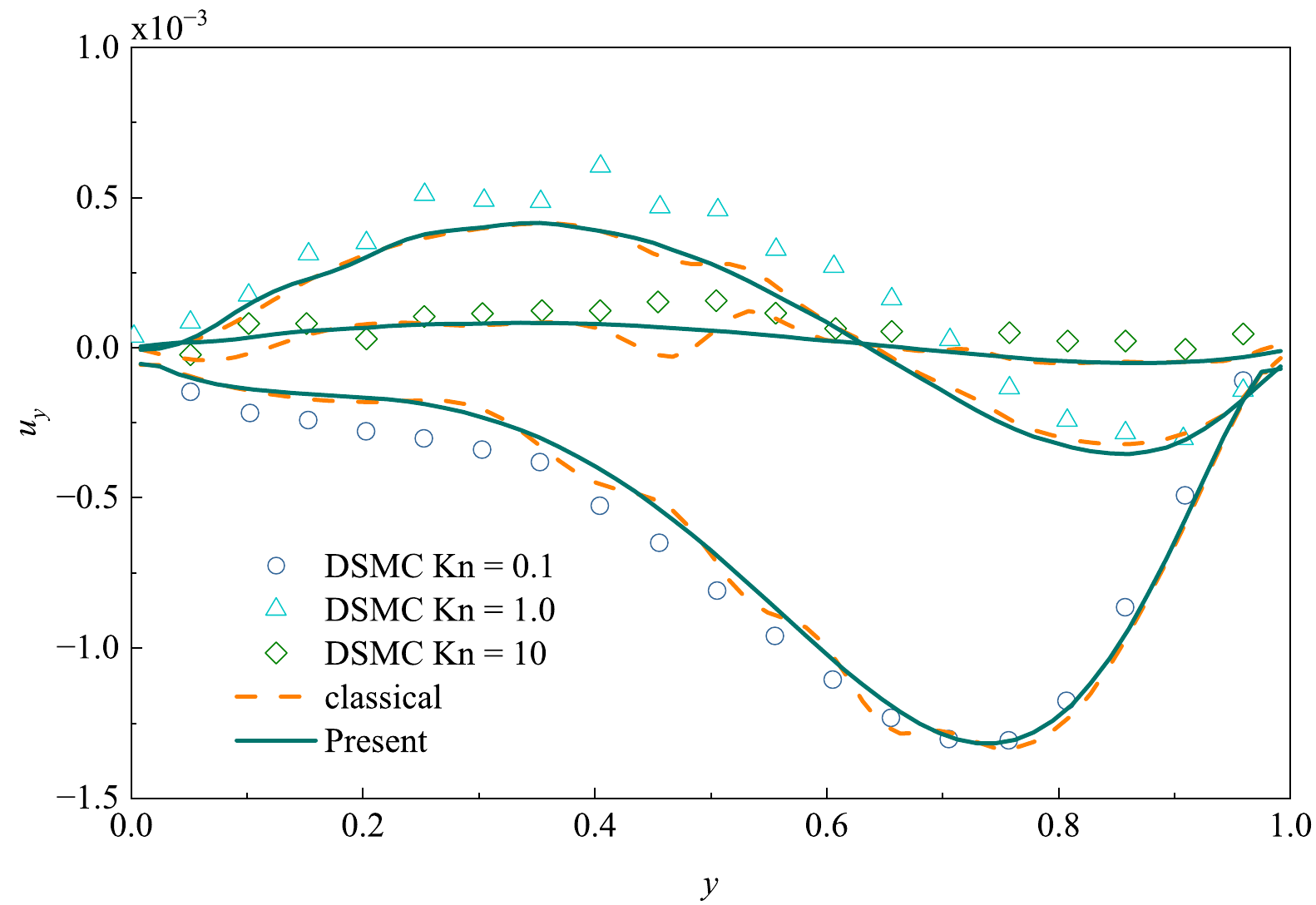}
	\caption{\label{tdiplot} \centering  Temperature and velocity profiles of the thermally driven cavity flow.}
\end{figure*}

\begin{table*}[!th]
	\caption{\label{tdidvm} Velocity discretization settings for the thermally driven cavity flow.}
	\centering
	\begin{tabular}{ccccc}
		\hline
		Kn & 0.001 & 0.1 & 1.0 & 10.0 \\
		\hline
		Classical & $8 \times 8$ half-range GH & $28 \times 28$ half-range GH & $161 \times 161$ NC & $201 \times 201$ NC \\
		Present & $4 \times 4$ GGJQ & $8 \times 45$ GGJQ & $8 \times 90$ GGJQ & $8 \times 120$ GGJQ\\
		Ratio & 4 &2.2 & 36 & 42  \\		
		\hline
	\end{tabular}
\end{table*}

In this section, we follow the approach of Zhu \textit{et al.}~\cite{TDCF} and assess the performance of the proposed 2D GGJQ through the benchmark problem of thermally driven cavity flow. The cavity has side length $L$, with the top wall maintained at $T_h$ and the remaining walls held at $T_c$. All boundaries are treated as fully diffusive thermal walls. For relatively small Knudsen numbers ($Kn=0.001,\,0.1$), Zhu \textit{et al.} employed the Half-range Gauss--Hermite (GH) quadrature, whereas for higher Knudsen numbers ($Kn=1.0,\,10.0$), the Newton--Cotes (NC) rule was utilized. In the present study, the GGJQ is applied consistently across all Knudsen number regimes. 

For $Kn=0.001$, the wall temperatures are set to $T_h=301\,\text{K}$ and $T_c=300\,\text{K}$. In this continuum regime, the normalized temperature, $\theta=(T-T_c)/(T_h-T_c)$, satisfies the analytical solution
\begin{equation}
	\theta(x, y) = \frac{2}{\pi} \sum_{n=1}^{\infty} \frac{(-1)^{n+1}+1}{n} 
	\sin\left(\frac{n \pi x}{L}\right) 
	\frac{\sinh(n \pi y / L)}{\sinh(n \pi)} .
\end{equation}

Since the particle distribution is close to equilibrium, we adopt \( \alpha = \beta = 1000 \) and employ a \( 4 \times 4 \) velocity discretization using the GGJQ, which closely approximates the Gaussian distribution. The \( 4 \times 4 \) half-range Gauss–Hermite (GH) quadrature diverges in this case; therefore, the \( 4 \times 4 \) GGJQ is compared with the \( 8 \times 8 \) half-range GH rule. Both approaches yield comparable accuracy, with \( L_2 \) errors of \( 2.17 \times 10^{-4} \) and \( 2.10 \times 10^{-4} \), respectively. Figure~\ref{tdi0.001} further compares the computed isothermal lines with the analytical solution. Both methods exhibit slight deviations from the exact solution, confirming that the GGJQ achieves nearly identical accuracy to the half-range GH quadrature while requiring only about one quarter of the computational cost.

For $Kn=0.1,\,1.0$, and $10.0$, we follow Zhu’s setup with wall temperatures $T_h=400\,\text{K}$ and $T_c=200\,\text{K}$. The Half-range GH quadrature employs $28\times 28$ discrete velocities, whereas the NC rule uses uniformly spaced velocity grids of $161\times 161$ and $201\times 201$ nodes over $[-4,4]\times[-4,4]$. For GGJQ, we set $\alpha=\beta=40$, yielding a maximum discrete velocity $\xi_{\max}\approx 4$, ensuring fair comparison with the NC rule; details are provided in Table~\ref{tdidvm}. Figures~\ref{tdi0.1}--\ref{tdi10.0} present the results for these Knudsen numbers. For $Kn=0.1$ and $Kn=1.0$, GGJQ produces results comparable to the Half-range GH and NC quadratures. In the free molecular regime ($Kn=10$), however, differences become pronounced. The NC rule relies on an extremely fine velocity grid but its solutions are still contaminated by unphysical vortices. The GGJQ, remarkably, achieves smooth solutions with only \textbf{2.3\%} of the discrete velocities needed by the NC rule.

Figure~\ref{tdiplot} further compares temperature and velocity profiles along the horizontal and vertical centerlines. In terms of temperature, GGJQ shows excellent agreement with both the Half-range GH and NC results. Its advantage is particularly evident in the velocity fields: classical quadrature rules suffer from oscillations that intensify with increasing Knudsen number, requiring increasingly fine velocity discretization. GGJQ effectively suppresses these non-physical fluctuations, providing consistently accurate results across the entire flow regime.

\subsection{2D Supersonic Cylinder flow}
\label{sec4.4}

\begin{figure*}[!th]
	\centering
	\subfigure[]{
		\label{cylindermesh}
		\includegraphics[width=6cm,height=5cm]{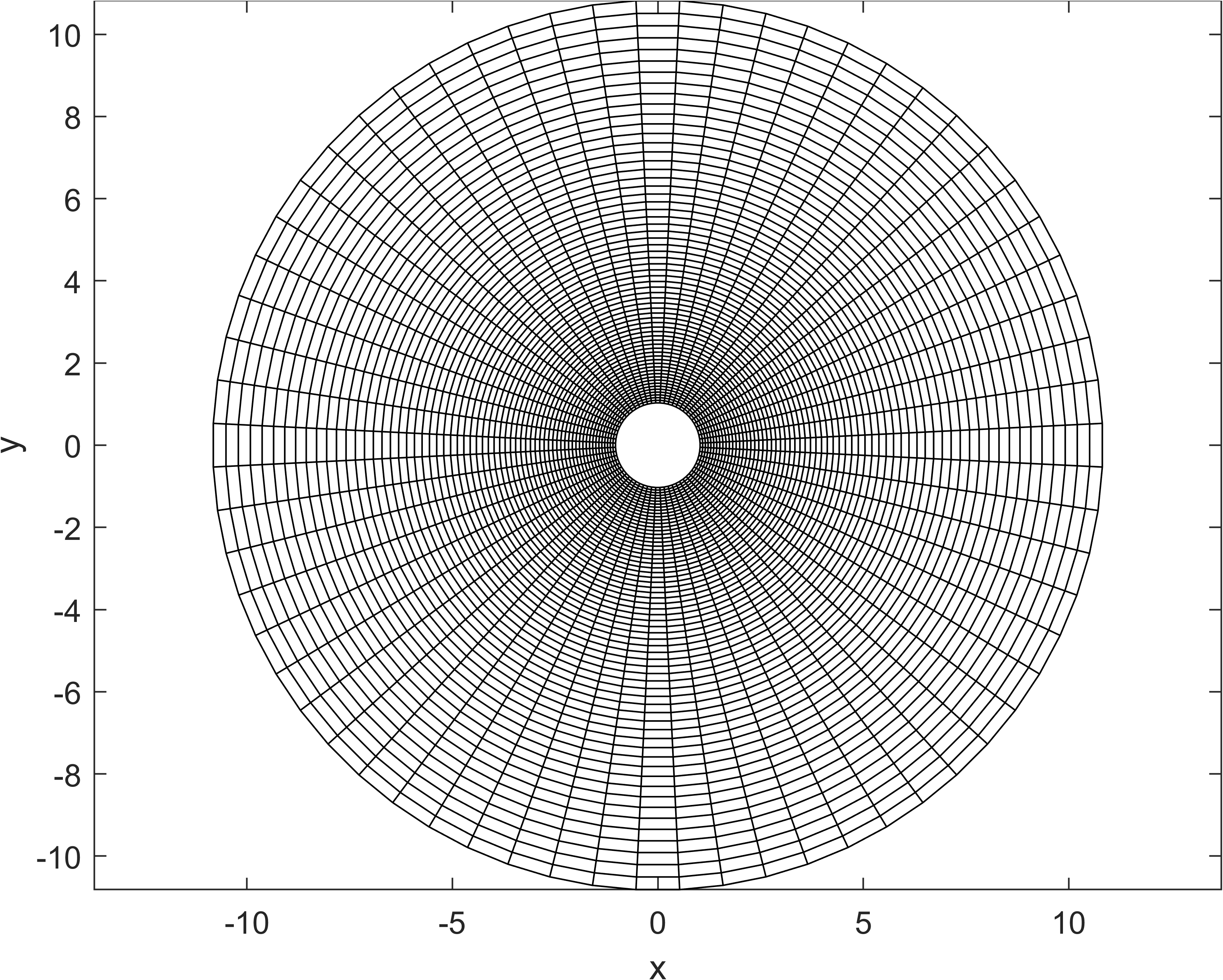}
	}
	\subfigure[]{
		\label{cylinderVelocity}
		\includegraphics[width=6cm,height=5cm]{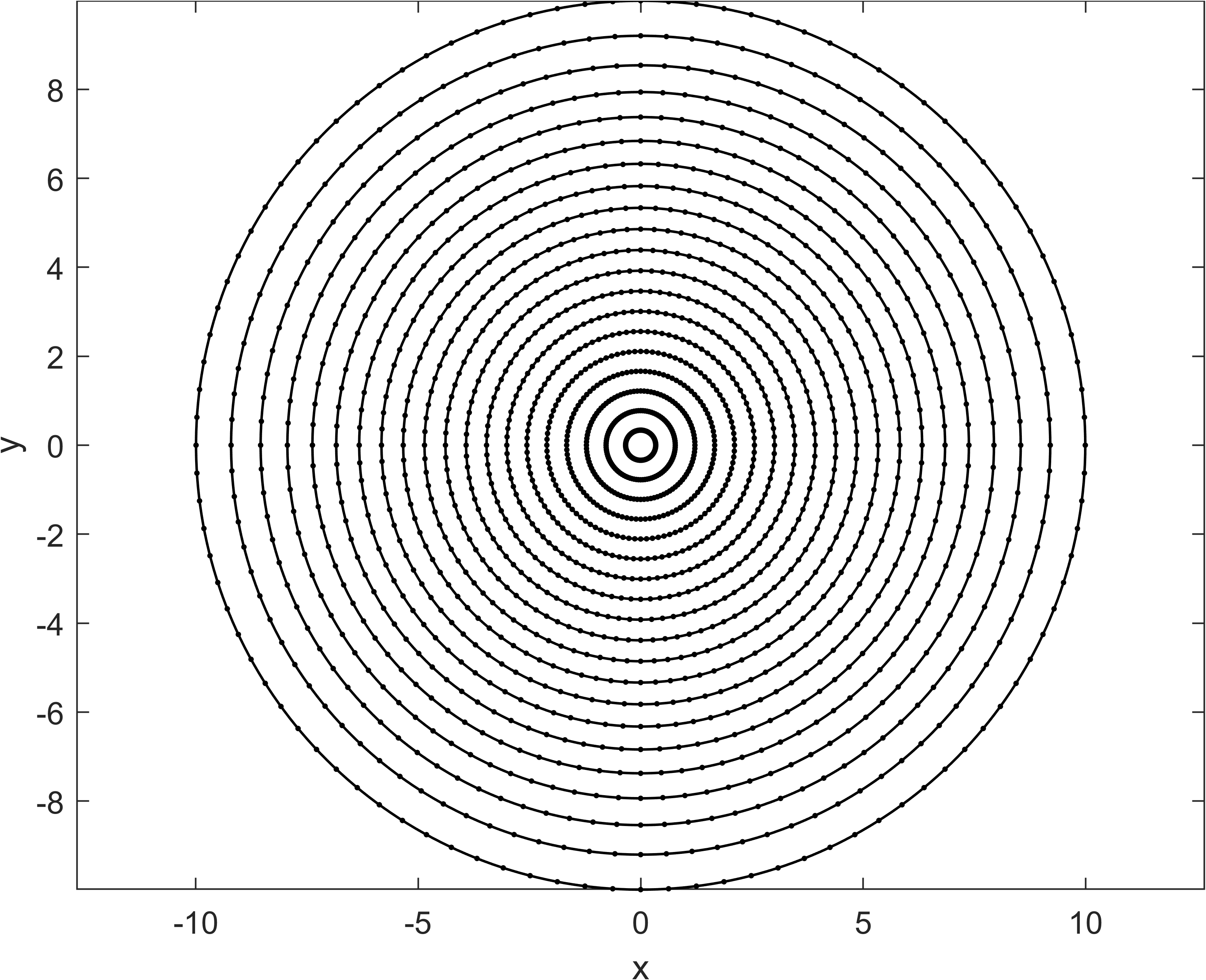}
	}
	\caption{\label{cylindermeshandVelocity} \centering  Computational mesh (a) and GGJQ-based distribution of discrete velocities (b) for 2D supersonic cylinder flow.}
\end{figure*}

\begin{figure*}[!th]
	\centering
	\includegraphics[width=5.5cm,height=4.5cm]{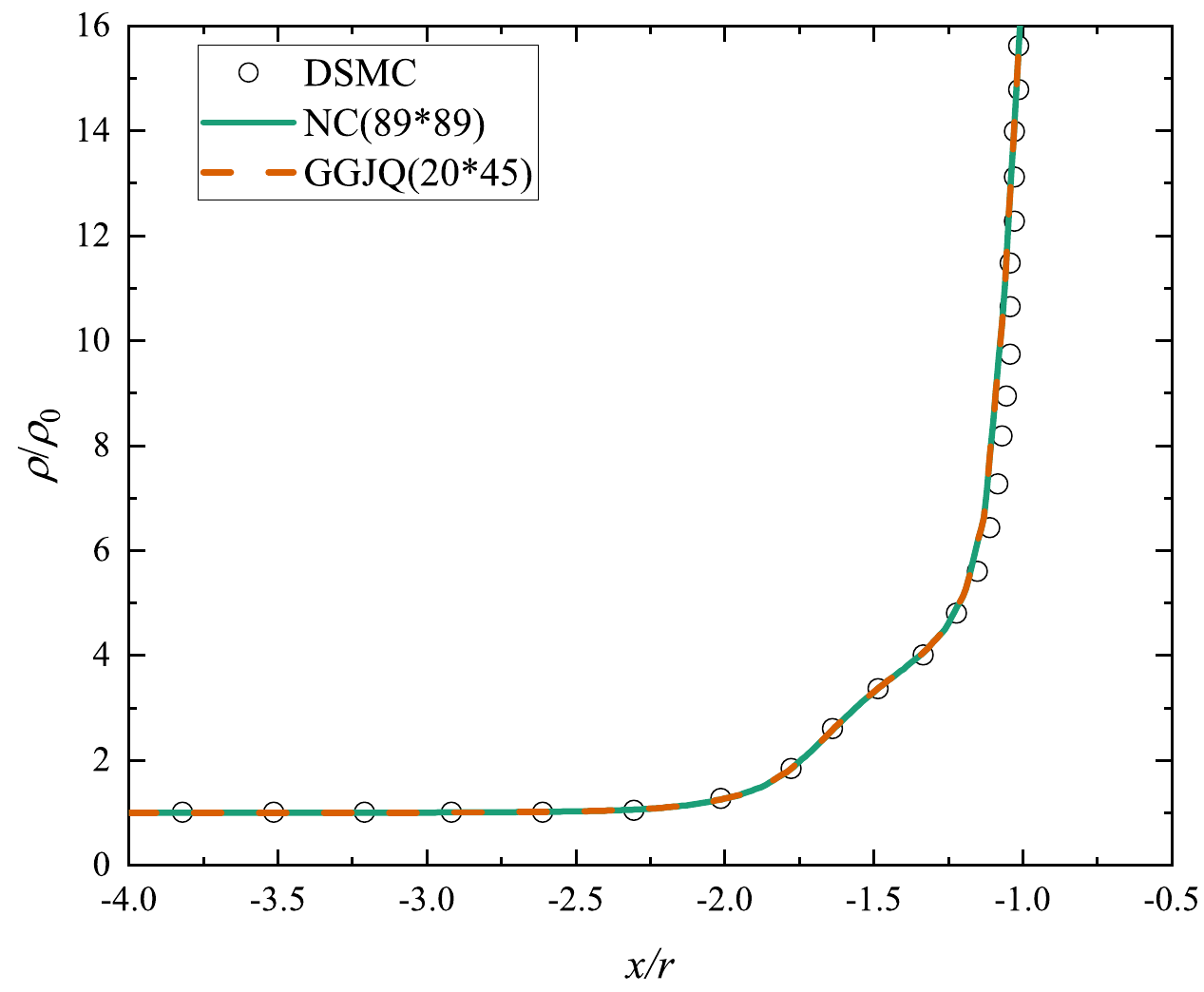}
	\includegraphics[width=5.5cm,height=4.5cm]{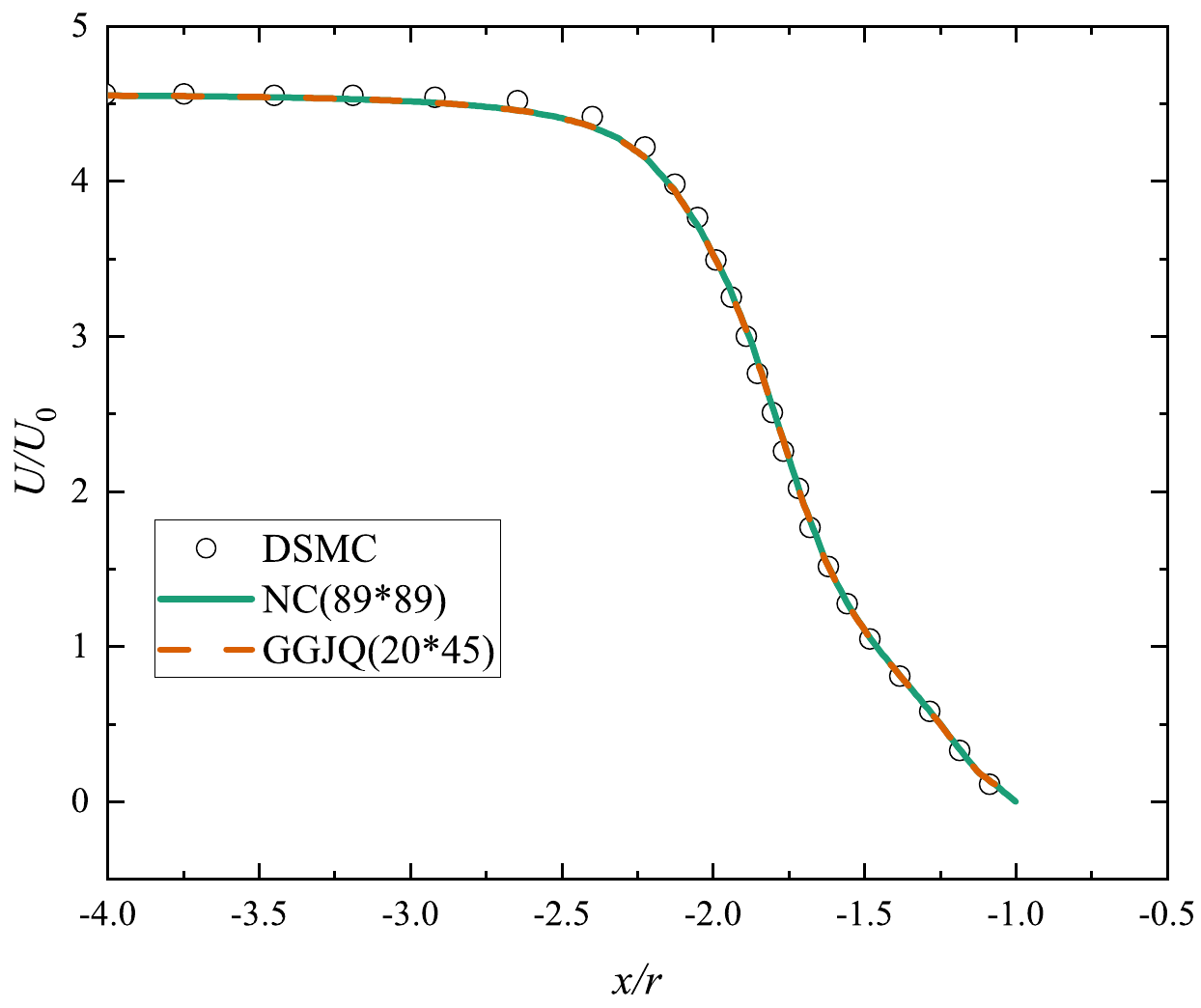}
	\includegraphics[width=5.5cm,height=4.5cm]{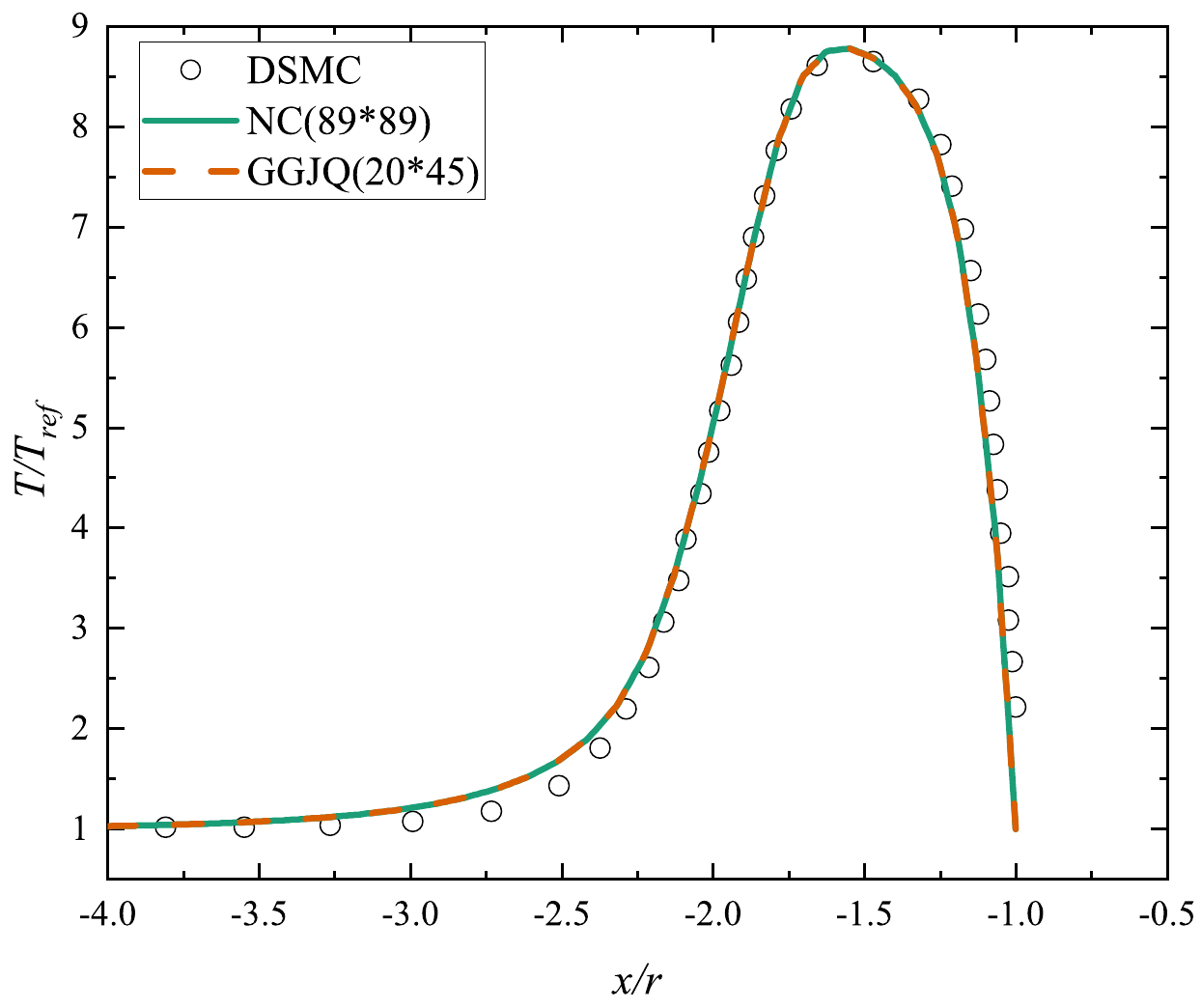}
	\caption{\label{cylinderPlot0.1} \centering  Density, u-velocity, and temperature profiles along the stagnation line of the supersonic cylinder flow for M = 5 and Kn = 0.1.}
\end{figure*}

\begin{figure*}[!th]
	\centering
	\includegraphics[width=5.5cm,height=4.5cm]{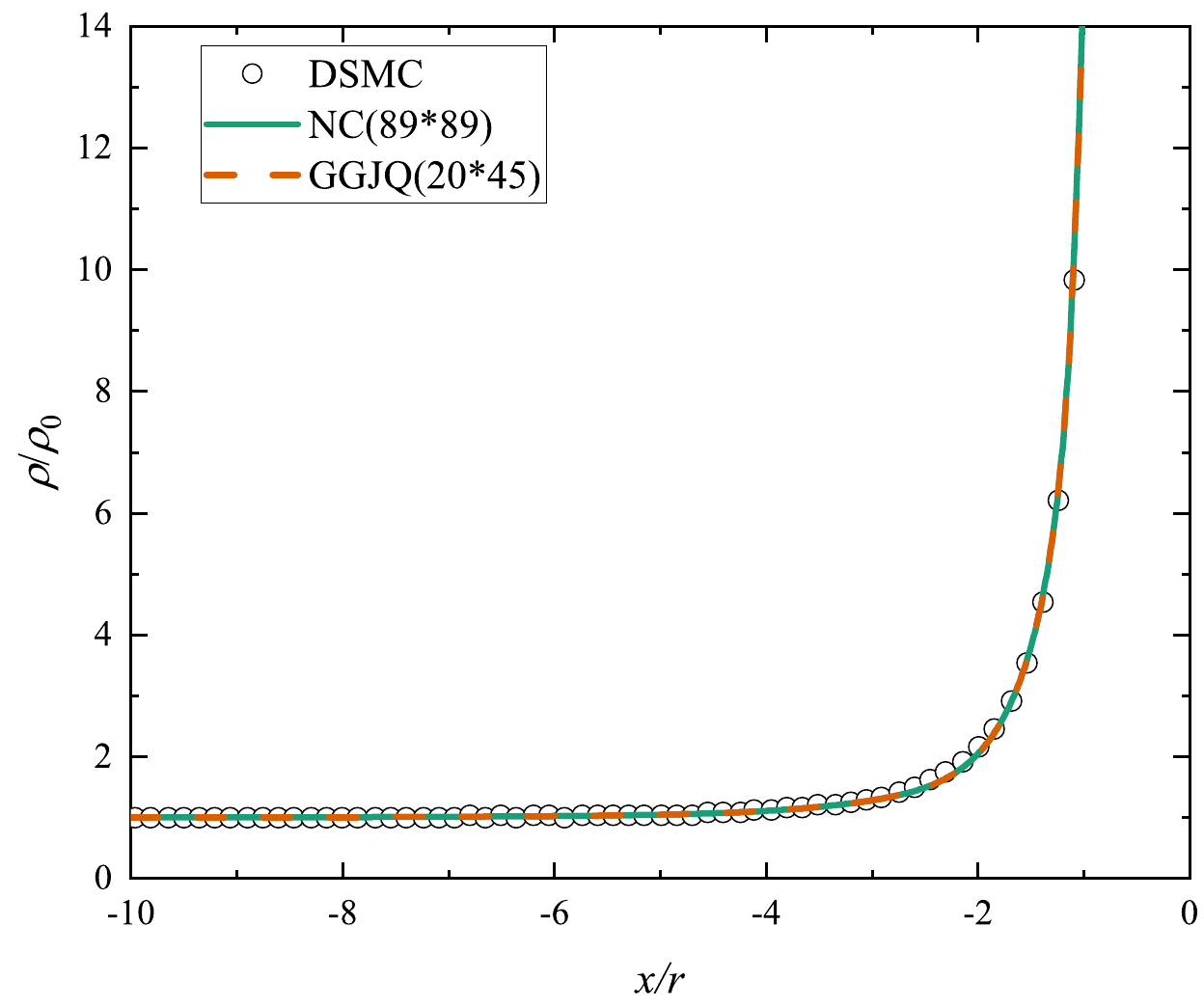}
	\includegraphics[width=5.5cm,height=4.5cm]{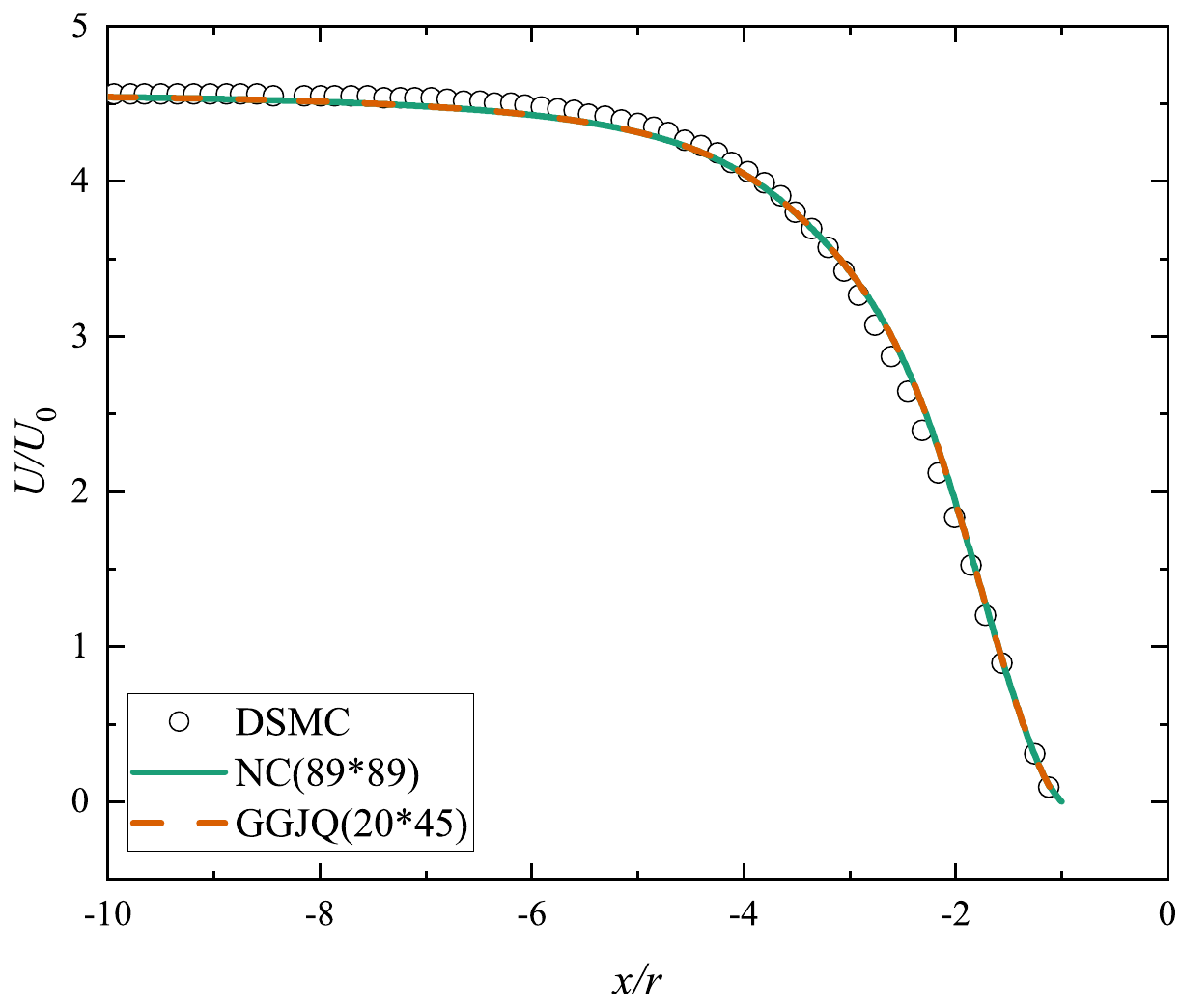}
	\includegraphics[width=5.5cm,height=4.5cm]{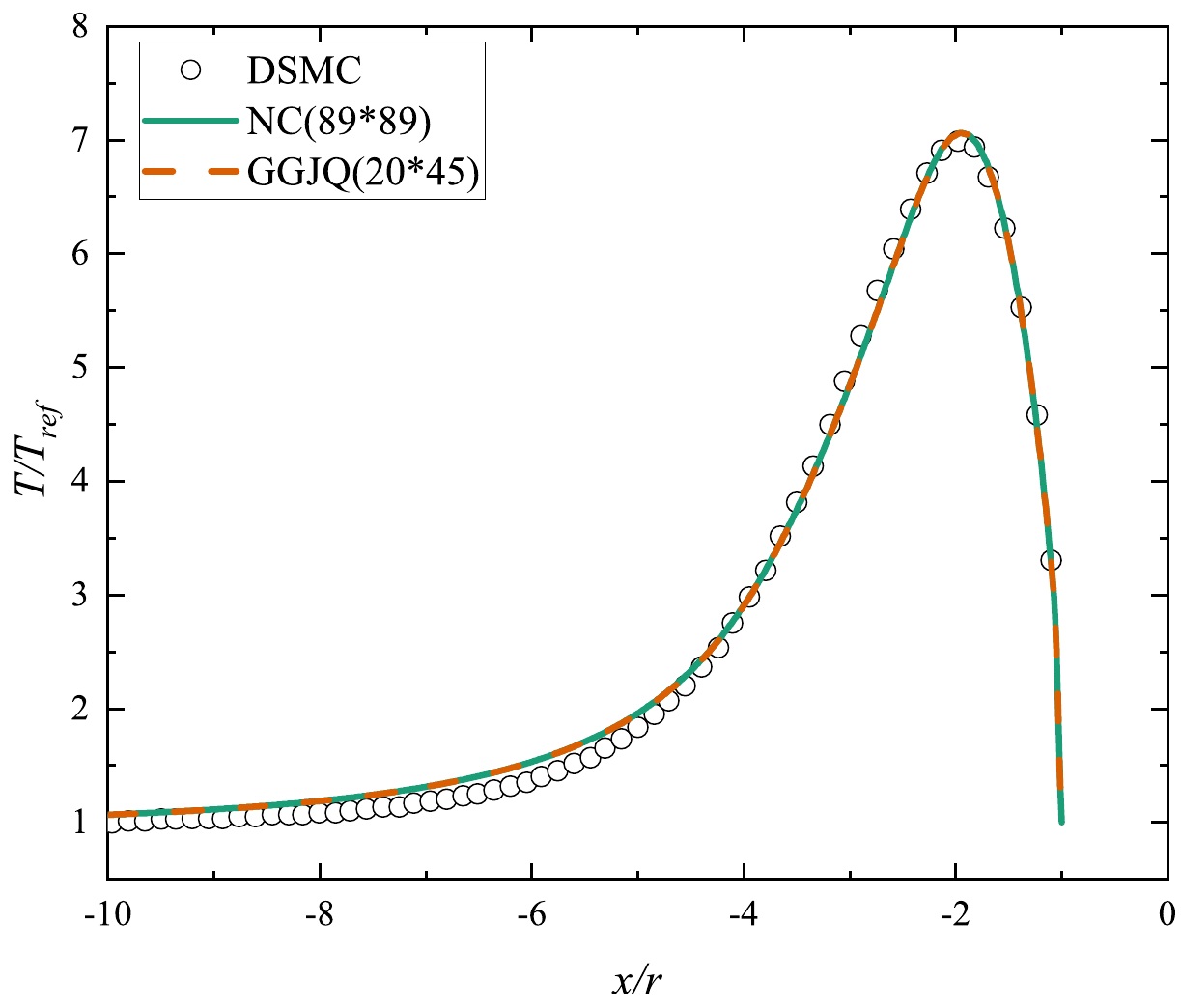}
	\caption{\label{cylinderPlot1.0} \centering  Density, u-velocity, and temperature profiles along the stagnation line of the supersonic cylinder flow for M = 5 and Kn = 1.0.}
\end{figure*}

\begin{figure*}[!th]
	\centering
	\subfigure{
		\includegraphics[width=7cm,height=7cm]{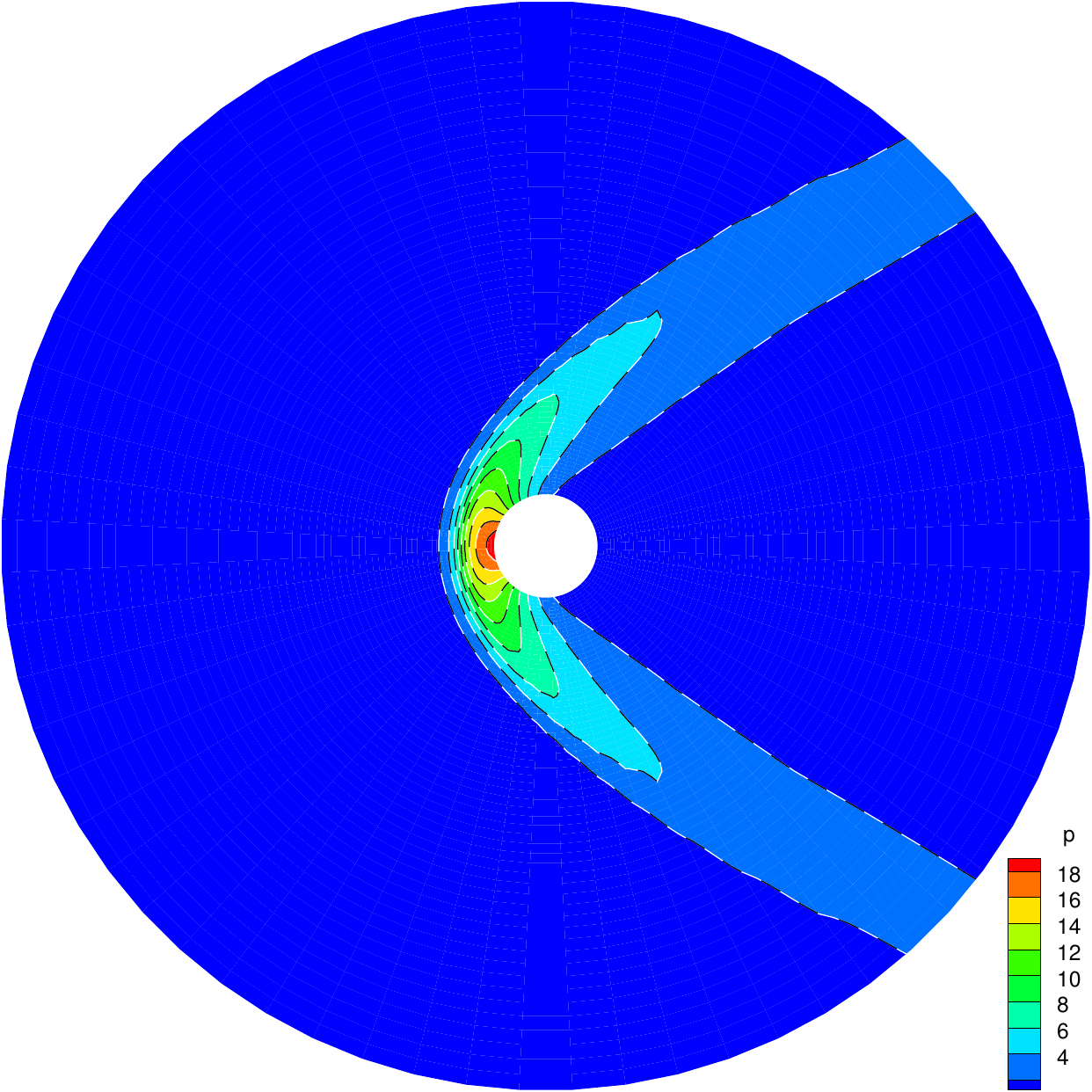}}
	\subfigure{
		\includegraphics[width=7cm,height=7cm]{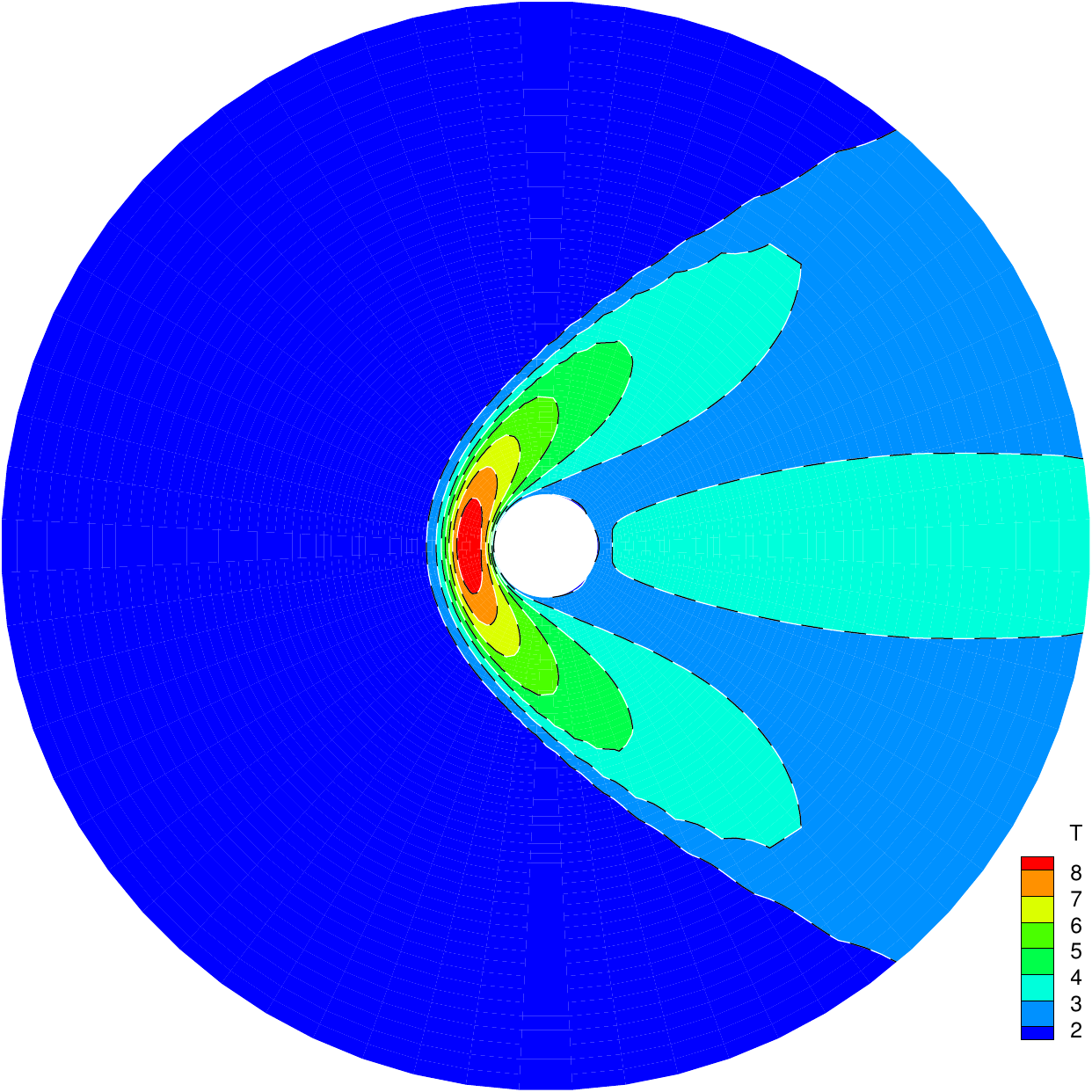}}
	\subfigure{
		\includegraphics[width=7cm,height=7cm]{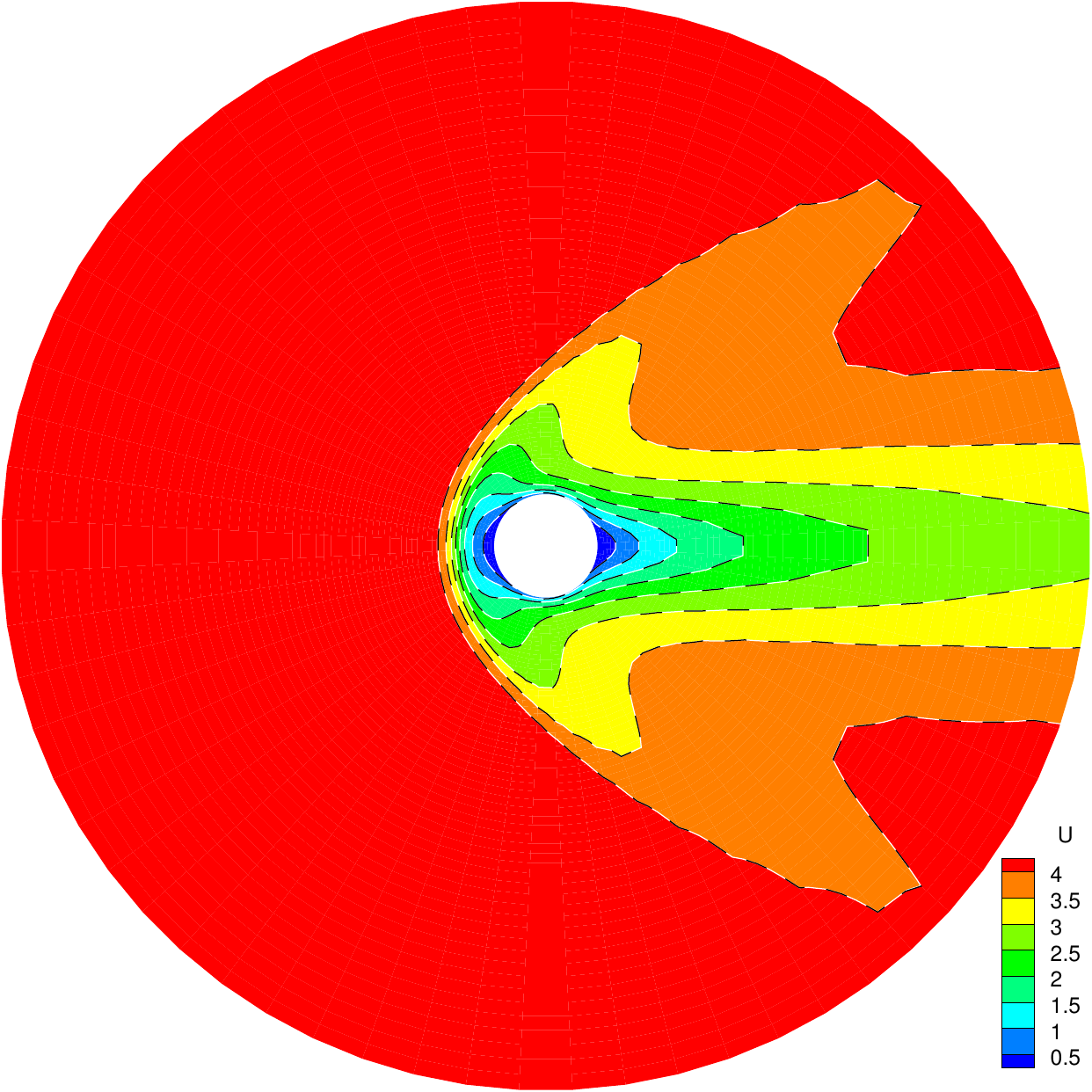}}
	\subfigure{
		\includegraphics[width=7cm,height=7cm]{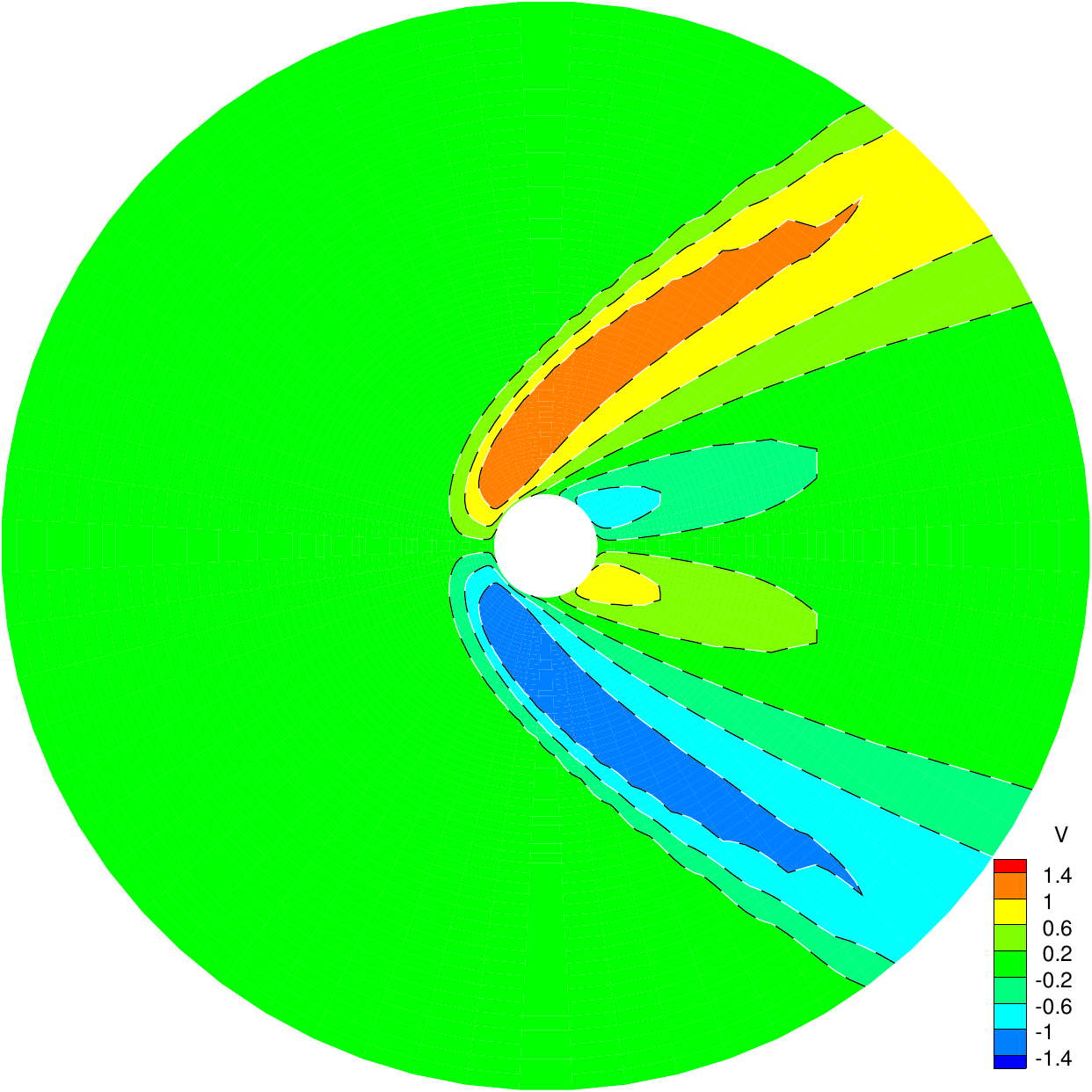}}
	\caption{\label{cylinder0_1} \centering  Comparison of pressure, temperature, u-velocity, and v-velocity contours for cylinder flow at Ma = 5 and Kn = 0.1 (NC: colored background with white solid line; GGJQ: black dashed line).}	
\end{figure*}

\begin{figure*}[!th]
	\centering
	\subfigure{
		\includegraphics[width=7cm,height=7cm]{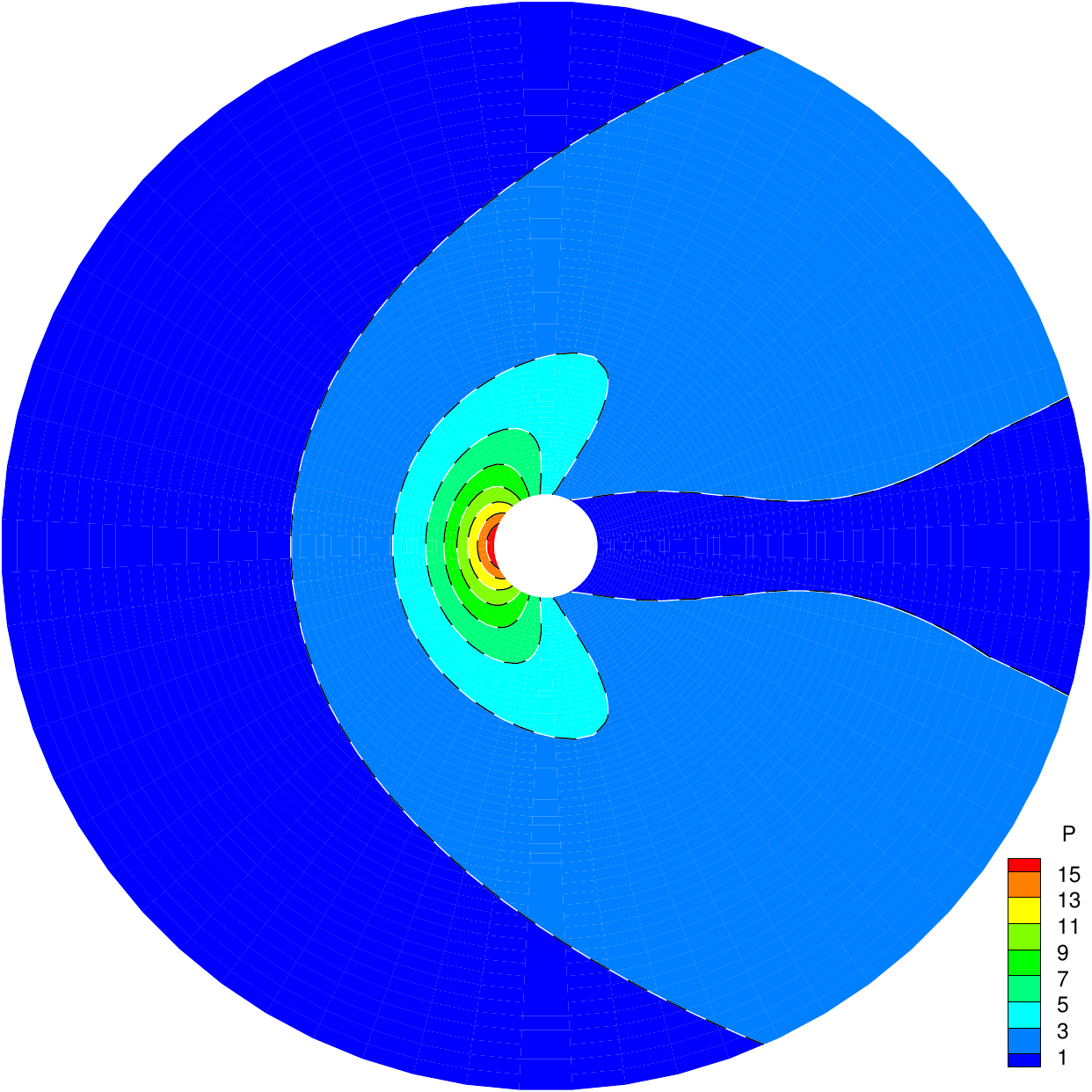}}
	\subfigure{
		\includegraphics[width=7cm,height=7cm]{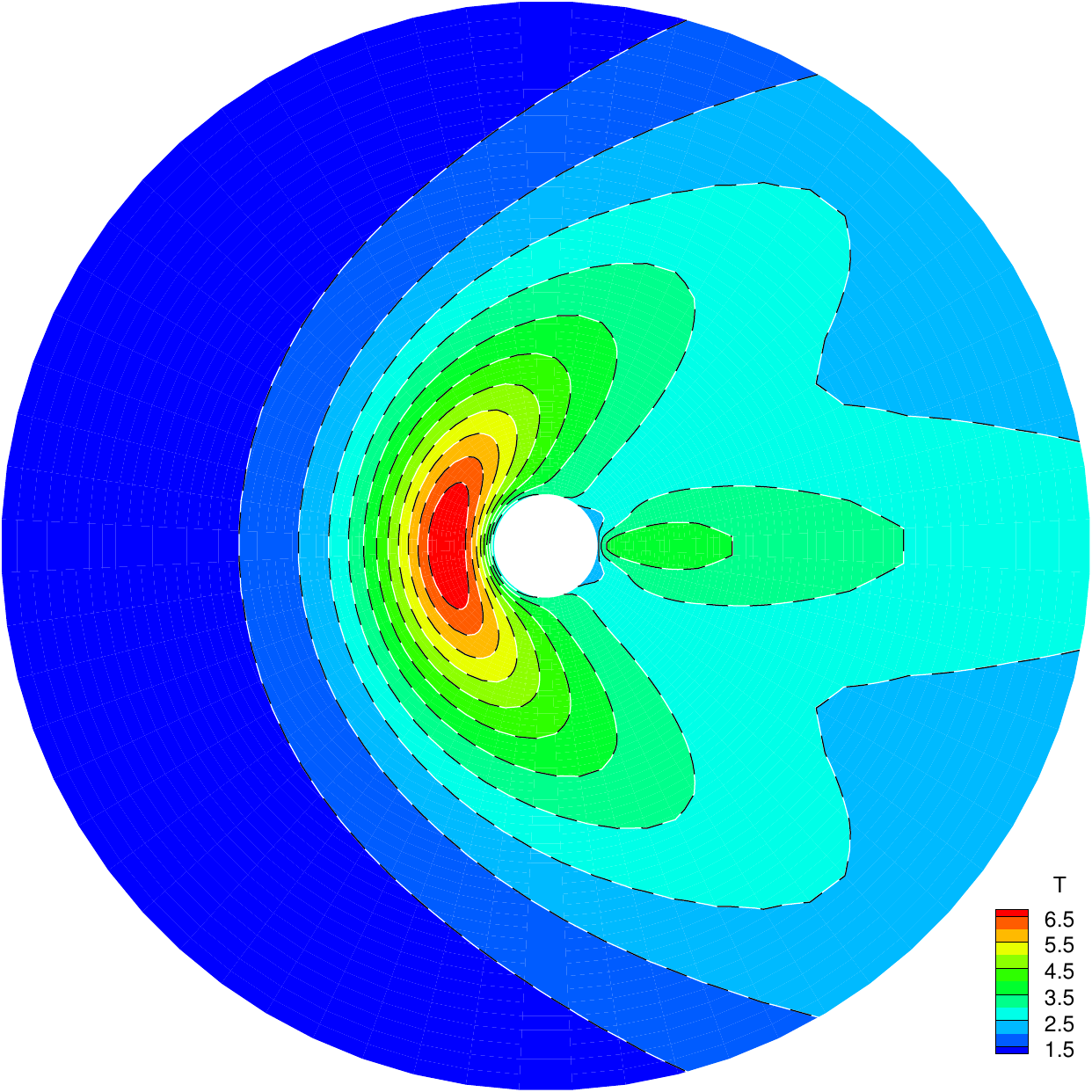}}
	\subfigure{
		\includegraphics[width=7cm,height=7cm]{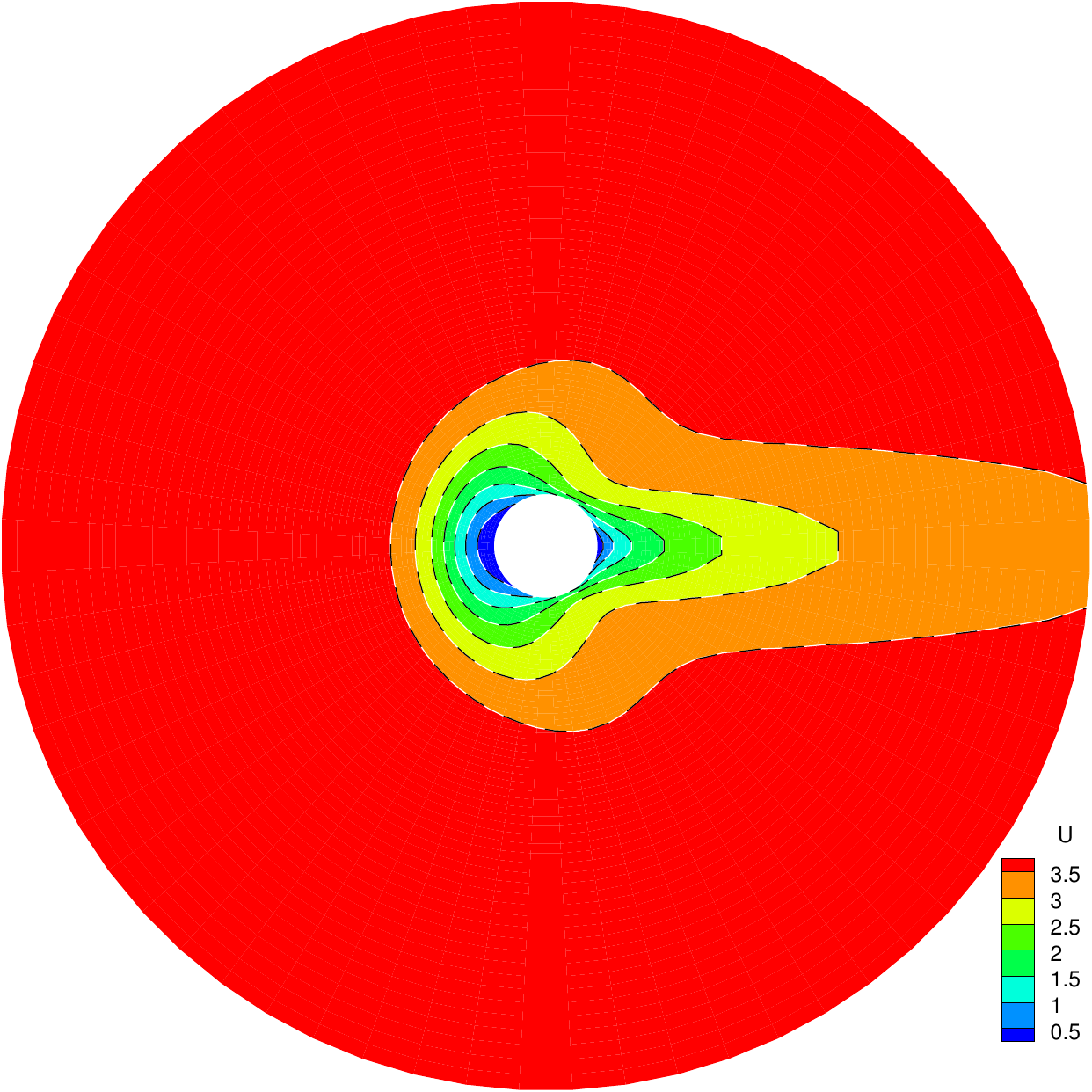}}
	\subfigure{
		\includegraphics[width=7cm,height=7cm]{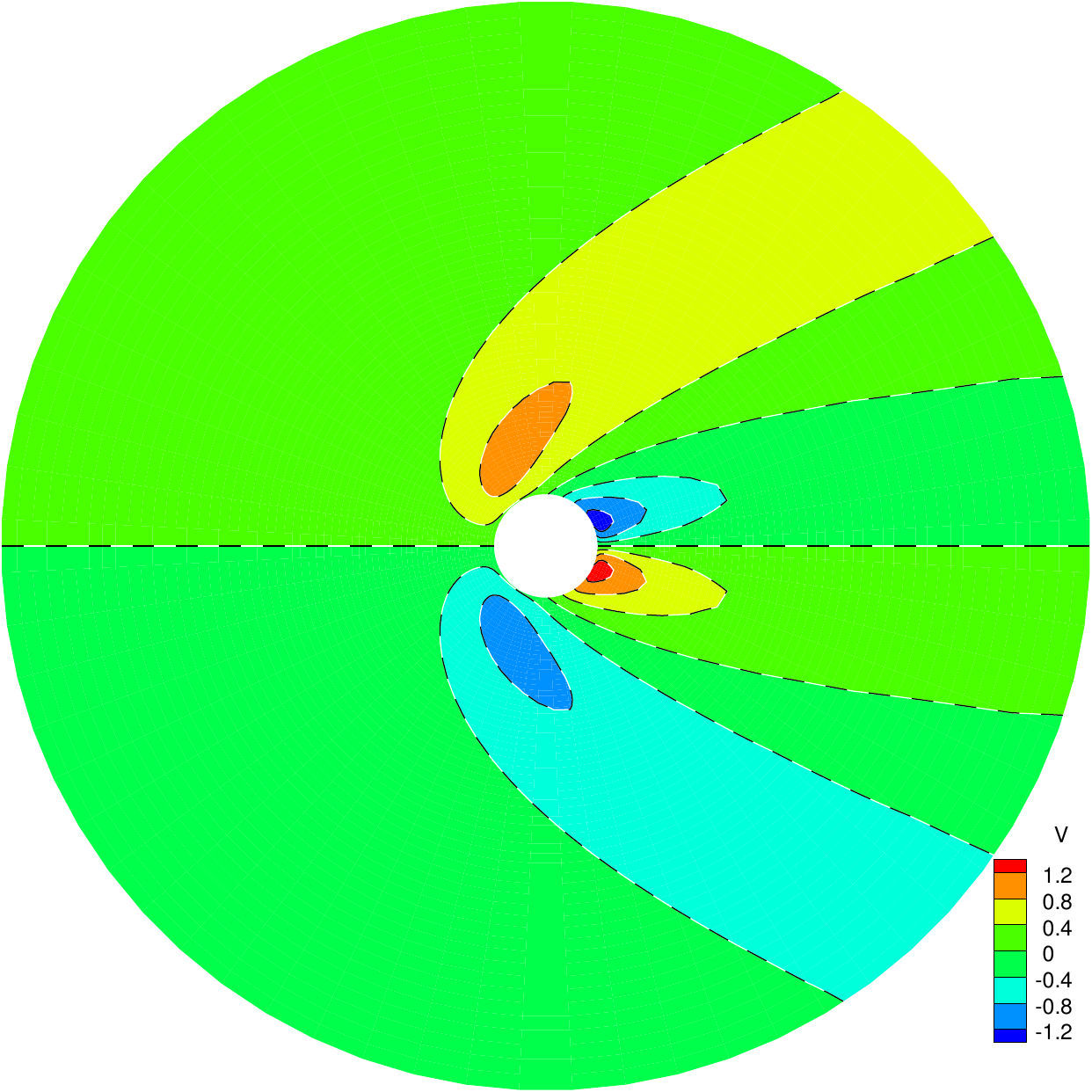}}
	\caption{\label{cylinder1_0} \centering  Comparison of pressure, temperature, u-velocity, and v-velocity contours for cylinder flow at Ma = 5 and Kn = 1.0 (NC: colored background with white solid line; GGJQ: black dashed line).}	
\end{figure*}

For all deterministic discrete velocity methods, the simulation of supersonic flows remains a highly challenging task. Traditional Gauss quadrature employs fixed abscissae, which makes it difficult to capture the wide range of velocity distributions inherent in supersonic flows. On the other hand, the Newton-Cotes (NC) rule requires a significantly larger number of discrete velocities to adequately cover the extended velocity domain.

In this section, the proposed GGJQ is applied to address this challenging problem. The benchmark case considered is the flow past a cylinder at a Mach number of $M=5$ with Knudsen numbers $Kn=0.1$ and $Kn=1.0$. The computational domain is an annulus bounded by $r=1$ and $R=11$, discretized with $64\times64$ cells. The radial grid is geometrically stretched with a maximum-to-minimum spacing ratio of 5, while the azimuthal direction is uniformly discretized, as shown in Figure~\ref{cylindermesh}. Boundary conditions follow Ref.~\cite{Chen2019}: the outer boundary is set as freestream with density $\rho_\infty=1$, velocity $u_\infty=4.56$, and temperature $T_\infty=1$; the inner cylinder wall is modeled as a stationary wall with $\rho_w=1$, $u_w=0$, and $T_w=1$.

The velocity space is discretized using both the Newton-Cotes (NC) rule and the present GGJQ method. For the NC rule, $89\times 89$ uniformly distributed discrete velocities are used in the interval $[-10,10]$. For GGJQ, the velocity discretization employs $20\times45$ points, with parameters $\alpha=600$ and $\beta=350$, such that the discrete velocities are distributed within a circular region of radius 10, as illustrated in Figure~\ref{cylinderVelocity}.

Figure~\ref{cylinderPlot0.1} and Figure~\ref{cylinderPlot1.0} present the density, velocity, and temperature distributions along the stagnation line for $Kn=0.1$ and $Kn=1.0$, respectively. Furthermore, Figure~\ref{cylinder0_1} and Figure~\ref{cylinder1_0} show the contour plots of pressure, temperature, and velocity for the two Knudsen numbers. In these figures, solid lines correspond to NC results, while dashed lines denote GGJQ results. Remarkably, the results obtained by the two methods are almost indistinguishable, whereas GGJQ requires only about \textbf{11\%} of the discrete velocities used in the NC rule. This clearly demonstrates the efficiency and effectiveness of the GGJQ method in simulating supersonic flows.

\subsection{3D lid-driven cavity flow}
\label{sec4.5}

\begin{figure*}[!th]
	\centering
	\includegraphics[width=8.5cm,height=7.5cm]{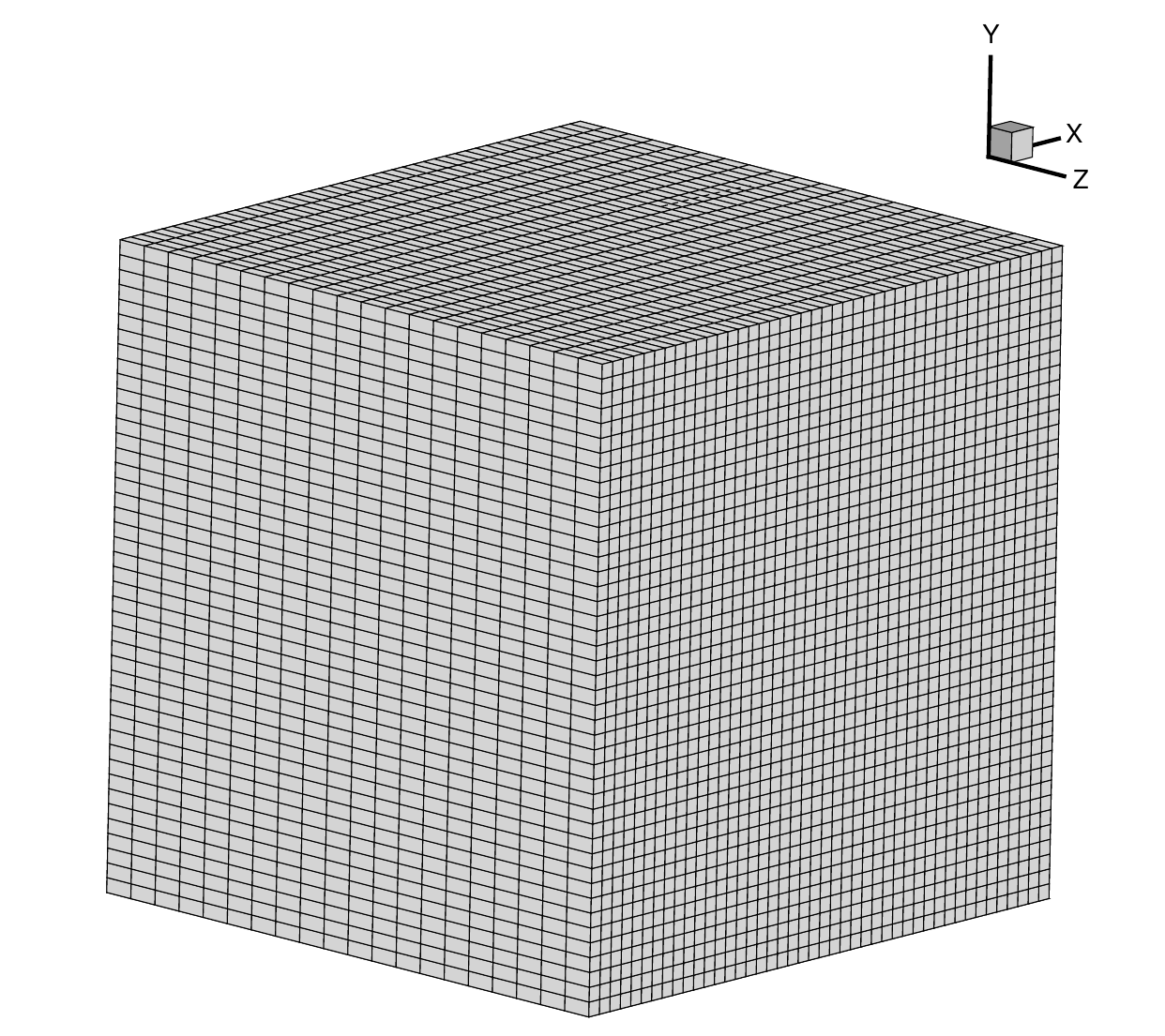}
	\caption{\label{3dmesh} \centering  Computational mesh for 3D lid-driven cavity flow.}
\end{figure*}

\begin{figure*}[!th]
	\centering
	\subfigure[]{\includegraphics[width=7cm,height=6cm]{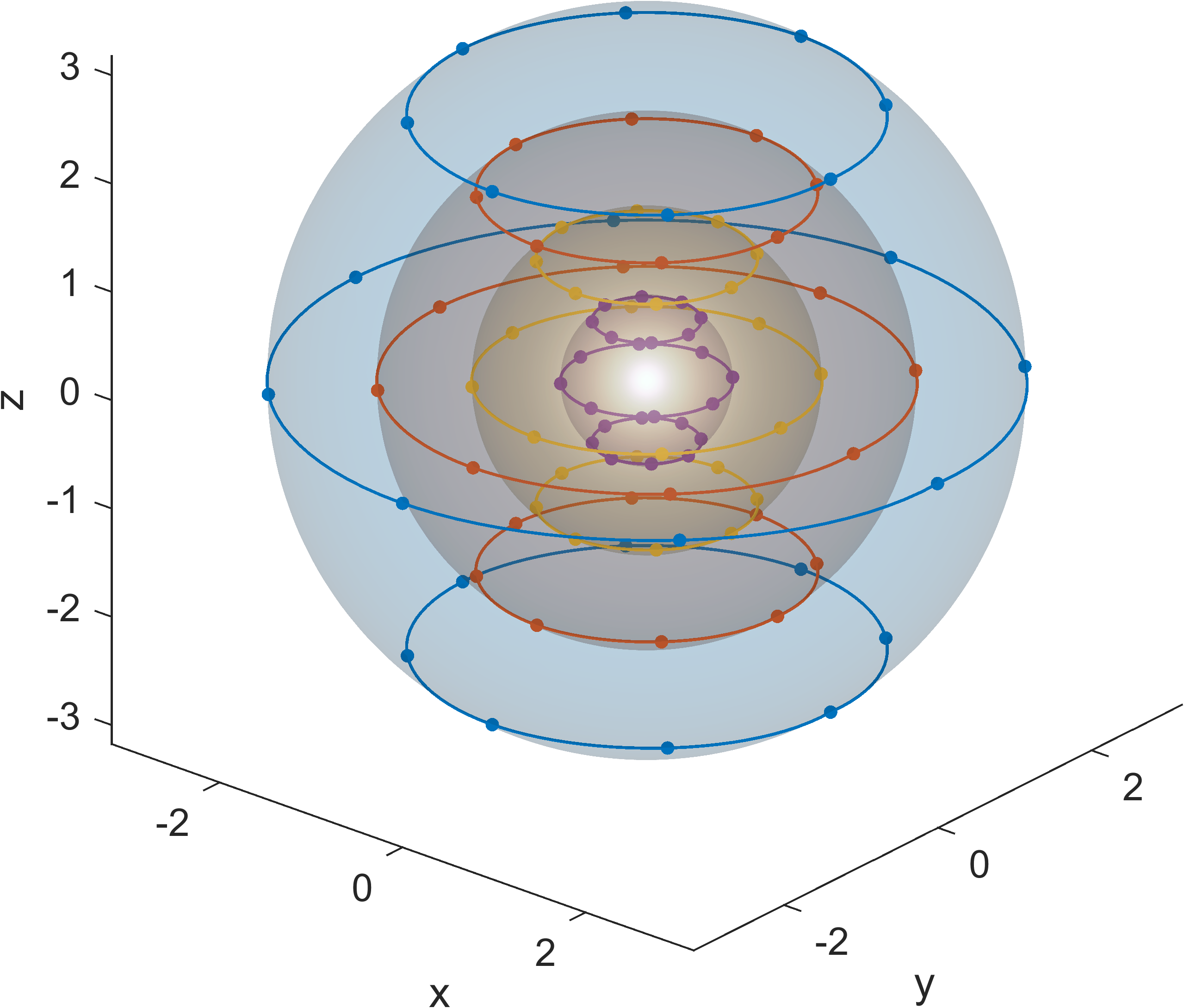}}
	\subfigure[]{\includegraphics[width=7cm,height=6cm]{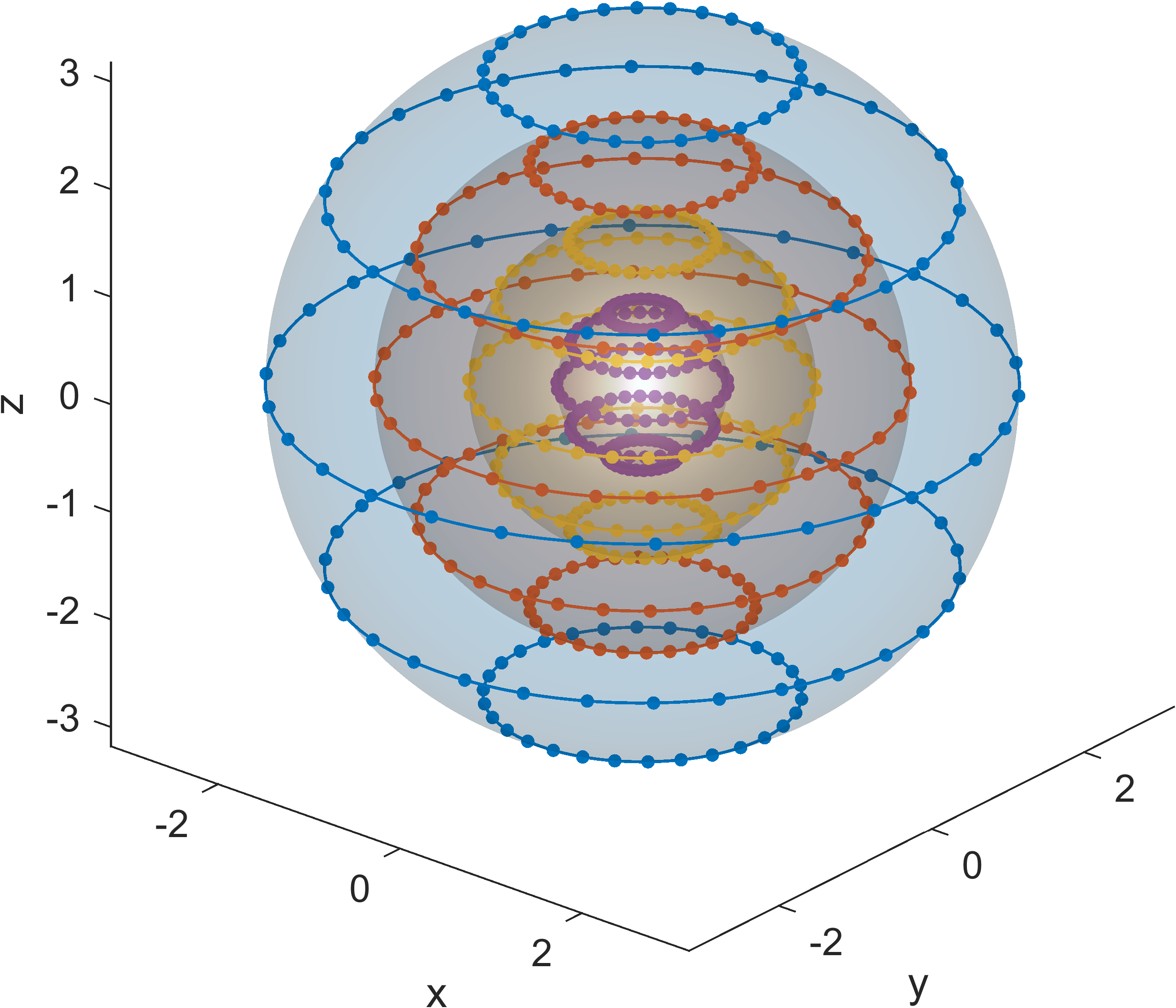}}
	\caption{\label{3dvolecitymesh} \centering  GGJQ-based 3D discrete velocity distribution for lid-driven cavity flow at different Knudsen numbers: (a) Kn = 0.01, 0.1; (b) Kn = 1.0, 10.0.}
\end{figure*}

\begin{figure*}[!th]
	\centering
	\subfigure[Kn=0.01]{
		\label{cavity3d0.01}
		\includegraphics[width=7cm,height=5.5cm]{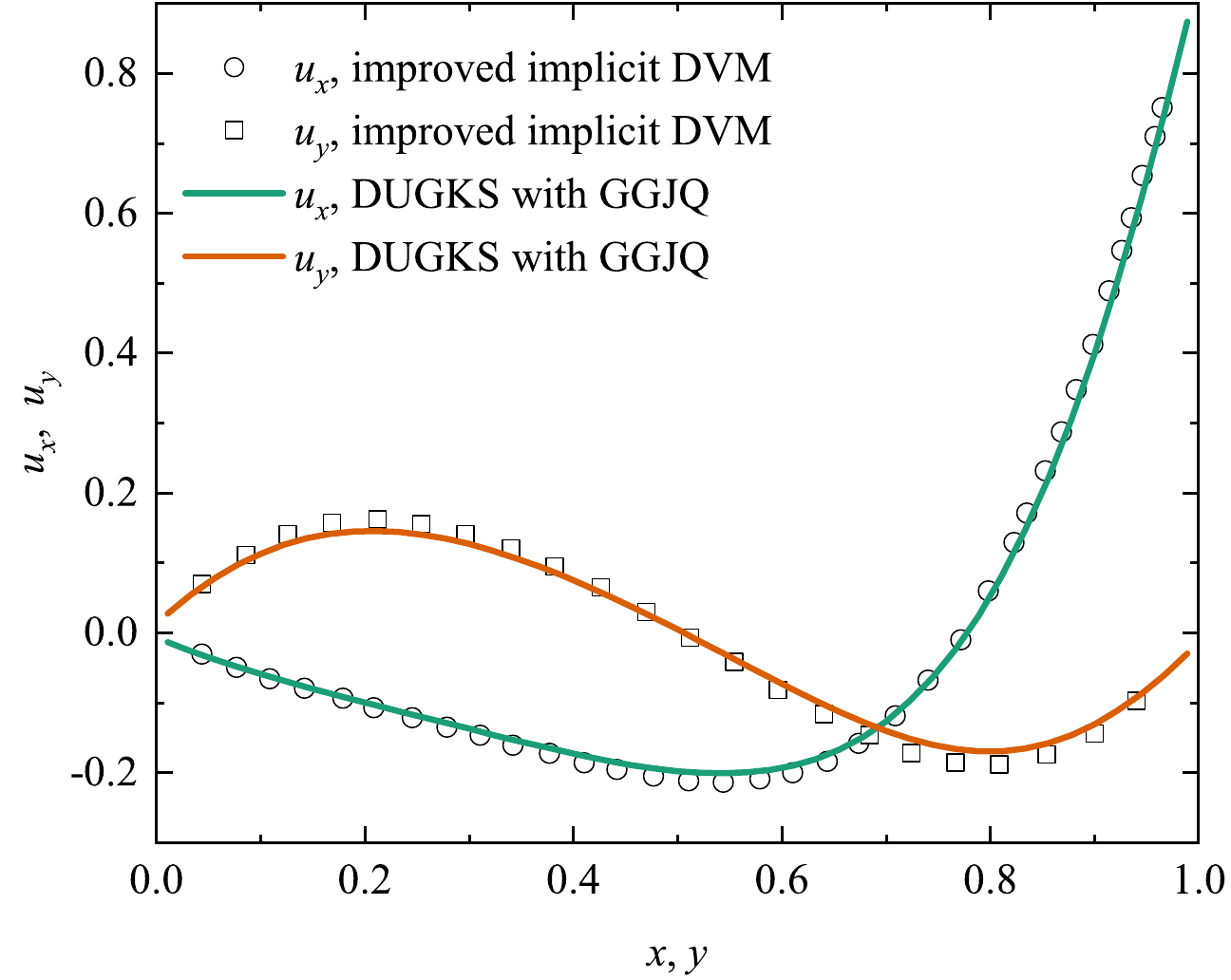}}
	\subfigure[Kn=0.1]{
		\label{cavity3di0.1}
		\includegraphics[width=7cm,height=5.5cm]{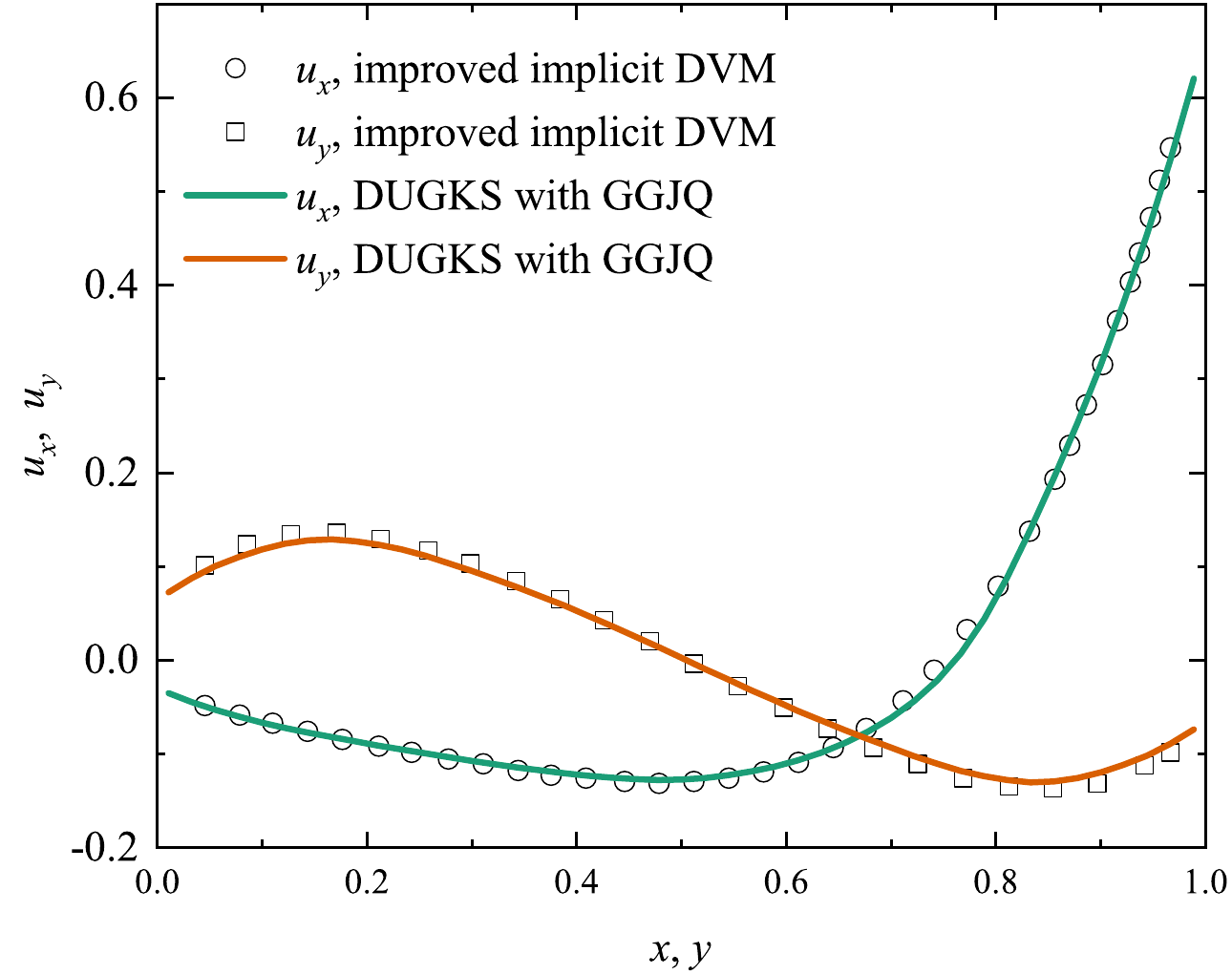}}
	\subfigure[Kn=1.0]{
		\label{cavity3d1.0}
		\includegraphics[width=7cm,height=5.5cm]{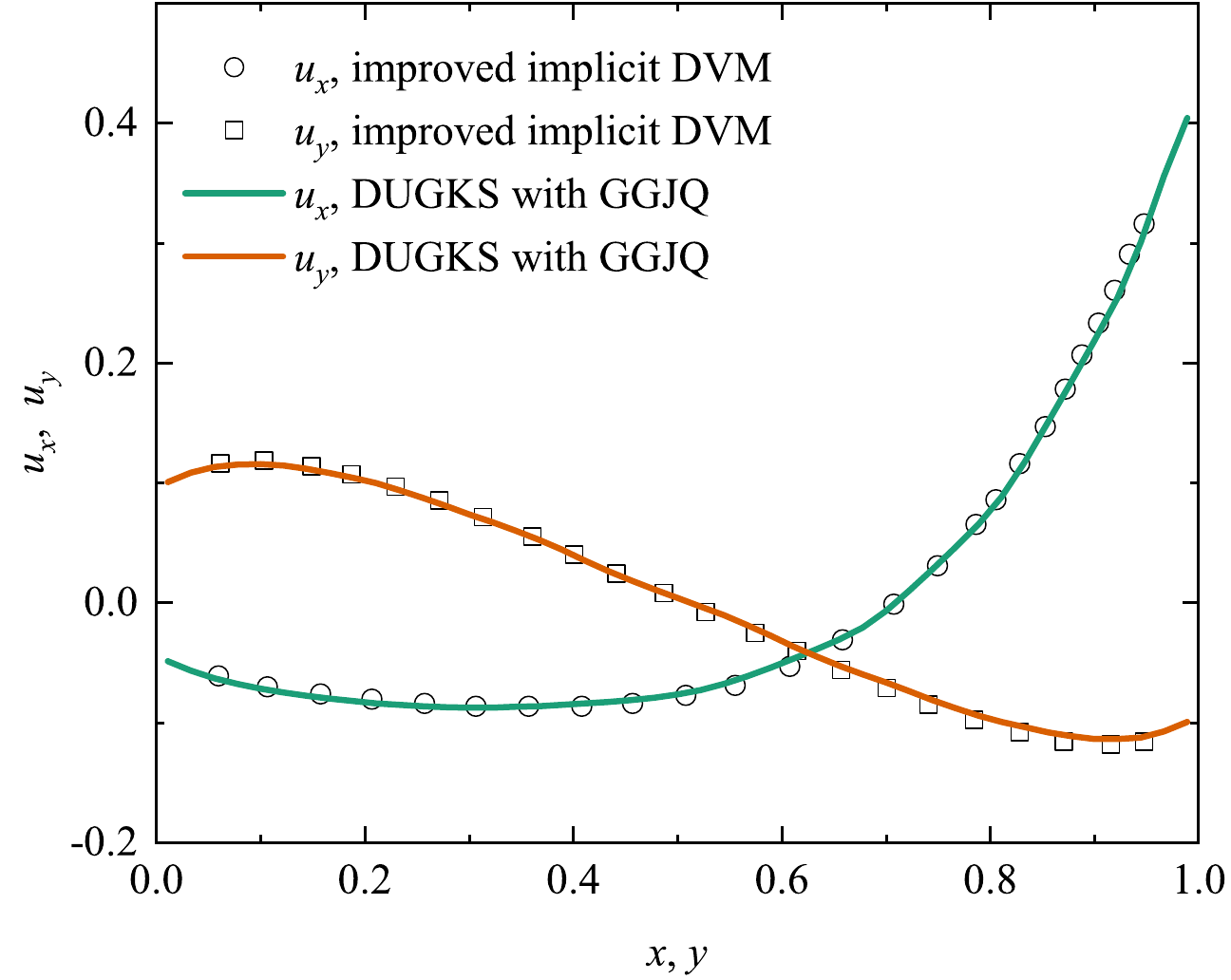}}
	\subfigure[Kn=10.0]{
		\label{cavity3d10.0}
		\includegraphics[width=7cm,height=5.5cm]{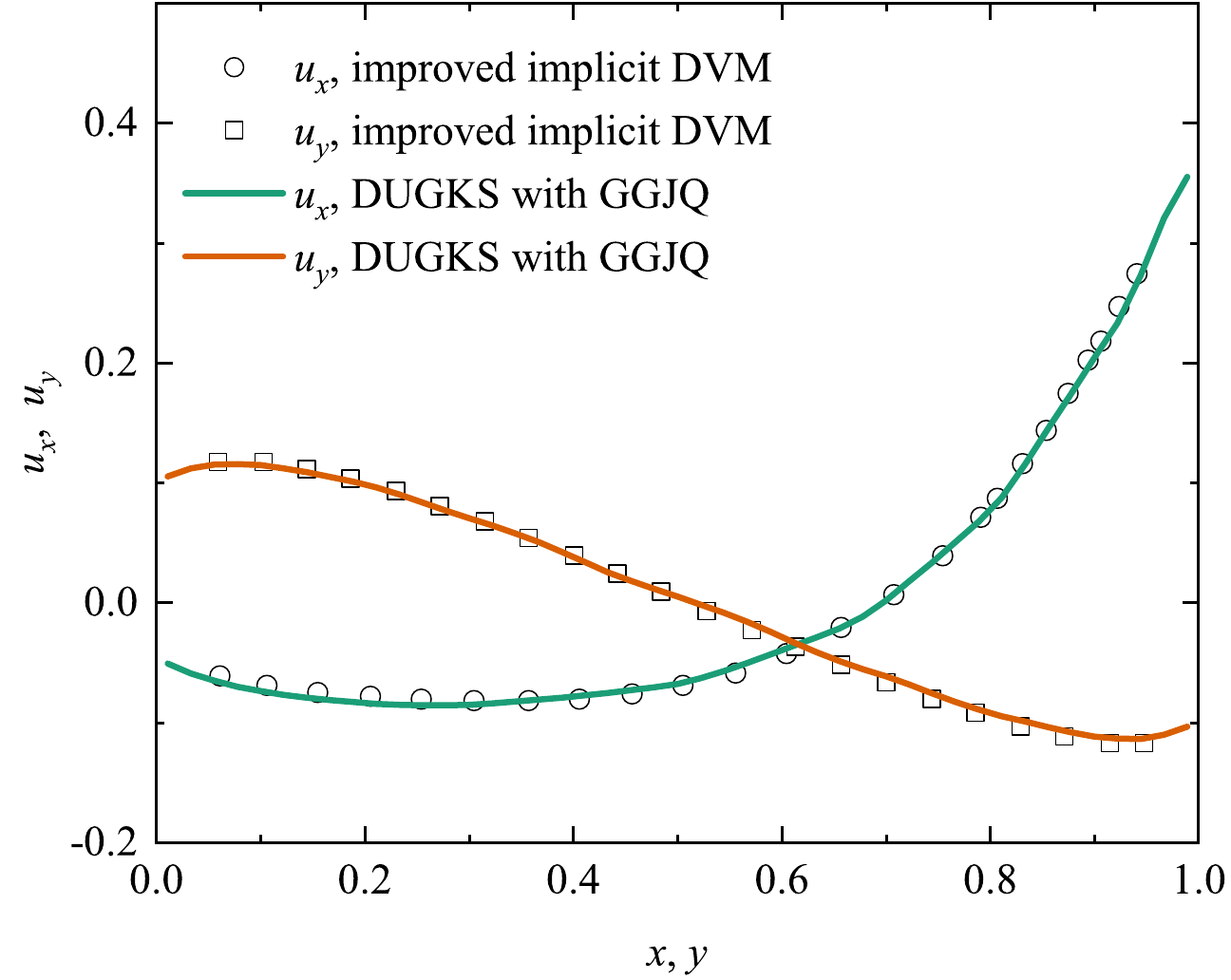}}
	\caption{\label{cavity3d} \centering  Velocity profiles of the 3D lid-driven cavity flow.}	
\end{figure*}

The three-dimensional lid-driven cavity flow is a well-established benchmark problem for assessing the accuracy and efficiency of mathematical models and numerical methods. In this section, the proposed 3D GGJQ is evaluated against the benchmark results obtained by Yang 
\textit{et al.} \citep{cavity3D} using the IDVM.

The computational domain is a cubic cavity with side length 
$L=1$, discretized by a uniform mesh of $45 \times 45 \times 21$ cells, as shown in Figure~\ref{3dmesh}. The top lid moves with a dimensionless velocity $u_w=0.15$ in the $x$-direction, while the other walls are stationary. All boundaries are maintained at a constant temperature.

For small Knudsen numbers ($Kn=0.01$ and $Kn=0.1$), Yang \textit{et al.} employed the $18^3$ Gauss-Hermite quadrature rule. In contrast, the present work applies the $4 \times 8 \times 3$ GGJQ. For higher Knudsen numbers ($Kn=1.0$ and $Kn=10.0$), Yang \textit{et al.} adopted the Newton-Cotes (NC) rule with $41^3$ uniformly distributed discrete velocities in the range $[-4,4]^3$. In comparison, the present simulations utilize the $4 \times 30 \times 5$ GGJQ, with parameters $\alpha=\beta=1000$. The discrete velocity distribution of GGJQ is shown in Figure~\ref{3dvolecitymesh}.

Figure~\ref{cavity3d} illustrates the velocity profiles along the horizontal and vertical centerlines of the cavity. The numerical results obtained with the proposed GGJQ show excellent agreement with the reference solutions of Yang \textit{et al.}. Remarkably, the number of discrete velocities employed in this work is only about \textbf{1.6\%} of that in the reference computations for small Knudsen numbers ($Kn=0.01,0.1$), and about \textbf{0.87\%} for large Knudsen numbers ($Kn=1.0,10.0$), clearly demonstrating the high efficiency of the proposed quadrature rule.

\subsection{ 3D Spherical Fourier flow}
\label{sec4.6}

\begin{figure*}[!th]
	\centering
	\subfigure[]{
		\label{sphermesh}
		\includegraphics[width=7cm,height=7cm]{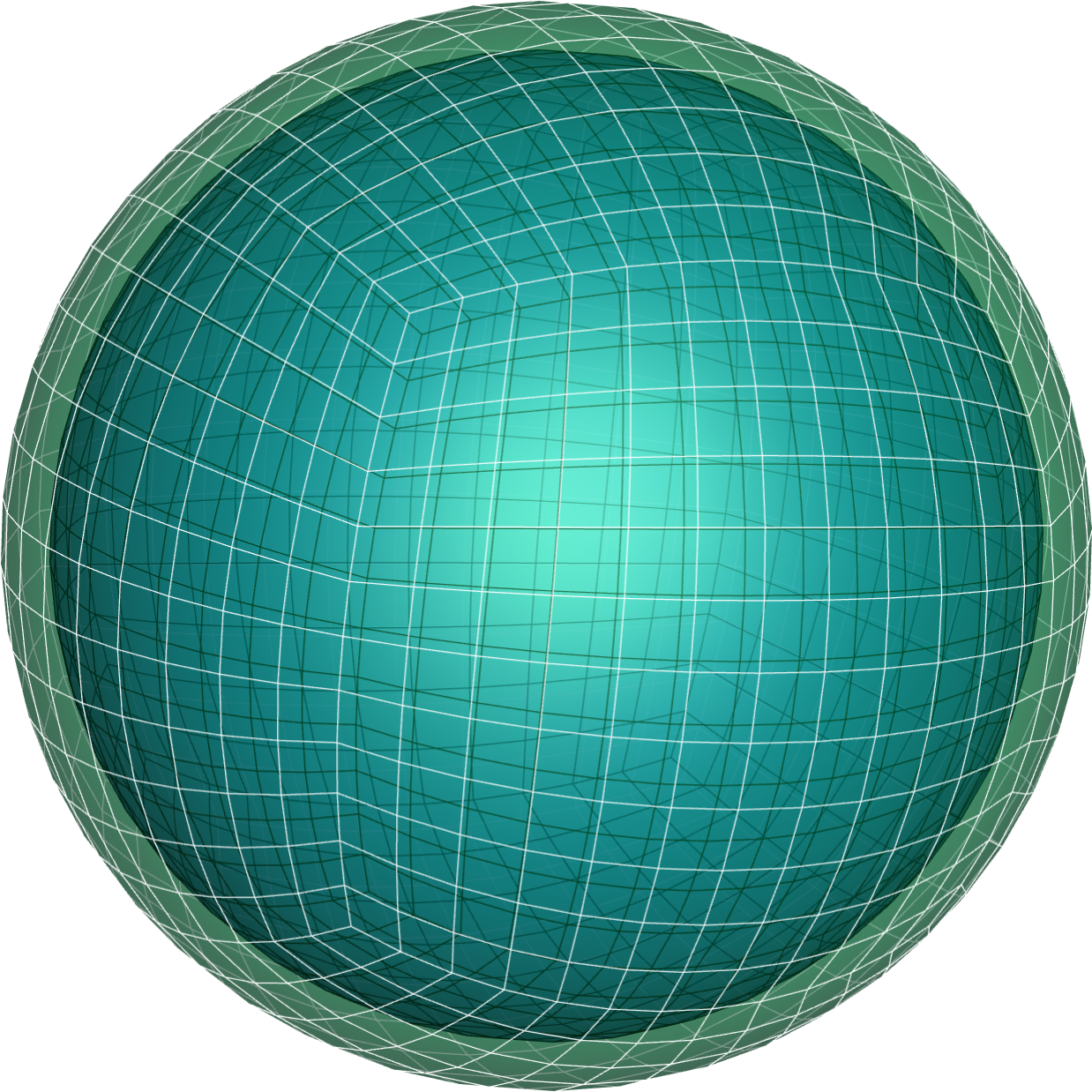}
	}
	\subfigure[]{
		\label{spherVelocity}
		\includegraphics[width=8.2cm,height=7cm]{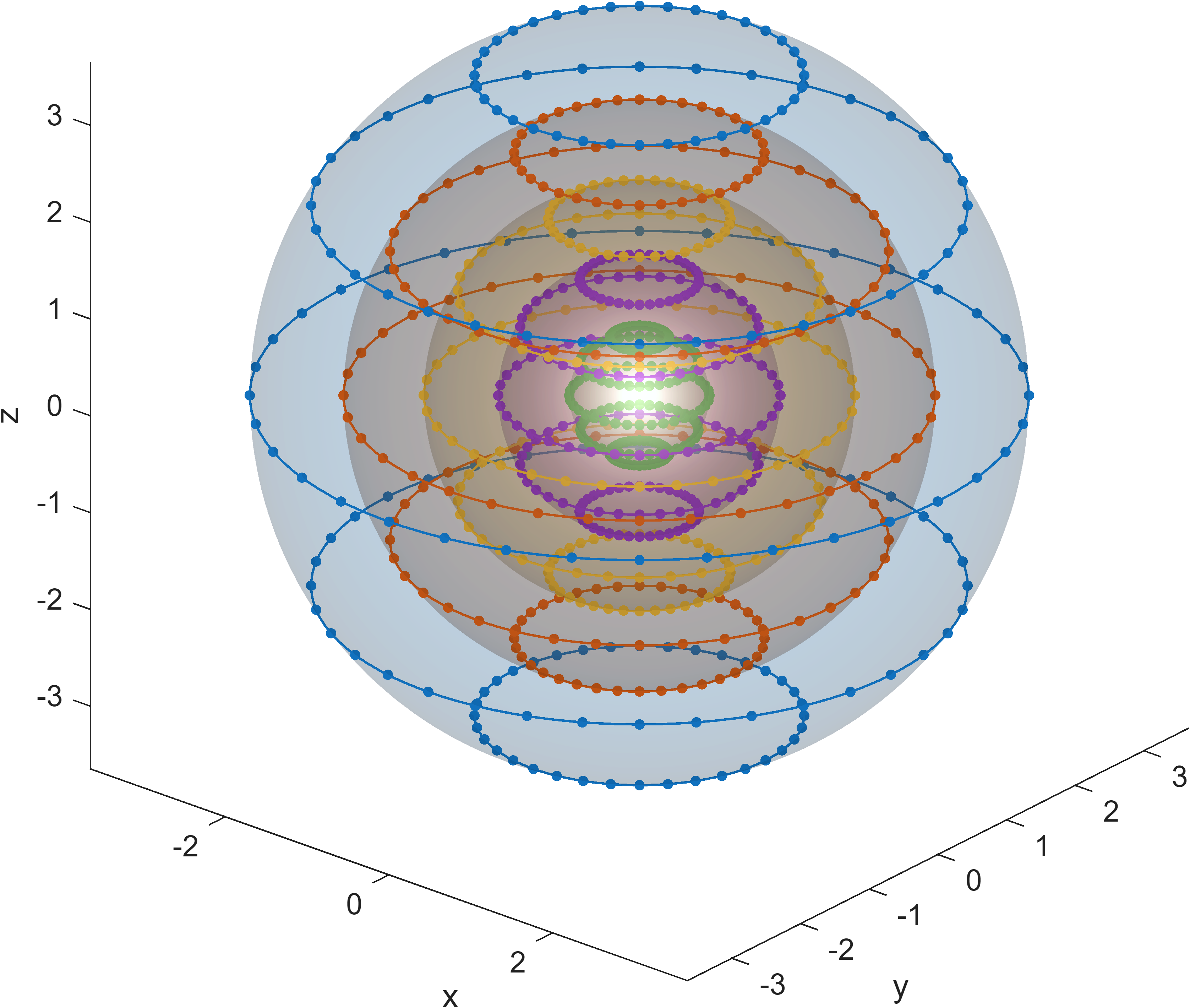}
	}
	\caption{\label{sphermeshandVelocity} \centering  Computational mesh (a) and discrete velocity distribution (b) for 3D spherical Fourier flow.}
\end{figure*}

\begin{figure*}[!th]
	\centering
	\subfigure[]{
		\label{spherrho}
		\includegraphics[width=8cm,height=6cm]{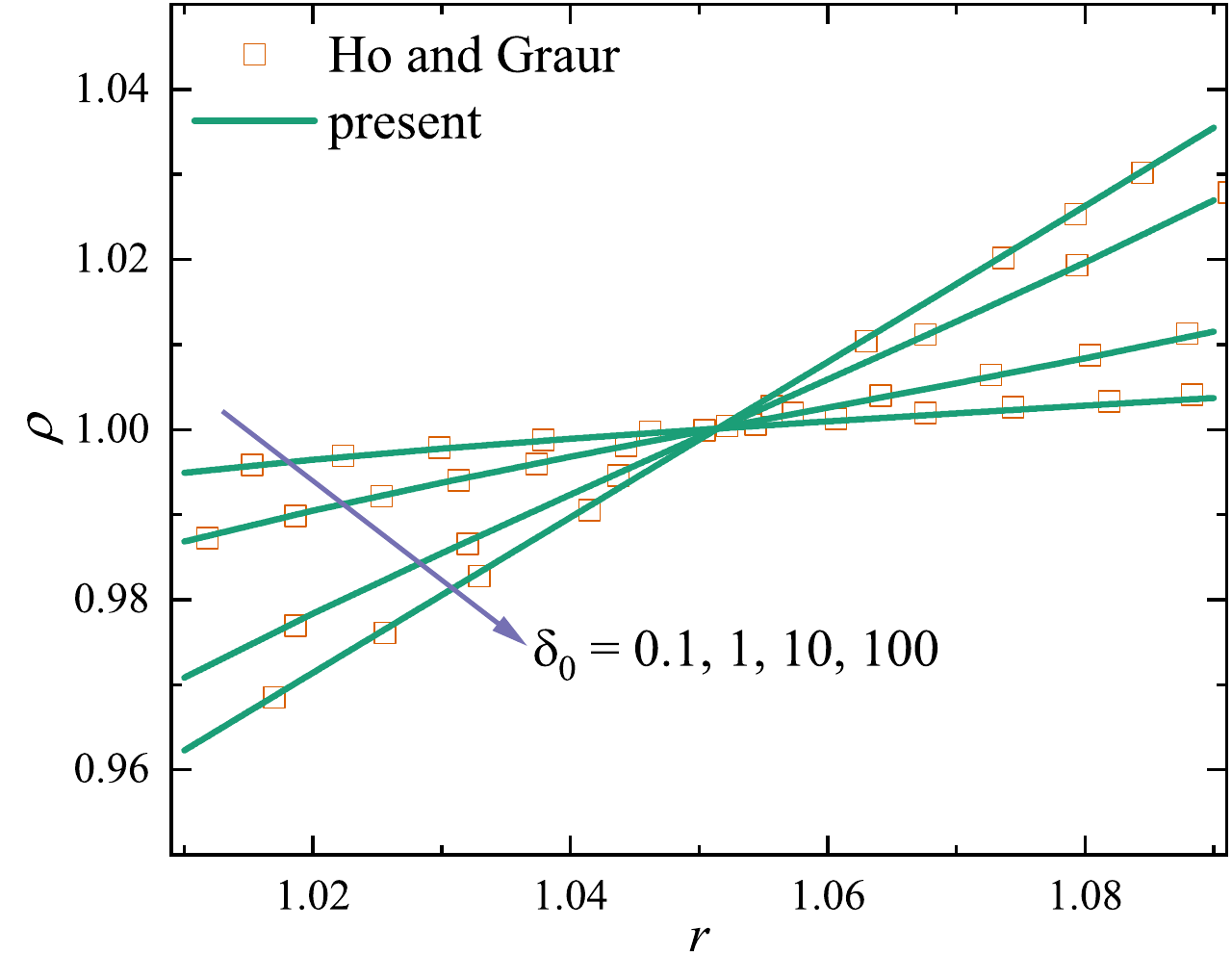}
	}
	\subfigure[]{
		\label{spherT}
		\includegraphics[width=8cm,height=6cm]{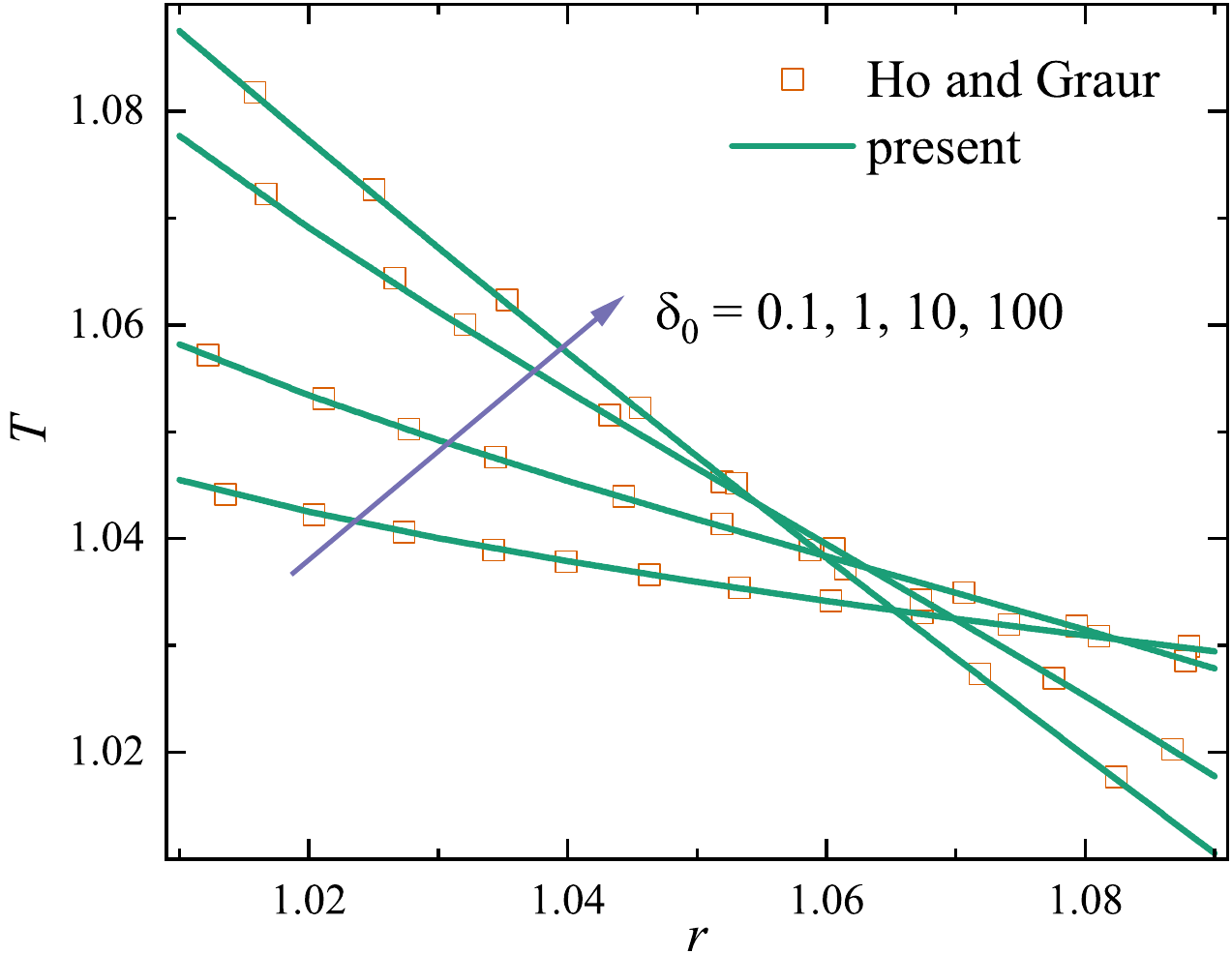}
	}
	\caption{\label{spherrhoT} \centering Density (a) and Temperature (b) profiles of 3D spherical Fourier flow.}
\end{figure*}

In this section, heat transfer between two concentric spheres is simulated to further verify the applicability of the proposed method to rarefied gas transport problems at different degrees of rarefaction. This problem is of practical relevance to spherical MEMS structures, thermal management of microelectronic shells, and hollow-sphere heat transfer. Owing to its strict geometric symmetry and well-defined boundary conditions, the flow degenerates into the classical Fourier heat conduction law in the continuum limit, while exhibiting pronounced non-equilibrium effects in the rarefied regime. Consequently, it is widely employed as a benchmark problem for kinetic models and numerical schemes.

The test case follows the setup given by Ho and Graur \citep{SFF}. The nondimensional temperatures of the inner and outer spherical surfaces are prescribed as $T_h=1.1$ and $T_c=1$, respectively, with radii $R_h=1$ and $R_c=1.1$. The computational domain, illustrated in Figure~\ref{sphermesh}, is divided into six identical segments. Each segment is discretized using $12\times12\times10$ cells, with only 10 grid points in the radial direction. For the velocity space discretization, five nodes of the GGJQ with parameters $\alpha=\beta=1000$ are employed in the radial direction, while 36 points are used in the azimuthal direction with a periodic trapezoidal rule. In the polar direction, a five-point Gauss–Legendre quadrature is adopted, as shown in Figure~\ref{spherVelocity}.

Figure~\ref{spherrhoT} presents the density and temperature distributions for different rarefaction parameters $\delta_0=R_0/\ell$, where $R_0=R_c-R_h$ and $\ell$ is the equivalent mean free path. Despite the use of relatively sparse spatial meshes and velocity discretization, the present simulation results show excellent agreement with the reference solutions for four values of $\delta_0$. This further confirms the robustness and accuracy of the proposed method in capturing multiscale flow phenomena in 3D rarefied gas dynamics.

\section{Conclusion}
\label{sec5}

A generalized Gauss–Jacobi quadrature (GGJQ) has been developed for one-, two-, and three-dimensional velocity spaces within the BGK–Shakhov kinetic framework. Focusing on single-peaked distribution functions, the GGJQ offers flexible and tunable velocity nodes that accurately capture both near-equilibrium and strongly nonequilibrium flows. Validation against six benchmark problems—including shock tubes, shock structures, and multidimensional cavity and spherical Fourier flows—demonstrates its accuracy, robustness, and versatility across a wide range of Knudsen and Mach numbers.

The results establish GGJQ as a systematic and efficient approach for constructing discrete velocity sets with adjustable resolution, bridging the gap between conventional fixed quadratures and the requirements of multiscale kinetic simulations. Future work will extend this framework to multi-peaked distribution functions and explore adaptive strategies for more complex nonequilibrium flows, further enhancing its applicability to practical gas dynamics problems.

\appendix

\section{Abscissas and weights for GGJQ}
\label{APPA}
\setcounter{table}{0}
\renewcommand\theequation{A.\arabic{equation}}
\lstset{
	language=Python,
	basicstyle=\ttfamily\small, 
	keywordstyle=\color{teal}\bfseries, 
	commentstyle=\color{gray}\itshape,
	stringstyle=\color{orange!80!black},
	numbers=left, 
	numberstyle=\tiny\color{gray},
	numbersep=6pt,
	frame=single,
	rulecolor=\color{gray!30},
	showstringspaces=false,
	breaklines=true, 
	tabsize=4,
	captionpos=b,
	backgroundcolor=\color{gray!2},
}

\begin{table}[b]
	\renewcommand{\arraystretch}{1.2}
	\small
	\centering
	\begin{threeparttable}
		\caption{Abscissas and weights for Generalized Gauss-Jacobi quadrature.}
		\label{tab:AandW}
		\begin{tabular}{ccc}
			\toprule
			\textbf{Dimension} & \textbf{Abscissas} & \textbf{Weights} \\
			\hline			
			1D & $\xi _i=\pm R_{\alpha \beta,i} $ & 
			$\small{\frac{\alpha^{\small{\frac{1}{2}}}}{2}}\small{\frac{W_i\left( \small{-\small{\frac{1}{2}},\beta -1} \right)}{w_{\alpha ,\beta}\left( \xi _i \right)}}$\\
			2D & 
			$\begin{cases}
				\xi _{x,ij}=R_{\alpha \beta,i} \cos \theta _j\\
				\xi _{y,ij}=R_{\alpha \beta,i} \sin \theta _j\\
			\end{cases}$ & 
			$\small{\frac{\alpha}{2}\frac{W_i\left( \small{0,\beta -1} \right)}{w_{\alpha ,\beta}\left( \boldsymbol{\xi }_{ij} \right)}}w_j$ \\
			3D & 
			$\begin{cases}
				\xi _{x,ijk}=R_{\alpha \beta,i} \cos \theta _j\sqrt{1-\varPhi _{\phi ,k}^{2}}\\
				\xi _{y,ijk}=R_{\alpha \beta,i} \sin \theta _j\sqrt{1-\varPhi _{\phi ,k}^{2}}\\
				\xi _{z,ijk}=\pm R_{\alpha \beta,i} \varPhi _{\phi ,k}\\
			\end{cases}$ & 
			$\small{\frac{\alpha ^{\small{\frac{3}{2}}}}{2}}\small{\frac{W_i\left( \small{\small{\frac{1}{2}},\beta -1} \right)}{w_{\alpha ,\beta}\left( \boldsymbol{\xi }_{ijk} \right)}w_j}\phi W_k\left( \phi -1,0 \right) $ \\
			\bottomrule
		\end{tabular}
		\begin{tablenotes}
			\tiny
			\item[a] $R_{\alpha \beta ,i}=\sqrt{\alpha \mathrm{arc}\tanh \left[ r_i\left( \small{\small{\frac{D}{2}}-1, \beta -1} \right) \right]}$,
			and $r_i\left( \small{a,~b} \right)$ and $W_i\left( \small{a,~b} \right)$ are the roots and weights of G-J rule with a weight function of $r^{a}\left( 1-r \right) ^{b}$ on the interval [0,1], respectively.
			\item[b] $\phi _{k}\left( \gamma -1,0 \right)$ and $W_{k}\left( \gamma -1,0 \right)$ are the roots and weights of G-J rule with a weight function of $\phi^{\gamma -1}$ on the interval [0,1] with the weigh, respectively.
		\end{tablenotes}
	\end{threeparttable}
\end{table}

In the velocity space $\boldsymbol{R}^D$, the GGJQ for an arbitrary function $\mathcal{F}$ can be expressed in the following form,
\begin{equation}
	I(\mathcal{F} )=\int_{\boldsymbol{R}^D}{\mathcal{F} (\boldsymbol{\xi })\,d\boldsymbol{\xi }}=\int_{\boldsymbol{R}^D}{w(\boldsymbol{\xi })\small{\frac{\mathcal{F} (\boldsymbol{\xi })}{w(\boldsymbol{\xi })}}\,d\boldsymbol{\xi }}=\sum_n{\small{\frac{\mathcal{W} _n}{w(\boldsymbol{\xi })}}\mathcal{F} (\boldsymbol{\xi }_n)},
	\label{eq:ggjqA}
\end{equation}
where \( \boldsymbol{\xi}_n \) are the integration nodes and \( \mathcal{W}_n / w(\boldsymbol{\xi}_n) \) are the corresponding quadrature weights, listed in Table~\ref{tab:AandW}.

The quadrature in the azimuthal angle \( \theta \) is given by Eq.~(23), while the radial and polar angles can be discretized directly using Python’s scientific computing libraries, as illustrated in Listing~\ref{lst:ggjq_nodes}.

\begin{lstlisting}[language=Python, caption={Example of generating GGJQ velocity nodes and weights in Python.}, label={lst:ggjq_nodes}]
import numpy as np
from scipy.special import roots_jacobi
import matplotlib.pyplot as plt

# Transform Gaussian quadrature nodes and weights
# from the interval [-1, 1] to [0, 1]
def gauss_jacobi(n, a, b):
  roots, weights = roots_jacobi(n, a, b)

  roots = 0.5 * (roots + 1)
  weights /= 2**(a + b + 1)
  return roots, weights

# Gauss quadrature nodes and weights in the radial direction
def r_GGJQ(D, T0, n, alpha, beta):
  a = alpha - 1
  b = D / 2 - 1    
  roots, weights = gauss_jacobi(n, a, b)

  # Compute radial discrete velocities  
  Rr = np.sqrt(alpha * T0 * np.arctanh(roots))

  # Compute radial weights including scaling factors
  fun0 = 0.5 * (alpha * T0)**(D / 2)
  fun1 = (1 - roots)**beta * (1 + roots)
  fun2 = (np.arctanh(roots) / roots)**b
  w_R = weights * fun0 / fun1 * fun2

  return Rr, w_R

# Gauss quadrature nodes and weights in the polar angle direction
def varPhi_GGJQ(n, phi):
  a = 0
  b = phi - 1
  roots, weights = gauss_jacobi(n, a, b)

  varPhi = roots**phi
  w_phi = phi * weights

  return varPhi, w_phi
\end{lstlisting}

\bibliographystyle{model1-num-names}
\bibliography{refs}
\end{document}